%% file: main_v2.tex
\documentclass[aps, prx, reprint, superscriptaddress, longbibliography, floatfix]{revtex4-2}

\usepackage{amsmath}
\usepackage{amssymb}
\usepackage{booktabs}
\usepackage{physics}
\usepackage{tikz}
\usepackage{quantikz}
\usepackage{hyperref}
\usepackage{comment}
\hypersetup{colorlinks=true, citecolor=blue, urlcolor=blue, linkcolor=blue, breaklinks=true}
\usepackage{makecell, array}

\date{Today}

\begin{document}

\title{Transversal architecture for megaquop-scale quantum simulation \\ 
with neutral atoms}

\author{Refaat Ismail}
\thanks{These authors contributed equally. Ordering by coin flip.\\
$^\dagger$Email address:
\href{mailto:czhao@quera.com}{czhao@quera.com};
\href{mailto:swang@quera.com}{swang@quera.com};
\href{mailto:sornborg@lanl.gov}{sornborg@lanl.gov};
\href{mailto:mkornjaca@quera.com}{mkornjaca@quera.com}}
~\affiliation{QuEra Computing Inc., 1284 Soldiers Field Road, Boston, MA, 02135, USA}
~\affiliation{Department of Physics and Astronomy, University of Kentucky, Lexington, KY, 40503 USA}

\author{I-Chi Chen}
\thanks{These authors contributed equally. Ordering by coin flip.\\
$^\dagger$Email address:
\href{mailto:czhao@quera.com}{czhao@quera.com};
\href{mailto:swang@quera.com}{swang@quera.com};
\href{mailto:sornborg@lanl.gov}{sornborg@lanl.gov};
\href{mailto:mkornjaca@quera.com}{mkornjaca@quera.com}}
~\affiliation{Los Alamos National Laboratory, Computer and Computational Sciences Division, Los Alamos, NM, USA}
~\affiliation{Department of Physics and Astronomy, Iowa State University, Ames, IA 50011, USA}

\author{Chen Zhao$^{\dagger}$}
~\affiliation{QuEra Computing Inc., 1284 Soldiers Field Road, Boston, MA, 02135, USA}

\author{Ronen Weiss}
~\affiliation{Department of Physics, Washington University in St. Louis, St. Louis, Missouri, 63130, USA}

\author{Fangli Liu}
~\affiliation{QuEra Computing Inc., 1284 Soldiers Field Road, Boston, MA, 02135, USA}

\author{Hengyun Zhou}
~\affiliation{QuEra Computing Inc., 1284 Soldiers Field Road, Boston, MA, 02135, USA}

\author{Sheng-Tao Wang$^{\dagger}$}
~\affiliation{QuEra Computing Inc., 1284 Soldiers Field Road, Boston, MA, 02135, USA}

\author{Andrew Sornborger$^{\dagger}$}
~\affiliation{Los Alamos National Laboratory, Computer and Computational Sciences Division, Los Alamos, NM, USA}

\author{Milan Kornja\v ca$^{\dagger}$}
~\affiliation{QuEra Computing Inc., 1284 Soldiers Field Road, Boston, MA, 02135, USA}

\date{\today}

\begin{abstract}

Quantum computing experiments have made remarkable progress in demonstrating key components of error-corrected quantum computing, a prerequisite for scalable quantum computation. While we anticipate the near-term arrival of early fault-tolerant quantum hardware capable of a million reliable quantum operations, the cost of preparing low-noise `magic resource states', which are necessary for performing universal quantum computation, presents a formidable challenge. The recently proposed partially-fault-tolerant architecture based on a space-time efficient analog rotation (STAR) approach attempts to address this challenge by using post-selection to prepare low-noise, arbitrary small-angle magic states. Its proposed physical implementation, however, assumes fixed qubit connectivity, resulting in implementation costs closer to leading fully-fault-tolerant approaches. In this work, we propose the transversal STAR architecture and co-design it with neutral-atom quantum hardware, deriving explicit and significant savings in logical layout, time, and space overhead. Through detailed circuit-level simulations, we derive the logical noise model for surface-code-based transversal STAR gadgets and verify their composability. At its limit, the transversal STAR architecture with neutral atoms can efficiently simulate local Hamiltonians with a total simulation volume exceeding 600, defined as the product of the number of logical qubits and Hamiltonian evolution timescales. Achieving this limit would require approximately 10,000 physical qubits at a physical error rate of $10^{-3}$. This is equivalent to a fully-fault-tolerant computation requiring over $10^6$--$10^7$ $T$ gates and a potential $>100\times$ saving on space-time volume for equivalent simulation over current, state-of-the-art protocols. Finally, we extend the transversal STAR architecture to high-rate quantum low-density parity-check (qLDPC) codes, demonstrating how a limited set of highly parallel transversal Clifford gates and generalized small-angle magic injection can be utilized for effective quantum simulation. We anticipate that the co-designed transversal STAR architecture could substantially reduce the physical resources necessary for early-fault-tolerant quantum simulation at the megaquop scale.
\end{abstract}

\maketitle

\section{Introduction} 
\label{sec:intro}

Quantum error correction (QEC) is a prerequisite for scalable quantum computation, allowing for the exponential suppression of errors at the cost of polylog space-time overhead~\cite{Shor1996, Campbell2017}. Recently, pioneering experiments have demonstrated individual key components of error correction and early computation with logical qubits~\cite{Bluvstein2024, Rodriguez2024, Paetznick2024, Vandam2024, Acharya2025, Bluvstein2025b}. Experimental advances have led to the anticipation of the new million-quantum-operation (``megaquop") era ~\cite{Preskill2018, Preskill2025}, providing resources adequate for material science and non-equilibrium quantum simulation applications~\cite{Campbell2021, Daley2022, Lin2022, Weiss2024}. Despite the hopeful outlook, the road to ``megaquop" hardware presents formidable challenges in terms of the space-time overhead of error-corrected quantum computation. 

At its core, two significant overheads challenge near-term megaquop-scale quantum applications: (1) the cost of preparing low-noise magic resource states necessary for universal quantum computation~\cite{Zhou2000, Eastin2009, Yoder2016, Campbell2017, Beverland2021} and (2) the synthesis of small-angle magic states from discrete ones~\cite{Ross2016, Kliuchnikov2016, Yin2025}. The preparation overhead arises because practical, universal quantum computation relies on magic states produced through distillation~\cite{Bravyi2005} or cultivation~\cite{Gidney2024, Chen2025} protocols. Distillation consumes many noisy logical inputs and introduces substantial costs in both storage and circuit depth~\cite{Zhou2000, Eastin2009, Yoder2016, Campbell2017, Beverland2021, Bravyi2005}. Even recent post-selection-based cultivation protocols~\cite{Gidney2024, Chen2025} still remain at least an order of magnitude less space-time efficient than the simplest non-magic-based logic --- transversal Clifford gates~\cite{Bravyi2013, Zhou2024}. The second overhead emerges for commonly proposed quantum applications that rely on Hamiltonian simulation, which natively require small-angle resource states. Since current protocols are optimized for large-angle discrete magic states, synthesizing the small-angle states from discrete resources introduces another order of magnitude in overhead~\cite{Ross2016, Kliuchnikov2016, Yin2025}.

A plausible solution for practical low-noise quantum simulation has recently been proposed in the Space-Time efficient Analog Rotation (STAR) architecture~\cite{Akahoshi2024, Toshio2024, Akahoshi2024b}. The essence of the proposal lies in the hybrid approach: performing more readily implementable Clifford gates with a fault-tolerant approach, while providing for non-fault-tolerant, but sufficiently low-noise, small-angle magic resource states. Critically, it relies on practical post-selection-based injection protocols that suppress the error rate of small-angle magic states directly proportional to the angle, $p_{L}(\theta) \sim |\theta| p_{\mathrm{ph}}$~\cite{Choi2023,Gavriel2023, Toshio2024, He2025}, where $p_{L}$ is the logical error rate, $p_{\mathrm{ph}}$ is the dominant physical error rate, and $\theta$ is the small rotation angle. While the linear scaling with the leading physical error limits the ultimate capabilities of the STAR approach, the angle suppression of the logical error rate hints at the potential for megaquop-scale quantum simulation applications~\cite{Akahoshi2024b}. Despite its promise, the cost and practical implementation potential of the proposed STAR architecture is, so far, limited by the singular experimental and QEC setting it was designed for --- 2D arrays of superconducting qubits with limited connectivity performing surface code QEC~\cite{Bravyi1998, Fowler2012, Stephens2014} with lattice surgery Clifford operations~\cite{Horsman2012, Litinski2019}. Given recent $T$-cultivation advances~\cite{Gidney2024}, $T$ gates of equivalent noise rate are available with an overhead similar to lattice surgery CNOTs, diminishing the appeal of the original fixed-connectivity STAR, which on average requires two lattice surgery CNOTs per teleportation. However, the wide variety of quantum hardware platforms with diverse capabilities and the QEC design space provide untapped opportunities for STAR architecture improvement through co-design and thus for fast-tracking megaquop-scale operation. 

Here, we propose the \textit{transversal STAR architecture} based on transversal Clifford operations with quantum simulation applications as a target and co-design the architecture with neutral-atom hardware. Neutral atom platforms~\cite{Bernien2017, Ebadi2021, Bluvstein2022, Wurtz2023, Chiu2025} are especially well-suited for early exploration of fault-tolerant quantum computing, as evidenced by successful demonstrations of transversal logical encoding, gates, logical sampling circuits, and magic state distillation~\cite{Bluvstein2024, Rodriguez2024, Bluvstein2025b}. The success of early logical demonstrations is anchored upon reconfigurable physical connectivity and a large degree of parallelism, core features that also influence our transversal STAR co-design methodology.  

The reconfigurable long-range connectivity~\cite{Bluvstein2022, Evered2023, Bluvstein2024, Bluvstein2025b}, together with correlated decoding~\cite{Cain2025}, allows for fast implementations of fold-transversal logical Clifford gates~\cite{Breuckmann2024, Zhou2024}. This, in turn, enables us to match the efficiency of transversal STAR magic state injection protocols with fast Clifford gates, solving the main lattice-surgery bottleneck of the fixed-connectivity STAR proposal~\cite{Toshio2024}. In order to assess transversal STAR capabilities, we perform detailed circuit-level simulations for both Clifford and magic angle injection and teleportation protocols for the surface-code case. These result in a logical gadget noise model for surface-code transversal STAR implementation, which we employ for quantifying space and time resource savings and, notably, verifying the composability of the relevant logical gadgets.  Describing the utility limits of the transversal STAR architecture, we estimate the total simulation volume of local Hamiltonians (the product of the number of modes or logical qubits and the number of characteristic Hamiltonian evolution timescales) in excess of 600. Achieving this scale would require approximately $10{,}000$ physical qubits at a physical error rate of $10^{-3}$ and correspond to a fully-fault-tolerant computation involving $\sim 10^7$ $T$-gates.

In addition, the hardware flexibility of neutral-atom platforms eliminates the routing overhead inherent in lattice-surgery-based STAR implementations, yielding additional space savings and also enabling low-overhead implementation of customized protocols. We showcase protocol customization on examples of distance-preserving fold-transversal $S$-gate~\cite{Chen2024} and straightforward transversal multi-rotation small-angle injection schemes~\cite{Gavriel2023, Toshio2024}. Moreover, the high degree of parallelism and dynamic long-range connectivity motivate us to extend transversal STAR to architectures based on high-rate quantum low-density parity-check (qLDPC) codes~\cite{Breuckmann2021, Lawrence2022, Quintavalle2023, Bravyi2024, Xu2024}. Specifically, we demonstrate how restricted, highly parallel transversal Clifford gates and generalized small-angle magic-state-injection protocols can be leveraged for efficient quantum simulation.

\section{Transversal STAR architecture overview}
\label{sec:STARoverview}

The STAR architecture, as originally proposed for superconducting quantum hardware~\cite{Akahoshi2024}, relies on the mix of short-distance fault-tolerant Clifford gates and non-fault-tolerant injection and teleportation of arbitrary angle rotations, while utilizing the surface code. For this architecture, the offline nature of magic generation allows the consideration of multiple post-selection-based protocols. 

Initially in Ref.~\cite{Akahoshi2024}, an angle-independent protocol was proposed that begins by injecting a magic state into a small error-correcting code block, which is then expanded to the final code distance through patch growth.
The protocol achieved an error rate of $2p_{\mathrm{ph}}/15$~\cite{Akahoshi2024} for magic angle injection, where $p_{\mathrm{ph}}$ is the total physical error rate in the symmetric Pauli noise model. Subsequently, a significant improvement for small-angle magic states, relevant for quantum simulation, has been proposed based on a generalization of the Choi et.~al.~\cite{Choi2023} protocol~\cite{Toshio2024}. The fidelity of the teleported small-angle rotation with the improved protocol scales linearly with angle and physical error rate $p_L(\theta)\approx \alpha |\theta|p_{\mathrm{ph}}$, where the prefactor $\alpha \approx 0.2d$ for two-qubit rotation-based injection protocols for the distance-$d$ surface code. This linear scaling of the logical error rate with the rotation angle has opened significant potential for quantum simulation applications, especially with algorithms based on Trotterized Hamiltonian simulation, as explored in Ref.~\cite{Akahoshi2024b, Toshio2024} targeting early-fault-tolerant quantum hardware. 

On one hand, error-corrected Cliffords allow the STAR architecture to overcome typical limitations of NISQ quantum hardware by extending Clifford circuit depth. Indeed, for high enough code distances, STAR puts a resource limitation only on the total rotation angle performed in a circuit, $\sum_a|\theta_a| \lesssim (\alpha p_{\mathrm{ph}})^{-1}$. In this regime, the STAR architecture still provides a significant advantage over standard fault-tolerant approaches by supplying comparatively resource-efficient small-angle magic that suffers no synthesis overhead. In contrast, magic state factories require sizable space-time volume~\cite{Campbell2017}. Magic state cultivation~\cite{Gidney2024} brings down these costs significantly for moderate-fidelity magic through extensive post-selection, but is likely still an order of magnitude more costly~\cite{Gidney2024, Chen2025} in terms of space-time volume compared to simple STAR injection. The second factor that compounds the total magic cost of the standard fault-tolerant approaches is the cost of small-angle synthesis. Small angles are ubiquitous in quantum simulation, but the typical cost of Solovay-Kitaev angle synthesis is $\sim 3\log_2 (1/\epsilon)$, where $\epsilon$ is the desired angular accuracy. In typical early-fault-tolerant applications, angle synthesis brings an additional factor of $10$--$50$ to a fully-fault-tolerant approach. 

\begin{figure*}[!htb]
\centering
\includegraphics[width=1.0\textwidth]{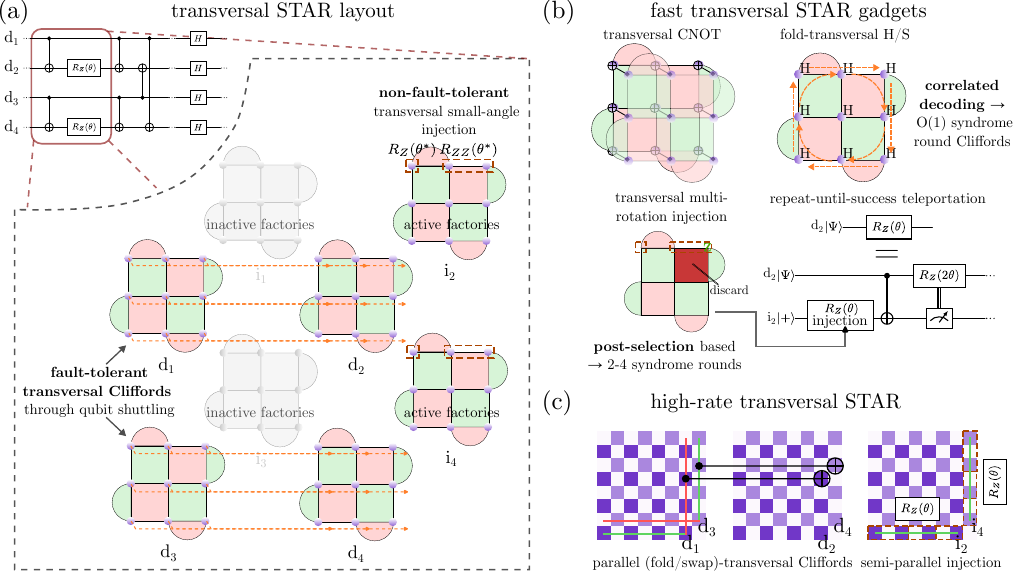}
\caption{\textbf{Overview of transversal STAR architecture.} Our transversal STAR architecture based on neutral atoms combines fast transversal fault-tolerant Cliffords~\cite{Cain2025, Zhou2024} performed by physical qubit shuttling with non-fault-tolerant small-angle magic created with post-selection-based transversal rotation protocols~\cite{Choi2023, Toshio2024}. a) The spatial layout of the surface-code-based transversal STAR contains data ($d_i$) and injection factory qubits ($i_i$), with no extra routing overhead. The layout shown realizes the four-qubit circuit shown in the inset. Inactive injection patches for the current circuit layer are grayed out.  b) Logical gadgets for our transversal STAR architecture include transversal CNOTs and fold-transversal single-qubit Cliffords ($H$ gate shown). The logical Clifford operations are followed by $1\text{--}2$ rounds of syndrome extraction. Successful decoding of this fast Clifford protocol has to account for error correlations arising from the circuit context~\cite{Cain2025}. Small-angle magic injection creates low error rate magic in the transversal post-selection protocol that involves single- and two-qubit rotation and $2\text{--}4$ total syndrome rounds. The injected magic is teleported in the stochastic repeat-until-success protocol, which on average succeeds after $2$ teleportation rounds.  c) The flexibility of the neutral-atom physical architecture and STAR injection protocols makes transversal STAR readily extensible to other classes of stabilizer codes, including higher-rate codes. Here, we show an example of a $[[32,2,4]]$ toric code --- data qubits are dark purple (two shades), ancillas light, while logical $X$/$Z$ operators have support on red/green lines. The gate set of transversal STAR is based on patch-parallel (fold/swap)-transversal Clifford, semi-parallel transversal injections on all the qubits within the patch, and code automorphisms~\cite{eberhardt2024}. This gateset enables effective co-design between the simulation problem instance and the QEC code.
}
\label{fig:Fig1_STAR}
\end{figure*}

While the original STAR proposal opens up significant utility potential in the early-fault-tolerant-hardware era, the ultimate utility of the architecture will depend on its careful hardware-aware evaluation and co-design. In particular, the limitations of fixed, planar qubit connectivity and surface-code-based computation leave plentiful space for improvements utilizing more flexible hardware architectures~\cite{Bluvstein2024, Moses2023} and more efficient quantum error-correcting codes~\cite{Breuckmann2021}. In addition, the original STAR proposals consider an idealized hardware-agnostic noise model, which directly begs the question of expected performance with more hardware-realistic noise. Here, we design and evaluate a transversal STAR architecture co-designed with neutral atoms addressing both the opportunity for sizable resource savings provided by hardware flexibility and the possibility for a more realistic evaluation through a hardware-derived physical noise model. We note, however, that the transversal STAR architecture and derived space-time advantages also apply to other hardware platforms with reconfigurable connectivity, provided sufficient operational parallelism is available.

\subsection{Design and advantages of transversal STAR architecture}
\label{sec:stardesign}

The transversal STAR architecture mirrors the original STAR in its mixed approach with fault-tolerant Cliffords and non-fault-tolerant small-angle magic. Our proposed concrete implementations are optimized for neutral-atom hardware, as shown in Fig.~\ref{fig:Fig1_STAR}. The flexible hardware connectivity allows us to consider transversal entangling Clifford gates implemented through qubit shuttling~\cite{Bluvstein2024}, shown in Fig.~\ref{fig:Fig1_STAR}(a). For the surface code implementation, this allows us to forgo the space overhead typically associated with the logical qubit layout, where ancilla patches are used to route the corresponding lattice surgery~\cite{Fowler2019} operations. The transversal STAR layout is simple: it contains only the data qubits directly required for the algorithm and factory qubits for injection. The relative density between the two qubit groups can be optimized for a given algorithm, with an additional flexibility of free dynamic reassignment of injection factories to any data qubit in the array through shuttling. Due to the possibility of eliminating routing logical ancillas with qubit shuttling, platforms with reconfigurable connectivity save a factor of $\sim 2$ compared to the original STAR proposal~\cite{Akahoshi2024} in data-qubit space volume, with the exact factor depending on the connectivity requirements of the algorithms. This is critical for the early-fault-tolerant quantum hardware that STAR is designed for, where the physical qubit volume is at a premium. 

In addition to the space savings, time savings can be obtained based on transversal and fold-tranversal logical Cliffords implemented by the physical operations shown in Fig.~\ref{fig:Fig1_STAR}(b), followed by $O(1)$ rounds of syndrome extraction~\cite{Cain2025, Zhou2024}. While the typical lattice-surgery-based logic considered originally for STAR requires $O(d)$ rounds of syndrome extraction, the deterministic nature of transversal logical operations has motivated a novel transversal algorithmic fault tolerance approach~\cite{Zhou2024} that can realize all transversal logic with $O(1)$ syndrome measurement rounds. The prerequisite for this sizable time saving is the ability to tackle correlated logical errors that spread within and between the patches during transversal Cliffords [see Fig.~\ref{fig:Fig1_STAR}(b)]~\cite{Cain2025}. While this increases the total volume of the decoding problem, recent works have shown that the classical decoding can still be performed efficiently~\cite{Cain2025, Cain2025b}. Together, layout space savings and algorithmic fault tolerance time savings provide a significant boost for transversal STAR and motivate us to further evaluate the exact resources through an effective logical noise model stemming from a hardware-derived physical noise model~\cite{Evered2023, Bluvstein2025b,  Rodriguez2024}. Beyond the extraction of the logical Clifford noise model itself, the transversal Cliffords with the correlated decoding approach, where each decoding window includes $O(d)$ past logical operations, naturally pose the important question of noise model composability in the circuit context. We therefore explicitly verify the composability in a deep random Clifford circuit.

In terms of magic injection and teleportation, a transversal multi-rotation injection protocol~\cite{Choi2023, Toshio2024} can be applied straightforwardly, see Fig.~\ref{fig:Fig1_STAR}(a)-(b). The protocol consists of a post-selection-based logical $\ket{+}$ initialization, a pattern of multi-qubit physical control-$Z$ rotations on the set of qubits corresponding to a logical $Z$ operator of the surface code, and the subsequent choice of the desired logical rotation through post-selection heralded by stabilizer values from post-protocol measurements. The dynamical connectivity allows us some simplifications to the protocol --- the transversal multi-rotation injection can be performed without ancillas as the qubits along the logical $Z$ support can be simply reconnected as necessary with qubit shuttling. Equivalently to the original STAR proposal, due to the arbitrary nature of the angles used, teleportation can be performed by a repeat-until-success protocol as each teleportation step has an equal chance of teleporting a $\pm \theta$ angle~\cite{Akahoshi2024}. The error rates of the injection and teleportation protocols are especially sensitive to the exact physical noise model, making the hardware-aware evaluation and subsequent optimization of the injection protocols paramount for assessing the practical utility of the transversal STAR architecture. Post-selection-based injection, after taking into account concrete post-selection rates achieved in the neutral atom setting, is a fast protocol requiring, on average, 2\text{--}4 syndrome rounds per injected magic state. The protocol speed thus uniquely matches our fast transversal Cliffords, for the typical Clifford-to-magic ratios required for quantum simulation.

Beyond the transversal multi-rotation injection simplification, reconfigurable qubit connectivity allows us, in general, to readily adapt protocol details. As a concrete example, we can employ the recently proposed fold-transversal $S$-gate protocols~\cite{Chen2024}; this brings the cost and performance of an $S$ gate in line with other Cliffords, mitigating a weakness in lattice-surgery-based STAR. The flexibility further allows much broader potential gains on the architecture level. We are able to consider versions of transversal STAR architecture beyond the surface code implementation. In particular, we engineer an efficient quantum simulation gate set for transversal STAR applied to special classes of qLDPC error correction codes~\cite{Breuckmann2021, Bravyi2024, Xu2024, Kovalev2013}. The injection protocol of STAR can be applied to any stabilizer code, as shown schematically on a toric code example in Fig.~\ref{fig:Fig1_STAR}(c), with the expected linear error scaling with the teleported angle. The hardware-native Clifford gate set still consists of fold or swap transversal Cliffords~\cite{Breuckmann2024}. These Clifford gates, however, are necessarily parallel to the logical qubits in the whole code patch. This is a significant limitation for general algorithmic compilation; nevertheless, we show that if the code and the gateset are co-designed with the Trotterized Hamiltonian simulation problem at hand, one can efficiently use the parallel transversal gateset. Our construction relies on the match between Hamiltonian locality and the underlying lattice-translation symmetry to the automorphisms of the quantum code. Given that the Trotterized simulation of local Hamiltonians is the main potential application of a STAR architecture, our transversal qLDPC-STAR proposal can potentially drastically reduce the overall physical space-time footprint of quantum simulation. Together, all transversal STAR design optimizations provide potentially $100$--$1000\times$ saving on space-time volume over current, state-of-the-art fully-fault-tolerant protocols.

\subsection{Paper overview}

The rest of the paper is organized as follows. We evaluate in detail the performance of the transversal STAR based on neutral atoms with extensive logical noise model calculations in Sec.~\ref{sec:logicalnoise}. We start with an overview of logical gadgets and the hardware-derived physical noise model in Sec.~\ref{sec:implementation_overview}, while the main results for the Clifford noise model and its composability follow in Sec.~\ref{sec:cliffordnoise}. The performance of small-angle injection and teleportation is presented in Sec.~\ref{sec:sm_inj_noise}. We summarize the limits of surface-code-based transversal STAR utility for Trotterized Hamiltonian dynamics in Sec.~\ref{sec:disc_utility}, while the details of transversal qLDPC-STAR construction are discussed in Sec.~\ref{sec:disc_qLDPC}. Finally, we provide an outlook in Sec.~\ref{sec:outlook}.

\section{Logical noise model for transversal STAR architecture}\label{sec:logicalnoise}

In this section, we analyze in detail the performance of the surface-code-based transversal STAR architecture based on neutral atoms. As established previously, the transversal STAR architecture leverages correlated decoding to execute Clifford gates transversally with $O(1)$ rounds of syndrome extraction between them, while implementing small-angle rotation gates through transversal multi-rotation  injection and teleportation protocols ~\cite{Choi2023, Akahoshi2024}. Our primary objective is to enable efficient error estimation for large circuits constructed with this architecture. Specifically, we aim to characterize the performance of qubits encoded in rotated surface codes running on neutral atom hardware~\cite{Bluvstein2022, Bluvstein2024, Rodriguez2024}. To facilitate this, we develop a method of extracting a reliable logical noise model characterized by an effective Pauli channel for each logical gate, allowing us to rapidly estimate circuit-level error rates through the composition of these individual error channels without needing full physical-level simulation.

The key challenge lies in developing this model within the correlated decoding framework, where the standard assumptions about error propagation and gate isolation break down. In bare circuits without any error correction, error estimation is straightforward: for a circuit composed of $N$ gates where each gate $i$ has error probability $p_i$, the total circuit error probability approximates to $p_{\text{total}} \approx \sum_{i=1}^{N} p_i$ (assuming independent errors and $p_i \ll 1$). Conventional fault tolerance alters this picture by actively removing errors before they affect the logical computation. Each logical gate is followed by $d$ rounds of syndrome extraction (where $d$ is the code distance), creating temporal isolation that allows one to treat each gate and its associated syndrome extraction as an independent ``gadget". Since the decoder operates approximately independently on each gadget—protecting against error chains up to weight $\lfloor d/2 \rfloor$—we can characterize each gadget with an independent logical error channel, and the total circuit error approximately becomes the sum of logical errors from all gadgets.

Correlated decoding~\cite{Cain2025, Cain2025b} introduces a paradigm shift: syndromes from different qubits are decoded together, with $O(1)$ rounds of syndrome extraction per logical gate independent of the code distance. This approach eliminates the natural temporal isolation of gates present in conventional schemes, introducing correlations that violate the standard independence assumptions. This raises some critical questions: can logical circuits with correlated decoding still be decomposed into manageable gadgets with independent error channels for the purpose of logical error estimation? If such decomposition is valid, what methods can effectively characterize the noise channel of the logical gadgets despite the correlations introduced by global decoding?

This section proceeds as follows: We first present the physical implementation of Clifford gates and our hardware-derived noise models. We demonstrate how to extract logical noise models for Clifford gates despite correlations due to correlated decoding and we validate our approach through systematic analysis. Finally, we show logical noise model results for small-angle rotation gates. Together, this demonstrates that neutral atom hardware can achieve the full potential of the transversal STAR architecture.

\subsection{Physical implementation and simulation setup}
\label{sec:implementation_overview}
\subsubsection{Logical qubit encoding and gate implementation}

We now turn to presenting implementation choices for the transversal STAR architecture, including the encoding scheme, gate implementations, noise modeling, and decoding strategy. We encode logical qubits in rotated surface code patches and implement Clifford gates transversally using the generating set $\{H$, $S$, CNOT$\}$. As shown in Fig.~\ref{fig:Fig1_STAR}(b), CNOT is implemented by applying physical CNOTs between corresponding data qubits of two logical patches, while Hadamard requires transversal $H$ gates on all data qubits followed by a patch rotation. The $S$ gate uses a fold-transversal implementation recently developed in Ref.~\cite{Chen2024}, which preserves the code distance and completes the set of distance-preserving transversal implementations for all Clifford generators in rotated surface codes. In physical implementation, we compile everything in terms of CZ gates as these are the native device gates. Full implementation details are provided in Appendix \ref{sec:Cliffords_implementation_details}.

For syndrome extraction (SE), we implement one round after each logical gate. Numerical benchmarks in Appendix \ref{sec:SE_benchmark} demonstrate that this choice achieves optimal error suppression for our gadgets (defined in the next section) while reducing computational runtime compared to multiple rounds. This extends previous results from the full-circuit level \cite{Cain2025, Zhou2025}, where one SE round per gate was shown to minimize overall circuit error rates for transversal Clifford circuits. Here, we validate this finding at the individual gadget level.

\subsubsection{Neutral-atom-hardware-derived noise model}
\label{sec:hardware_noise_model}

Our simulations employ a circuit-level noise model derived from experimentally characterized error mechanisms in current neutral-atom quantum processors~\cite{Rodriguez2024, Bluvstein2024, Bluvstein2025b}. The model is designed to capture the dominant physical noise processes relevant to neutral-atom hardware while remaining amenable to efficient classical simulation through Pauli error channels.

In particular, we incorporate several key features that distinguish neutral-atom platforms. These noise processes are motivated by the zoned neutral-atom architecture demonstrated in Ref.~\cite{Bluvstein2024}, in which atoms are dynamically transported between storage, entangling, and readout regions while maintaining coherence. While such architectures enable flexible connectivity and efficient control, they also introduce characteristic error channels, particularly transport- and control-related. First, we include $Z$-biased noise strongly during both single- and two-qubit gate operations, reflecting the dominant dephasing and control errors observed experimentally~\cite{Evered2023}. Second, we account for transport-induced errors arising from the physical movement of atoms required to realize two-qubit interactions. Third, we model spectator errors caused by global laser pulses used to implement entangling gates ($CZ$), which can inadvertently affect other qubits present in the entangling region. Finally, we include atom loss events occurring during Rydberg-mediated gate operations, which arise when optical trapping potentials are temporarily disabled~\cite{Evered2023, Bluvstein2024}. In our simulations, we model these loss events using an effective Pauli-$Y$ error channel, yielding a conservative, implementation-agnostic upper bound on the logical error rate. The rationale for modeling atom loss using an effective Pauli-$Y$ channel, as well as a quantitative validation of the resulting logical error-rate bound against explicit-loss simulations~\cite{baranes2025}, is presented in Appendix~\ref{sec:loss_bound_validation}.

We parameterize the overall noise strength by the total error rate associated with CZ gate implementation, defined in Eq.~\eqref{eq:physical_error_rate}, which includes both the intrinsic gate error and the accompanying transport error. This provides a convenient and experimentally meaningful metric for comparing different noise regimes. For current-generation hardware, this corresponds to an effective physical error rate of approximately $p_{\mathrm{ph}} \approx 7.3 \times 10^{-3}$. A complete specification of all error channels, rates, and modeling assumptions is provided in Appendix~\ref{sec:physicalnoise}.

\subsubsection{Decoding strategy}

We decode all circuits using the recently developed Minimum Weight Parity Factor (MWPF) decoder \cite{wu2025mwpf}. For MWPF, the key hyperparameter that balances speed and accuracy is the cluster node limit, where 0 provides the fastest decoding and infinity provides optimal accuracy. For our circuits, we set the cluster node limit to 500 to optimize the balance between error suppression and computational feasibility for our target code distances $d$ = $ \{ 3, 5, 7\}$ across large sample sizes (up to $5 \times 10^7$ shots). MWPF is designed to handle errors that trigger more than two detectors (hyperedges in the decoding graph), which commonly arise in transversal Clifford circuits. The decoder attempts to certify optimality while maintaining practical scalability. Appendix \ref{sec:decoders} presents comprehensive benchmarks comparing MWPF against alternative decoding algorithms, including belief propagation variants (BP-OSD, BP-LSD) \cite{Roffe_2020, Roffe_LDPC_tools, delfosse2021} and the recently developed logical matching decoder \cite{Cain2025b, serraperalta2025}, which extends standard matching to arbitrary surface-code transversal Clifford circuits. 

Our benchmarks reveal trade-offs between error suppression and computational efficiency. While logical error rates vary by only one order of magnitude across all decoders, computational costs span several orders of magnitude. The runtime advantage of matching-based approaches becomes increasingly pronounced with the code distance: at near-term physical error rates $p_{\mathrm{ph}} = 7.3 \times 10^{-3}$, the gap between logical matching and the next fastest decoder, MWPF with cluster node limit 0, grows from one order of magnitude at $d=3$ to three orders of magnitude at $d=7$. This scaling suggests that logical matching may be the only computationally viable option for distances beyond $d=7$ among current decoders.

Notably, MWPF variants with lower cluster node limits (0 and 50) outperform both BP-OSD and BP-LSD in both error suppression and runtime, offering superior performance without compromise.  We emphasize that our comparisons use unoptimized BP-OSD and BP-LSD with default hyperparameters as configured in the LDPC package \cite{Roffe_LDPC_tools}. Note that BP-OSD on non-topological codes might perform better as shown in \cite{wu2025mwpf}. Also, note that these results are specific to the transversal Clifford circuits studied here, and decoder performance may be circuit dependent. Future work could explore both this circuit dependence and potential improvements to logical matching through belief propagation preprocessing, following the success of belief-matching approaches in improving thresholds for standard surface codes \cite{higgott2023}.

\subsection{Logical noise model for Clifford gates}
\label{sec:cliffordnoise}

\subsubsection{Overview of logical noise model extraction}
\label{sec:extraction}

To characterize the behavior of logical Clifford gadgets inside deep fault-tolerant circuits, we extract an effective logical noise model for each gadget. In our framework, a \emph{gadget} consists of a logical Clifford gate followed by a constant number of syndrome extraction (SE) rounds, independent of the code distance. This construction reflects the fundamental building blocks appearing repeatedly in deep Clifford circuits.

Our goal is to represent each gadget by an equivalent Pauli channel acting after an ideal logical gate, capturing its net effect on logical information when used in a correlated decoding setting. Two challenges must be addressed to achieve this: (i) gadgets with only a constant number of SE rounds do not, by themselves, provide sufficient syndrome data for reliable decoding, and (ii) gadgets located near temporal circuit boundaries can behave differently from those embedded deep within a circuit.

We overcome these challenges by embedding the target gadget into specially designed test circuits that mimic deep-circuit conditions and amplify the gadget’s error signal through repetition. By comparing the behavior of circuits with and without the target gadget block, we isolate the logical error contribution of the gadget and fit it to an effective Pauli noise model. This procedure enables a consistent and reproducible extraction of logical noise parameters for all Clifford gates considered in this work.

The full extraction protocol—including circuit constructions, stabilizer-based error identification, and the linear solver used to infer Pauli channel parameters—is described in detail in Appendix~\ref{app:extraction}. Figure~\ref{fig:Fig2_noise_circuit_sketch} provides a schematic overview of the extraction procedure.

\begin{figure}[t]
\centering
\includegraphics[width=0.5\textwidth]{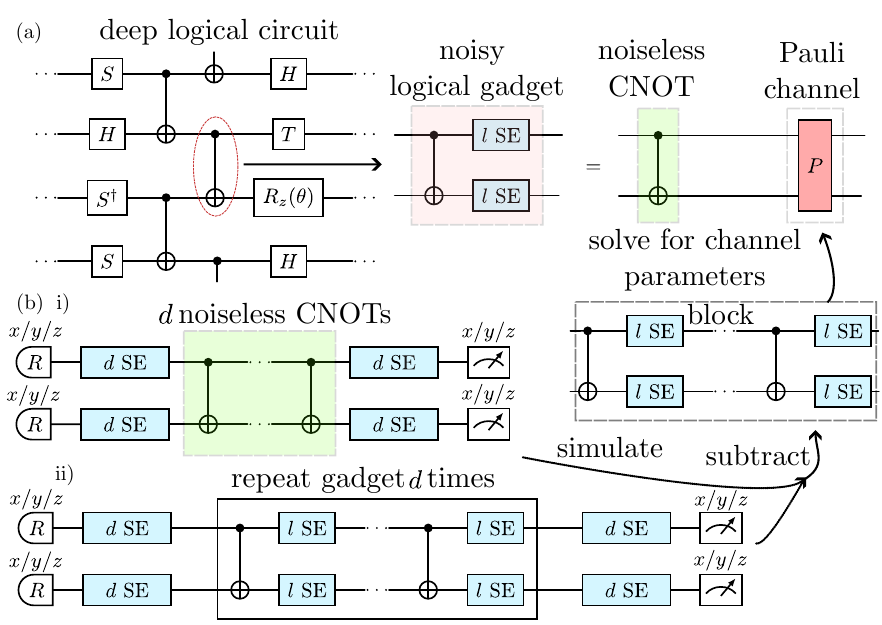}
\caption{\textbf{Extraction of logical noise model in correlated decoding setting.} The process of extracting a logical noise model for logical gadgets in deep Clifford circuits is illustrated. (a) \textit{Gadget noise modeling:} A deep Clifford circuit (left) contains logical gates that, when implemented with fault-tolerant protocols, become logical gadgets consisting of the physical-level implementation of the gate followed by $l$ rounds of syndrome extraction (SE), where $l$ is independent of the code distance $d$. To extract a logical noise model,  we model the effect of gadget noise on the overall circuit using a noiseless gate followed by a Pauli channel, that effectively captures the gadget's noise behavior in deep Clifford circuit contexts. (b) \textit{Extraction protocol:} Two circuits are used for parameter extraction. Circuit (i) serves as the base circuit: two logical qubits are initialized in $X$, $Y$, or $Z$ basis, followed by $d$ rounds of SE, $d$ noiseless CNOT gates, another $d$ rounds of SE, and measurement in the appropriate basis. Circuit (ii) replaces the $d$ noiseless CNOT gates with $d$ repetitions of the logical gadget (CNOT + following SE), capturing the base circuit noise plus the noise from the $d$-repetition block. Full simulation using hardware-aware physical-level noise models is performed on both circuits. By subtracting the noise characteristics of circuit (i) from circuit (ii), the noise contribution of the $d$-repetition block is isolated. This information is then used to solve for the Pauli channel parameters, completing the logical noise model extraction for the gadget (see Appendix~\ref{app:extraction} for more details).
}
\label{fig:Fig2_noise_circuit_sketch}
\end{figure}

\subsubsection{Clifford gadget noise}
\label{sec:clfford_noise}

\begin{figure}[t]
\centering
\includegraphics[width=1.0\columnwidth]{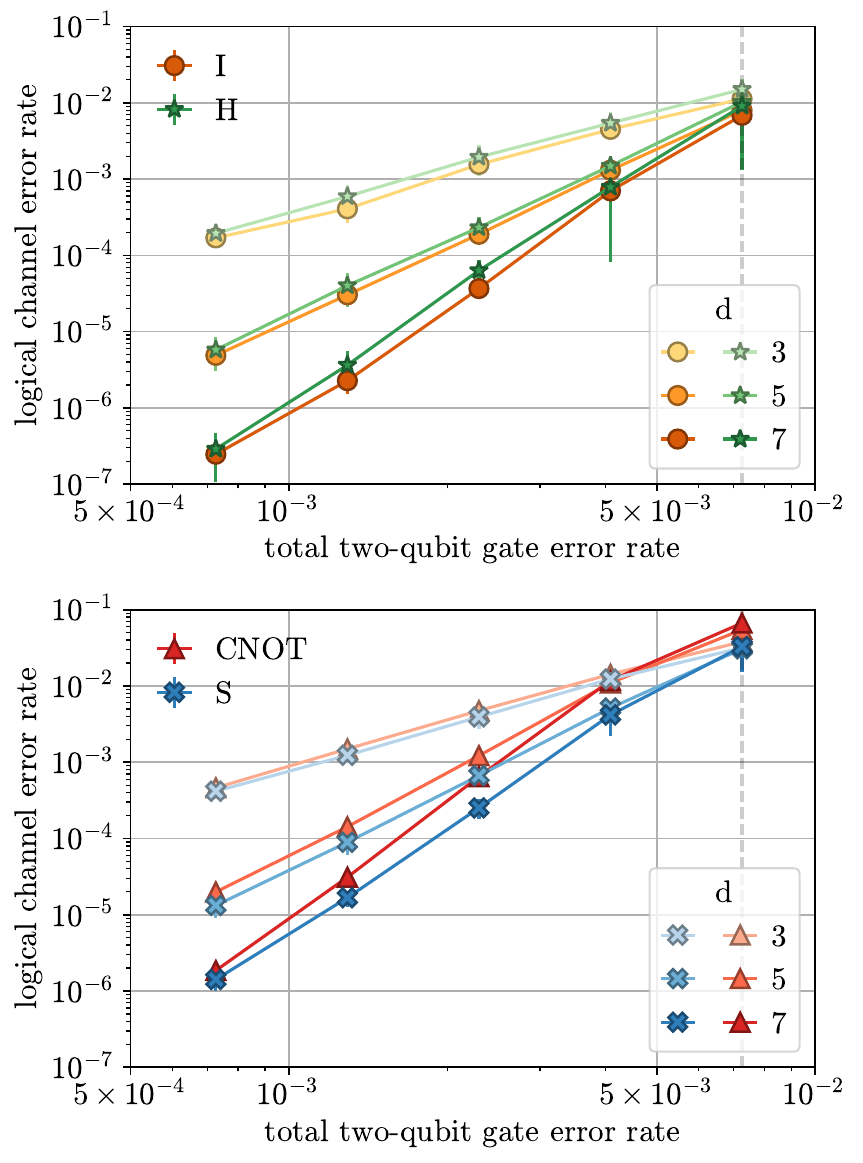}
\caption{\textbf{Logical error rates of transversal Clifford gadgets.} Extracted Pauli channel error rates for four surface-code-encoded Clifford gadgets ($I$, $H$, $S$, CNOT) across code distances $d \in \{3,5,7\}$ using a neutral-atom hardware-aware noise model. Physical error rates $p_{\mathrm{ph}}$ span from current neutral atom capabilities ($7.3 \times 10^{-3}$, grey dashed line) to anticipated $10\times$ improvements~\cite{Bluvstein2025b}. The x-axis shows the total error rate associated with physical two-qubit gate operation (this accounts for various implementation details, see Sec.~\ref{sec:hardware_noise_model}); the y-axis shows the logical channel error rate, which is calculated as the sum of channel parameters. Logical channels are extracted as described in Fig.~\ref{fig:Fig2_noise_circuit_sketch}. Full extracted channel parameters are provided in Table~\ref{tab:Tab2_logical_noise} in the appendix.
}
\label{fig:Fig3_Clifford_noise}
\end{figure}

We apply the above extraction protocol to four transversal Clifford gadgets: $\{I$, $H$, $S$, CNOT$\}$. Figure~\ref{fig:Fig3_Clifford_noise} presents the extracted logical error rates across code distances $d \in \{3,5,7\}$ and physical error rates ranging from current neutral atom capabilities (total effective two-qubit gate error of $p_{\mathrm{ph}} = 7.3 \times 10^{-3}$) to anticipated $10\times$ improvements in future devices, as discussed in the recent experimental work~\cite{Bluvstein2025b}. Each gadget implementation includes one SE round following the gate operation. For the $S$ gate, we employ the Intra-SE implementation (Fig.~\ref{fig:Fig11_transversal_gates}), where the gadget consists of an SE round with the gate applied mid-extraction plus an additional SE round. All results use the MWPF decoder~\cite{wu2025mwpf} with the cluster node limit hyperparameter set to 500.

Several key observations emerge from these results. Most notably, distance-7 surface code achieves logical error rates approaching $10^{-6}$ at the anticipated $10\times$ improved physical error rates, demonstrating the potential for early fault-tolerant quantum computation. All gadgets exhibit clear error suppression with increasing code distances below a threshold near $10^{-2}$, consistent with our expectations. Additionally, distance-7 encoding achieves a break-even point near $3 \times 10^{-3}$ where logical gadgets outperform their unencoded counterparts (not accounting for potential improvements for unencoded gates from error mitigation techniques), marking the regime where fault tolerance becomes advantageous.

The logical gadgets demonstrate a clear performance hierarchy reflecting their underlying complexity. The identity and Hadamard gadgets show the lowest error rates, with syndrome extraction operations dominating their error budgets. The $S$ gadget exhibits higher error rates due to correlations between $X$ and $Z$ checks that create challenging hyperedge errors, while the CNOT gadget shows the highest error rates from both complex hyperedge decoding and errors measured across two logical qubits. We verify the accuracy of these extracted error rates through circuit composition in the following subsection.

We note an apparent artifact in the CNOT gadget results where the $d=5$ and $d=7$ curves intersect at lower physical error rates compared to other gadgets. This behavior likely stems from decoder limitations near threshold rather than fundamental properties of the gadget. Near threshold, the MWPF decoder requires substantially higher cluster node limits to maintain performance, particularly for CNOT circuits which operate over two logical qubits and naturally generate larger clusters than single-qubit gadgets. Combined with the increased syndrome density near threshold, this leads to clusters that may exceed our chosen limit of 500 nodes, causing performance degradation. While increasing the cluster node limit would likely address this issue, it results in prohibitive runtime increases. We verified that this artifact diminishes when using alternative decoders such as logical matching or BP-based decoders. However, these decoders do not achieve the same level of error suppression at low physical error rates. Since our primary focus is to evaluate near-term devices with anticipated $10\times$ improved error rates ($p_{\mathrm{ph}} \lesssim 10^{-3})$—where MWPF with a cluster node limit 500 achieves the best performance, approaching $10^{-6}$ logical error rates with manageable runtimes—we maintain this decoder configuration throughout our analysis.

\subsubsection{Logical noise composability}
\label{sec:compose}

\begin{figure*}[t]
\centering
\includegraphics[width=1.0\textwidth]{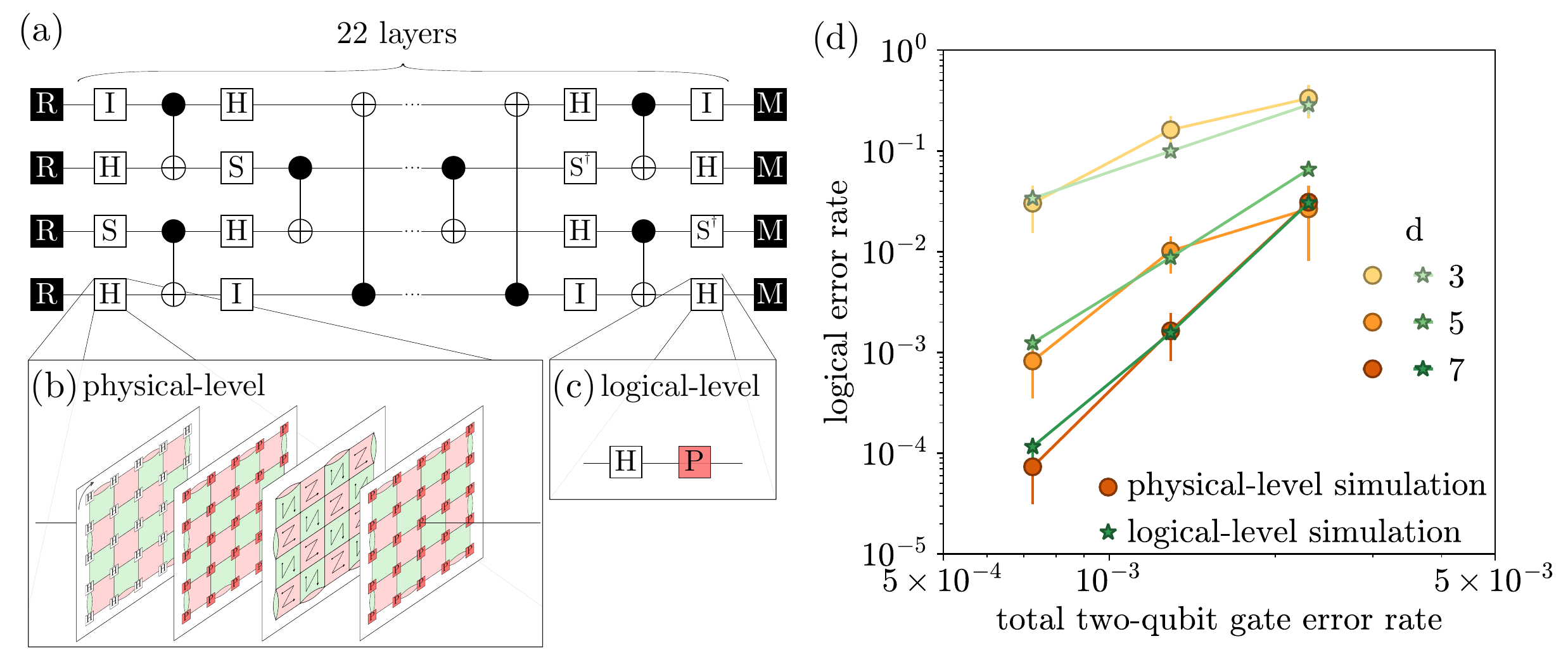}
\caption{\textbf{Clifford logical noise model composability in the correlated decoding setting.} The composability of transversal Clifford gadget noise model is tested by comparing logical- and physical-level simulations on a deep Clifford circuit. (a) Test circuit consisting of 4 logical qubits with 22 layers of random Clifford gates. Each layer consists of nearest-neighbor CNOTs (alternating between pairing each qubit with its left/right neighbor) followed by single-qubit gates randomly chosen from \{$I$, $S$, $H$\} on each qubit. The circuit applies a sequence of Clifford operations followed by their inverse, returning the system to a product state that enables deterministic, uncorrelated measurements on individual qubits. (b) Physical-level simulation approach: for each gate in the circuit, we simulate the full surface code implementation—encoding the logical qubit in distance-$d$ patches, inserting physical errors according to the chosen hardware model. (c) Logical-level simulation approach: for each gate in the circuit, we apply the ideal logical gate followed by the corresponding extracted logical error channel obtained via our gadget noise extraction method (Fig.~\ref{fig:Fig2_noise_circuit_sketch}).  (d) Comparison of logical error rates between physical-level (orange) and logical-level (green) simulations as a function of total two-qubit gate error rate, shown for multiple code distances. The strong agreement validates logical noise model composability. 
}
\label{fig:Fig4_Clifford_composability}
\end{figure*}

To verify that our extracted logical noise models accurately predict the behavior of circuits under correlated decoding \cite{Cain2025, Cain2025b, serraperalta2025}, we perform a systematic validation experiment. We construct a test circuit with 4 logical qubits and 22 layers of random Clifford operations, where each layer consists of nearest-neighbor CNOTs followed by randomly selected single-qubit gates from the set $\{I$, $S$, $H\}$. We compare two simulation approaches: 1) physical-level simulation where each logical qubit is expanded into a full surface code patch of physical qubits, implementing transversal gates, applying the complete hardware noise model, and performing syndrome extraction (computationally intensive but accurate); and 2) logical-level simulation where we operate directly on the 4 logical qubits by applying ideal logical gates followed by our extracted Pauli error channels. The results are shown in Fig.~\ref{fig:Fig4_Clifford_composability}.

Notably, the strong agreement between logical error rates at low $p_{\mathrm{ph}}$ predicted by both approaches across multiple code distances and physical error rates demonstrates that our gadget-level error channels compose correctly to predict overall circuit error rates, validating that logical circuits with correlated decoding can be decomposed into quasi-independent gadgets despite their joint syndrome processing. This enables rapid error estimation for large circuits with hundreds or thousands of logical gates that would be computationally prohibitive to simulate at the physical level—we can extract gadget parameters once and compose them as needed while maintaining reasonable accuracy. The success of this approach shows that gadgets in correlated decoding settings remain composable for error rate prediction—preserving the computational advantages of gadget-based analysis alongside the efficiency and error suppression gains from reduced syndrome extraction.

\subsection{Logical noise model for small-angle rotations}
\label{sec:sm_inj_noise}

To efficiently implement non-Clifford logical operations, we consider non-fault-tolerant small-angle rotations based on the STAR architecture~\cite{Choi2023, Toshio2024}, which introduce limited logical noise. In Ref.~\cite{Choi2023}, the protocol applies transversal single-qubit $R_Z(\theta^*)$ gates to $\ket{+}_L$, followed by postselection to discard error-corrupted outcomes. This effectively realizes a logical rotation gate combined with Clifford operations on $\ket{+}_L$, though the success probability diminishes with increasing code distances. To address this limitation, Ref.~\cite{Toshio2024} instead employs transversal multi-rotation (TMR) gates; the resulting success rate is a function of the number of such gates applied. Here, we utilize this TMR  protocol~\cite{Toshio2024, Choi2023} to inject a small-angle rotation into the logical $\ket{+}_L$ state. We aim to estimate logical error rates for the small-angle injection and teleportation tailored to the surface-code-based transversal STAR architecture, with the physical noise model relevant to neutral-atom hardware.

\subsubsection{Transversal multi-rotation (TMR) protocol with neutral atoms}
\label{sec:magicnoise}

The TMR protocol \cite{Toshio2024,Choi2023} prepares the small-angle magic state $ \ket{m_\theta}_L=R_{Z,L}(\theta)\ket{+}_L$. The protocol proceeds by initializing $\ket{+}_L$ and then applying $k$ multi-qubit-$Z$-rotation gates on the physical qubits such that the rotated qubits lie in the logical $Z$ operator support, as shown in Fig.~\ref{fig:Fig1_STAR}. Thus, $k$ can be chosen between $1$ and $d$. The corresponding logical state after TMR becomes
\begin{align}
\begin{split}
\label{eq:R_z}
    &\quad\prod_{i}^{k} R_{Z_s,i}(\theta^*)\ket{+}_L\\
   & =\cos^k\left(\frac{\theta^*}{2}\right)\ket{+}_L +(-i)^k \sin^k\left(\frac{\theta^*}{2}\right)\ket{-}_L\\
   &+(\text{$Z$-error terms})\\&
    = {\sqrt{P_{\text{ideal}}}} \left [ \cos(\frac{\theta}{2})\ket{+}_L +(-i)^k \sin(\frac{\theta}{2})\ket{-}_L \right ]+\\&
    (\text{$Z$-error terms})
\end{split}
\end{align}
where $R_{Z_s,i}$ is a multi-qubit or single-qubit physical $\theta^*$ rotation gate, with $Z_s$ being a $Z$ Pauli string characterizing the gate support. After postselection removing all $Z$-error terms, one obtains $\ket{m_\theta}_L$ logical state with $\frac{\theta}{2} =\arcsin(\frac{\sin^k(\theta^*/2)}{\sqrt{\sin^{2k}(\theta^*/2)+\cos^{2k}(\theta^*/2)}})$, and the ideal success rate in the noiseless case of $P_{\text{ideal}}=\cos^{2k}\left ( \frac{\theta^*}{2} \right )+\sin^{2k}\left ( \frac{\theta^*}{2} \right )$. 

\begin{figure}[t]
\centering
\includegraphics[width=1.0\columnwidth]{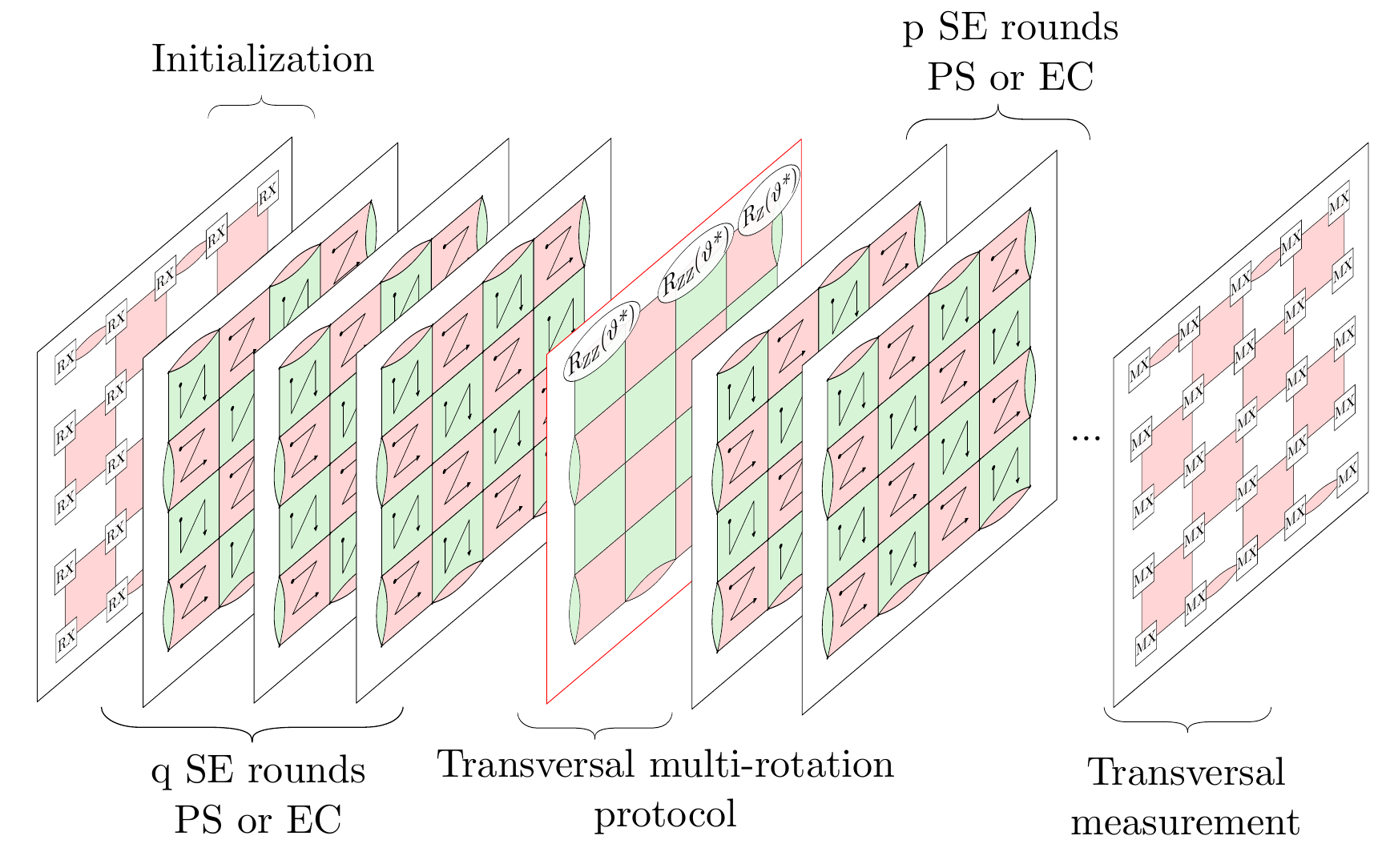}
\caption{ \textbf{The surface-code STAR transversal multi-rotation protocol.} The surface-code transversal STAR injection circuit is composed of $q$ rounds of initialization, transversal multi-rotation layer, and $p$ rounds of syndrome extraction (SE) for the final transversal measurement. For the initialization, after the first stabilizer-setting round, $q-1$ (typically 1--2 saturates our error estimates) rounds of stabilizer postselection (PS) or partial error correction (EC) are required to avoid the dangerous errors occurring at the critical injection qubits where rotation gates are applied. In the middle of the circuit, $k$ multi-qubit control-$Z$ rotation gates are applied to the qubits that support the logical $Z$ gate. After the multi-rotation, the $p$ rounds (typically 2) of syndrome extraction with full post-selection or partial error correction are implemented to filter out $Z$ error terms in Eq.~\eqref{eq:R_z}. The transversal measurement, without postselection or error correction, is then appended to estimate logical error rates.
}
\label{fig:Fig5_injection_circuit_sketch}
\end{figure}

In the realistic noisy case, the physical circuit errors can corrupt the postselection and thus introduce the logical error in the final state. The most dangerous errors flip the parity of the expected $Z$-error terms, effectively inducing a coherent logical misrotation error. To suppress the circuit noise, several rounds of postselection (or partial error correction, see Appendix~\ref{sec:fid_est_TMR}), are applied after the protocol, as shown in Fig.~\ref{fig:Fig5_injection_circuit_sketch}. Postselection, done typically for two syndrome rounds, filters out the results with syndrome errors, including the $Z$-error terms that herald logical misrotation. Moreover, circuit errors that occur in the initialization stage before the implementation of TMR could induce logical errors by a similar mechanism. Hence, one either initializes the $\ket{+}_L$ fault-tolerantly (requiring $d$ rounds) or, more efficiently, by full or partial post-selection in several (2--3) rounds.  Given the large $Z$-error bias of our physical noise model and the sensitivity of the TMR protocol to $Z$ errors, the performance of the TMR protocol requires some detailed reevaluation compared to the simple noise model estimates in Ref.~\cite{Choi2023, Toshio2024}.

The TMR protocol can be applied to any CSS code~\cite{Calderbank1997} as described. Since a neutral-atom system has dynamically reconfigurable long-range connectivity, it can implement variants of TMR more readily, regardless of the quantum error correction code or the TMR connectivity. In contrast, for quantum hardware platforms with fixed connectivity, based on the code and specific TMR protocol used, multirotation gates could require extra overhead and thus require a lower physical error budget. For example, for the surface code TMR protocol that conveniently attempts to use two-qubit rotations with $k=\lceil d/2 \rceil $, hardware that has nearest-neighbor square lattice connectivity~\cite{Acharya2024} would require ancillas and additional gate overhead to perform the TMR rotation gates. However, neutral-atom hardware can directly implement the TMR gates through native two-qubit gates between data qubits.

\subsubsection{Fidelity calculation methodology}
\label{sec:fid_Rz}

In order to estimate the total error rate of the teleported arbitrary-angle rotation, we need to evaluate the logical Clifford gadget errors and TMR injection errors. These error channels are then employed for the simulation of the repeat-until-success (RUS) teleportation protocol. The non-Clifford nature of the TMR injection protocol represents a challenge for conventional stabilizer simulations, which we circumvent by employing the methods of ensemble Clifford circuit simulations. The ensemble simulations, as proposed by~\cite{Choi2023,Toshio2024}, decompose the total problem into separate logical Pauli-$Z$ basis state simulations (see details in Appendix~\ref{sec:fid_est_TMR}) and replace the injected rotation angle in each separate simulation by zero. In Appendix~\ref{sec:fid_est_TMR}, we provide a validation of this protocol on conventionally classically simulable instances.

Generically, the logical noise model for TMR can be characterized as follows: 
\begin{align}
\label{eq:inf_2}
      &1-\mathcal{F}  \approx P^{(t)}_{\theta^*} \epsilon_{Z,t}\\ 
      &+\sum_j^{\left \lfloor\frac{k}{2}\right \rfloor} P^{(n)}_{\theta^*,j}\left [ (1-\epsilon_{\text{tot},j})\sin^2(\Delta_j)+\epsilon_{Z,j}\cos^2(\Delta_j)\right ]\notag
\end{align}
where $P^{(t)}_{\theta^*}$ ($P^{(n)}_{\theta^*,j}$) is the probability of obtaining the logical target (wrong) angle, $\epsilon_{Z,t}$ ($\epsilon_{Z,j}$) is the logical phase flip error rate for the target (wrong) angle, $\epsilon_{\text{tot},j}$ is the total error rate of wrong-angle rotation, and $\Delta_j=(\theta-\theta_j)/2$ is the misrotation amount of the specific wrong angle. Each of the corresponding probabilities can be directly estimated from an ensemble Clifford simulation. The underlying noise model of the TMR injection is obtained by simulating the entire process of $\ket{m_\theta}_L$ state preparation, including the noise effects of the realistic $q=3$ initialization and $p=2$ postselection rounds (see Fig.~\ref{fig:Fig5_injection_circuit_sketch}). We make sure not to use any transversal measurement information for postselection. In Appendix~\ref{sec:fid_est_TMR}, we provide in addition error estimates for the isolated TMR injection gadget itself, with an approach similar to one described in subsection~\ref{sec:extraction}.

In practice, in Eq.~\ref{eq:inf_2}, the first term is the probability that a logical $Z$ error occurs for the target angle, which scales as $O(p_{\mathrm{ph}}^k)$ with the physical error rate $p_{\mathrm{ph}}$, as a string of $k$ such physical errors is needed to introduce a logical error. The second term, scaling as $O( p_{\mathrm{ph}}\theta^{2(1-1/k)})$, is the main term contributing to infidelity and it stems from the coherent misrotation error. The rate of misrotation is directly proportional to the physical error rate as the physical $Z$ error interferes with the wrong-angle postselection. The amount of misrotation, however, is proportional to the size of the logical angle injected. The third term scales again as $O(p_{\mathrm{ph}}^k)$, which is the probability of the logical $Z$ error in the wrong-angle rotation case. When $k$ is sufficiently large, the logical angle $\theta$ is small, and the physical noise $p_{\mathrm{ph}}$ is low, the first and third terms become subleading compared to the second, misrotation term, as discussed in detail in Appendix~\ref{sec:fid_est_TMR} and Ref.~\cite{Toshio2024}. Given that the lowest order error term is a coherent misrotation, injection fidelity underestimates its impact, as it represents average case performance for a single injection. As shown in Ref.~\cite{Toshio2024}, trace-distance-based metrics better quantify the application-relevant worst-case coherent misrotation error rates. In the small-angle and physical error rate limit, the characteristic $\sim p_{\mathrm{ph}}\theta$ of the trace-distance-based logical error rate is observed regardless of $k$. This is the scaling we expect to observe for the teleported small-angle magic states.

\subsubsection{Arbitrary angle injection fidelity}
\label{sec:inj_fid}

We now present the optimized performance of the TMR injection across code distances and physical error rates informed by the neutral-atom hardware noise model. Based on the evaluation of the result from different TMR initialization protocols in Appendix~\ref{sec:fid_est_TMR}, we apply $q=2$ rounds of initialization with full postselection and $k=3$ TMR protocol (see Fig.~\ref{fig:Fig5_injection_circuit_sketch}) for different code distances, extracting the error rate and postselection rate.

\begin{figure}[t]
\centering
\includegraphics[width=1\linewidth]{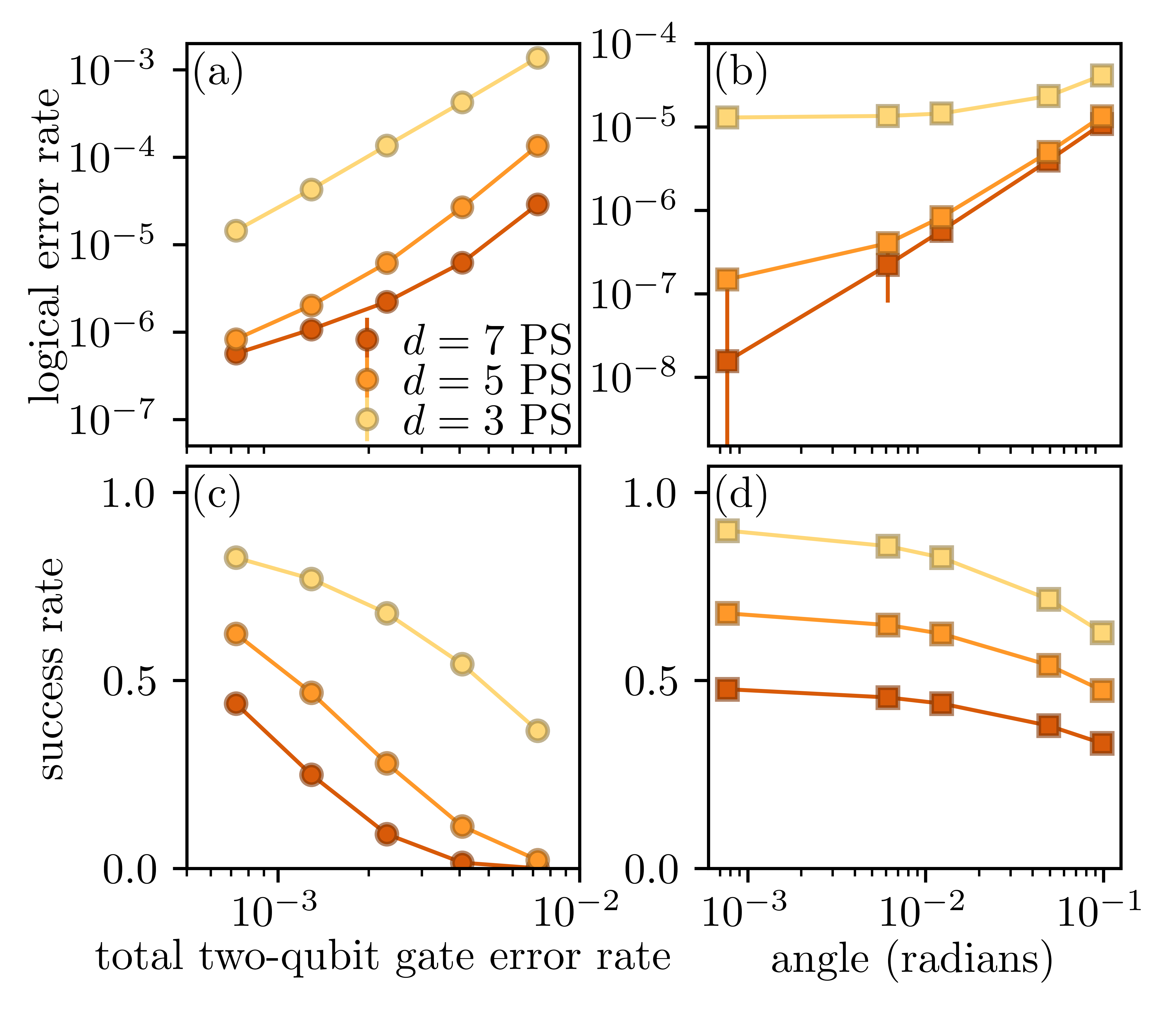}
\caption{ \textbf{The logical small-angle injection fidelity and success rate.} Upper panels: Logical error rate of end-to-end injection for various code distances ($d = 3, 5, 7$) using the $k = 3$ full postselection (PS) TMR protocol from Fig.~\ref{fig:Fig5_injection_circuit_sketch}, shown (a) as a function of the total two-qubit gate error rate at a fixed injection angle of $10^{-2}$ radians, and (b) at a fixed total two-qubit gate error rate of $7.3 \times 10^{-4}$.
Lower panels: Injection success rate at different code distances shown (c) at a fixed angle of $10^{-2}$ radians as a function of total two-qubit gate error rate, and (d) as a function of injected angle at total two-qubit gate error rate of $7.3 \times 10^{-4}$.
}
\label{fig:Fig6_injection_fidelity}
\end{figure}

Fig.~\ref{fig:Fig6_injection_fidelity} visualizes the success rate and logical infidelity for different code distances. The code distance protection is used throughout end-to-end injection, and thus smaller code distances ($d=3, 5$) show significantly degraded performance at small angles. However, already at $d=7$, the expected transversal STAR performance saturates for the expected rotation angles and physical error rates, including the linear scaling with physical error rate and angle in the small-angle limit. No further increase in code distance will be beneficial for injection error rates, due to non-fault-tolerant nature of the protocol. The error rates reached are significantly below $p_{\mathrm{ph}}\theta^{4/3}$ (coherent misrotation contribution for $k=3$) at $10\times$ lower than current physical error rates, showcasing the potential of the isolated TMR gadget for megaquop quantum operations. We note that achieving the optimal transversal STAR injection performance at $d=7$ required optimization of the protocol for the neutral-atom relevant noise model, which included $q=2$ rounds, that have proved sufficient to overcome the negative effects of $Z$-noise bias. 

The corresponding success rate at the current physical noise are around $0.5\%$ even at $d=7$, comparable to recent magic state distillation demonstration on neutral-atom hardware~\cite{Rodriguez2024}. This suggests an opportunity for demonstrating the basic principles of transversal STAR operations using current generation of neutral-atom hardware, including error rate scaling, for $d=3\text{--}7$. The background error subtraction described in Appendix~\ref{sec:fid_est_TMR}, together with a choice of non-fault-tolerant initialization approaches~\cite{Bluvstein2024, Rodriguez2024}, could further enhance the feasibility of such experimental demonstrations in the near term. At the expected $10\times$ smaller error rate, the postselection rate remains around $0.5$ even at $d=7$, providing an efficient injection factory.

\subsubsection{Arbitrary angle teleportation fidelity}
\label{sec:teleportation_fid}

We proceed to estimate the total logical error rate of $R_{Z,L}(\theta)$, teleported via repeat-until-success protocol, as shown in Fig.~\ref{fig:Fig1_STAR}. Our methodology for evaluating the infidelity of transversal STAR teleportation relies on the noise model of the logical Clifford gates, shown in Sec.~\ref{sec:clfford_noise}, and its composability, demonstrated in Sec.~\ref{sec:compose}, as well as end-to-end estimates of logical injection fidelity at a range of angles described in~\ref{sec:inj_fid}. As a result, we can observe a clear separation between Clifford and intrinsic injection contributions to teleported magic angle error rates.  

In practice, we perform the logical-level simulation that takes into account the extracted logical Clifford and end-to-end TMR injection noise channels, with final teleportation fidelity evaluated by Hilbert-Schmidt test:
\begin{align}\label{eq:h_schmidt}
    \mathcal{F}=\frac{\text{Tr} (R^\dagger_{Z,I}(\theta)R_{Z,\mathcal{N}}(\theta))}{2}
\end{align}
with $R^\dagger_{Z,I}(\theta)$ ($R_{Z,\mathcal{N}}(\theta)$) being the ideal (noisy) rotation gate. Each teleportation is simulated as a full  RUS protocol.  We select to evaluate teleportation fidelity at rotation angles $\theta=\frac{\pi}{2^n}$, which allows the protocol to complete with the logical $S$ gate, while still well approximating fidelity at intermediate angles~\cite{Toshio2024}. In this setting, the total teleporation error rate, $p_L(\theta)$, can be estimated from the infidelities of each length-$m$ RUS teleportation chain, $\mathcal{E}\left ( \theta=\frac{\pi}{2^{n-m}} \right )$, weighted by the corresponding success probability
\begin{align}\label{eq:teleport}
    p_L\left ( \theta=\frac{\pi}{2^{n}} \right )=\frac{1}{2^{n-1}}\mathcal{E}_S+\sum_{m=0}^{n-2} \frac{1}{2^m} \mathcal{E}\left ( \theta=\frac{\pi}{2^{n-m}} \right )
\end{align}
where $\mathcal{E}_S$ is the infidelity of the logical $S$ gate. We note that on average, teleportation will succeed after $m=2$ attempts, which, together with the number of injection syndrome rounds, ensures overall good protocol speed. This also means that the teleportation fidelity will include the effect of two logical CNOT gates.  We optionally optimize our protocol by applying the TMR parameter that switches the protocol from $k=3$ to $k=1$ at the larger rotation angles, governed by the angle at which the infidelity of $k=3$ end-to-end injection crosses over with that of $ k=1$. This, in practice, has diminishing returns for the TMR with already optimized TMR hyperparameters (Appendix~\ref{app:hyperparameters_TMR}). Some additional minor improvement in teleportation error rates is possible if direct $k=1$ logical rotation is replaced by the original fixed-angle STAR protocol~\cite{Akahoshi2024}. However, except in exponentially rare late stages of teleportation, the fixed-angle protocol with its $\sim 2p_{\mathrm{ph}}/15$ error scaling has a drastic impact on the circuit error budget (even more so at the realistic $p_{\mathrm{ph}}\sim 10^{-3}$), and is thus not considered for the transversal STAR gateset in our current study. 

\begin{figure}[t]
\centering
\includegraphics[width=1\linewidth]{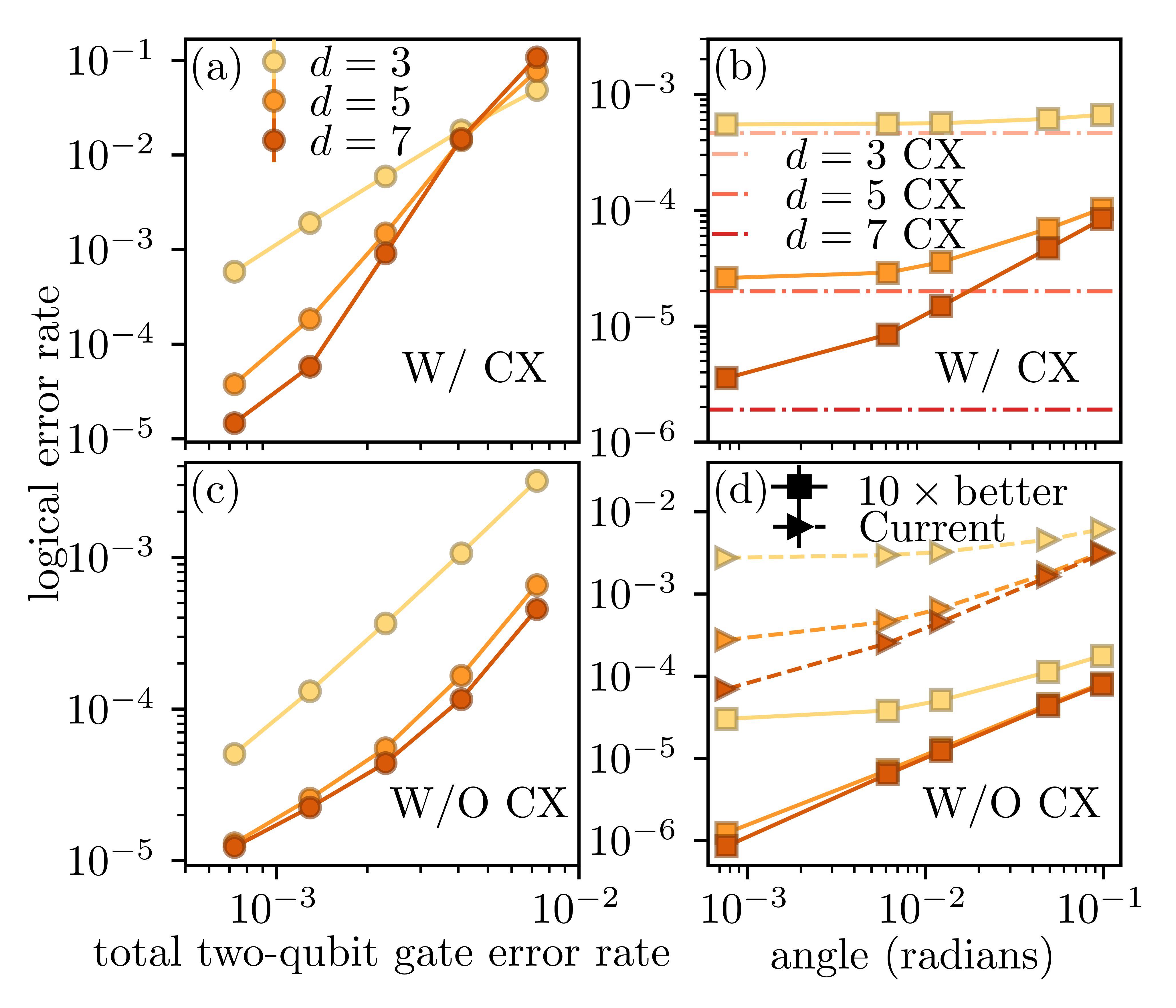}
\caption{ \textbf{The total small-angle teleportation error rates.} The upper panels: the total repeat-until-success teleportation fidelity calculated from injection error rates from Fig.~\ref{fig:Fig6_injection_fidelity}(a)-(b) and logical Clifford error model from Fig.~\ref{fig:Fig3_Clifford_noise}, specifically: (a) logical error rate for several code distances at fixed angle of $10^{-2}$ radians and (b) varying angles at $7.3\times 10^{-4}$  physical noise scale. The horizontal dot-dashed lines in (b) denote $2\times$ CNOT Clifford error rate. The bottom panels, (c)-(d): the teleportation error rate assuming noiseless logical Cliffords in the repeat-until-success protocol between $7.3\times 10^{-4}$ [square with full lines in (d)] and $7.3\times 10^{-3}$ physical error rate [triangles with dashed lines in (d)]. The results presented in (a)-(b) include the noise stemming from teleportation CNOT gate error of appropriate distance (labeled with ``W/ CX''), while panels (c)-(d) showcase noiseless teleportation CNOT gate results (``W/O CX'') that better showcase the intrinsic analog magic-related limits of STAR architecture.} 
\label{fig:Fig7_teleportation_fidelity}
\end{figure}

The upper panels of Fig.~\ref{fig:Fig7_teleportation_fidelity} show the total infidelities of the logical $R_{Z,L}(\theta)$. Both in the large noise scale regime [Fig.~\ref{fig:Fig7_teleportation_fidelity}(a)], and the small-angle regime [Fig.~\ref{fig:Fig7_teleportation_fidelity}(b)], the main contribution of infidelity comes from the logical CNOT errors. In fact, for small angles, the teleportation fidelity approaches the fidelity of two logical CNOT gates needed for the protocol on average. As the rotation angle increases, the fidelities of $d=5$ and $7$ teleportation protocols converge as they become dominated by the TMR noise.

To explore the ultimate limits of the transversal STAR architecture in the regimes where logical Clifford noise is subleading, we also show logical $R_{Z,L}(\theta)$ infidelity calculated with noiseless CNOTs in the bottom panels of Fig.~\ref{fig:Fig7_teleportation_fidelity}. At small physical noise and small angles, we recover the expected approximate scaling of the total logical error rate $p_L(\theta)\sim \alpha p_{\mathrm{ph}}\theta$~\cite{Toshio2024} in both the target angle $\theta$ and the physical error rate $p_{\mathrm{ph}}$. We measure the prefactor $\alpha \approx1.5$ for $d=7$, which is competitive with the previous results~\cite{Toshio2024}. We observed an $O(1)$ prefactor despite the strong dangerous $Z$ bias noise in our physical noise model, a consequence of the TMR injection optimization made. Surface code $d=7$ has sufficient protection against noise for the TMR protocol itself, as manifested by the saturation of the teleporation error rates in  Fig.~\ref{fig:Fig7_teleportation_fidelity} (c)-(d). Better Clifford fidelities at $d=9$ (extrapolated from fitting $d=3,5,7$ data; see Appendix \ref{sec:logical_noise} for fitting details) would allow for achieving the optimal transversal STAR best performance at the physical error rate of around $10^{-3}$ and angles up to $10^{-3}$ and thus achieving magaquop-scale Hamiltonian simulation with the transversal STAR architecture. 

\section{Utility of transversal STAR architecture for Trotterized Hamiltonian dynamics}
\label{sec:disc_utility}

 Equipped with the concrete surface-code-based construction and the associated hardware-aware logical noise model, we proceed to discuss the limits and extensions of transversal STAR utility. In particular, we discuss conditions in which the architecture reaches the effective simulation power of a megaquop quantum computer for the task of Trotterized quantum simulation of local Hamiltonians.

 \subsection{Transversal STAR simulation capacity}
\label{sec:sim_capacity}

\begin{figure*}[t]
\centering
\includegraphics[width=1.0\textwidth]{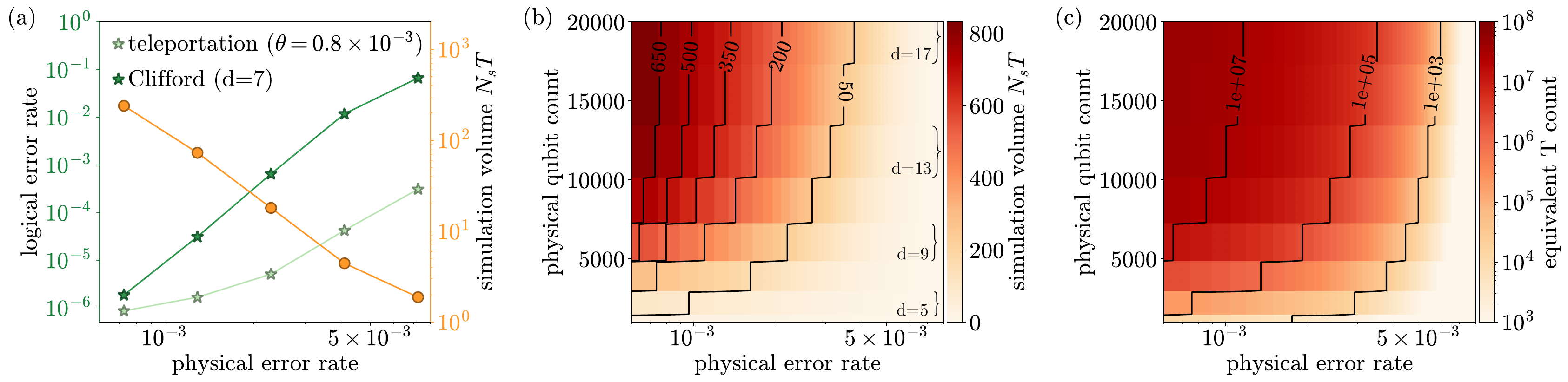}
\caption{\textbf{Trotterized dynamics utility of the transversal STAR architecture.}
The limits of transversal STAR architecture utility for Trotterized quantum simulation are presented according to analysis in Sec.~\ref{sec:disc_utility}. (a) The Clifford CNOT error rates ($d=7$) from Fig.~\ref{fig:Fig3_Clifford_noise} and small-angle rotation error rates ($\theta=0.8\times 10^{-3}$) from Fig.~\ref{fig:Fig7_teleportation_fidelity}, determine the effective local ($\omega\sim l_1\sim 1$, $n_C\sim 4$) Hamiltonian simulation volume ($N_sT$). The volume is estimated as a function of physical error rate for the first-order Trotter simulation. (b) The Hamiltonian simulation volume as a function of number of physical qubits and the physical error rate. The simulation volume saturates around $10,000$ physical qubits for $10 \times$ lower physical error rate than current, as Clifford error becomes subleading at $d=9$. The Clifford error is extrapolated from Fig.~\ref{fig:Fig3_Clifford_noise}, as covered in Appendix~\ref{sec:logical_noise}, while at least $N_s=30$ logical data qubits were required in simulation volume estimates. (c) The equivalent $T$ count as a function of the number of physical qubits and the physical error rate under the same assumptions as in (b) and angle synthesis and precision assumptions described in Sec.~\ref{sec:disc_utility}. The limit of transversal STAR is manifest in the saturation regime with similar parameters to the simulation volume, and the resulting equivalent $T$-counts are in the $10^7$ range.
}
\label{fig:Fig8_STAR_limits}
\end{figure*}

 The linear scaling of logical fidelity with physical gate infidelity for the teleported magic angle, $p_L(\theta) \approx \alpha p_{\mathrm{ph}}\theta$, ultimately constrains STAR architecture utility for Trotterized Hamiltonian simulation~\cite{Akahoshi2024}. Importantly, the limit of simulation power depends on the total logical angle performed in the circuit and not on the number of Trotter steps (circuit depth). Assuming a local Hamiltonian, the total $L_1$ norm will take the form $\lvert\lvert H \rvert \rvert_1 =l_1 N_s $, where $N_s$ is the total number of sites (qubits, spins, or fermionic modes), and $l_1$ is a site-number-independent constant with units of relevant Hamiltonian energy scale ($J$). The total logical angle performed ($\theta_T$) directly depends on the norm and total evolution time $T$ (units of relevant Hamiltonian time-scale, $1/J$), with $\theta_T=l_1 N_s T$. Thus, in the case of large enough code distances where Clifford error rate is subleading, we get an ultimate limit of local Hamiltonian simulation with STAR~\cite{Toshio2024}:
\begin{equation}\label{eq:star_limit}
    p_L(\theta_T) \approx \alpha p_{\mathrm{ph}} \theta_T  \lesssim 1, \qquad
    N_s T  \lesssim \frac{1}{l_1 \alpha p_{\mathrm{ph}}}.
\end{equation}
We refer to $N_s T$ as the simulation volume. In practice, for many relevant local spin and fermionic Hamiltonians, the norm per site $l_1$ is $O(1)$~\cite{Campbell2021}, so the limits with the expected near-term neutral-atom noise rates ($p_{\mathrm{ph}} \lesssim 10^{-3}$) might look something like 30--100 logical qubits evolved for $T=10$. From our current simulations, presented in Sec.~\ref{sec:logicalnoise} and summarized in Fig.~\ref{fig:Fig8_STAR_limits}(a), together with their effective simulation volume, this utility limit of transversal STAR can be reached at $d=9$ for physical error rate $10 \times$ better than current.

In order to realize such simulations, Trotter error also has to be controlled. In fact, matching the circuit error given by the total logical angle to Trotter error will ensure optimal circuit depth. Assuming for simplicity first-order Trotter decomposition for local Hamiltonian simulation, the total Trotter error will be 
\begin{equation}
p_{\mathrm{tr}} \approx wN_s(\Delta t)^2 N_{\mathrm{tr}}, \qquad 
\Delta t =T/N_{\mathrm{tr}}.
\end{equation}
Here, $\Delta t$ is the time per Trotter step, $N_{\mathrm{tr}}$ is the total number of Trotter steps (comparable to total circuit depth), and $w$ represents the relevant Trotter norm per site derived from the commutators of Hamiltonian terms participating in the Trotter decomposition~\cite{Campbell2019} --- for local Hamiltonians, $w$ is a site-number-independent constant with units of $J^2$. Matching the circuit error rate, $p_L(\theta_T)$, dominated by the small-angle magic and the Trotter error rate $p_{\mathrm{tr}}$, as well as STAR total simulation time limit of Eq.~\eqref{eq:star_limit}, one obtains the estimate of the effective number of Trotter steps (circuit depth) and total STAR circuit volume ($N_s N_{\mathrm{tr}}$):
\begin{align}\label{eq:star_depth_volume_larged}
    p_{\mathrm{tr}} & \sim  p_L(\theta_T) \lesssim 1,  
     \nonumber \\
    N_{\mathrm{tr}} & \sim \dfrac{w T}{l_1\alpha p_{\mathrm{ph}}}  \sim \frac{w}{N_s\left(l_1\alpha p_{\mathrm{ph}}\right)^2}, \\
    N_s N_{\mathrm{tr}}&  \sim \frac{w}{\left(l_1\alpha p_{\mathrm{ph}}\right)^2}.
\end{align}
The second line uses the simulation volume limit of Eq.~\eqref{eq:star_limit}. Note that the total circuit depth ($\sim N_{\mathrm{tr}}$) and volume ($\sim N_s  N_{\mathrm{tr}}$) scale as $1/p_{\mathrm{ph}}^2$, with a clear advantage over the $1/p_{\mathrm{ph}}$ scaling for NISQ circuits. The concrete scaling form depends on the Trotterization order chosen ($q$), with $1/p_{\mathrm{ph}}^{1+1/q}$, but it remains favorable compared to NISQ scaling. In practice, constraints based on the Clifford error rate, maximum practical transversal STAR code distance in terms of injection postselection rates, and growing Trotter norm $w$ would limit relevant applications to low-order Trotter simulation (first and second order Trotter), as further decrease of Trotter errors would compete against Clifford error rates. At the same time, the transversal STAR benefits over the NISQ approach, at $p_{\mathrm{ph}}\approx 10^{-3}$, are the most pronounced in the low-order Trotter simulation regime.

Based on the practical error rates we obtain for the teleported small-angle magic in Sec.~\ref{sec:cliffordnoise} and the extrapolation to higher code-distance, and the typical local Hamiltonians where $w \sim l_1^2$, Eq.~\eqref{eq:star_depth_volume_larged} evaluates to circuit volume of $N_s N_{\mathrm{tr}}\approx 10^{6}$ with circuit depth $N_{\mathrm{tr}}\approx 10^4$. By making Clifford error rate subleading, the code distance that would enable such a circuit volume corresponds to $d=9\text{--}11$ with exact details depending on the application. Fermionic simulation applications would in particular require higher code distances due to the need for fermionic encoding and the corresponding fermionic mode reordering, which, while not affecting overall $L_1$ norm and magic requirement, increases relative Clifford requirements~\cite{Kivlichan2018, Campbell2021}. Based on $d=9\text{--}11$, the effective data-qubit volume necessary to support transversal STAR simulations would be $10,000\text{--}20,000$ physical qubits, assuming a typical factor of 2 for the injection factory space, based on the typical Hamiltonian simulation and the expected post-selection rates from Sec.~\ref{sec:inj_fid}. Fig.~\ref{fig:Fig8_STAR_limits}(b) summarizes the dependence of Hamiltonian simulation volume as a function of the number of physical qubits and physical error rate, including the saturation of transversal STAR simulation power at $d=9$ and physical error rate $10\times$ better than current.

\subsection{Comparison with fully fault-tolerant approaches}
\label{sec:comparison_T}

 While the transversal STAR architecture has obvious advantage over NISQ approaches in terms of more effective use of limited two-qubit gate fidelity, another useful comparison is the equivalent $T$-gate numbers necessary to run a circuit equivalent to the STAR limits. As the basis for this comparison, we take typical angles that the $T$-gate-based approach would attempt to synthesize $\theta \approx l_1\alpha p_{\mathrm{ph}}/w \sim \alpha p_{\mathrm{ph}}/l_1$, and the cost of $T$-synthesis of angle $\theta$ with accuracy $\delta^2$ to be $N_{\mathrm{syn}}\approx - 3\log_2{\delta}$~\cite{Ross2016}. In order to obtain a relevant $T$-gate equivalent, we can match the effective STAR small-angle magic error rates with synthesis error~~\cite{Yin2025}: 
 \begin{equation}
 \delta(\theta)\approx \sqrt{p_L(\theta)},
 \end{equation}
 leading to $N_{\mathrm{syn}} \approx -3\log_2{(\alpha p_{\mathrm{ph}}/\sqrt{l_1})}\sim 30$ $T$-gates per typical small-angle injection. Note that the logarithm involved makes this estimate only weakly dependent on the order of the Trotter product formula used and Hamiltonian and Trotter norms, although somewhat different estimates might be derived in algorithm-specific settings, such as in terms of quantum phase estimation accuracy and error budgets~\cite{Bocharov2015, Kivlichan2018}. Still, at least in terms of Trotterized simulation of quantum dynamics for local, low-norm Hamiltonians, transversal STAR has the potential to reach an effective simulation power of $\sim 10^7$ $ T$-gate circuit, as shown in Fig.~\ref{fig:Fig8_STAR_limits}(c).
 
 The clear advantage of transversal STAR over fully fault-tolerant approaches in this megaquop-characteristic simulation regime is further extended by low-space overhead for the required number of $T$-factories based on reported and expected postselection rate. To quantify this space-time advantage, we compare with the closest competitors in the megaquop regime --- the recently proposed $T$-cultivation methods~\cite{Gidney2024, Chen2025} coupled with the approaches that employ the characteristic structure of parallel rotation layers to reduce overall $T$-gate synthesis cost, especially hamming weight phasing and catalysis~\cite{Gidney2018, Kivlichan2018, Kan2024}. In terms of neutral atom related approaches, recent cultivation improvements~\cite{Chen2025} for dynamically reconfigurable architectures apply, potentially bringing down effective space-time costs of $10^{-6}$ error rate $T$-factories to only $2\text{--}4\times$ more expensive than STAR injection. The synthesis overhead, however, still applies and necessitates higher quality $T$-gates, which, due to the sharp step up between two known cultivation protocols, require $10\text{--}20\times$ space-time volume over STAR injection per each $T$. Together with $\sim 30\times$ direct synthesis overhead, the resulting total space-time disadvantage over the transversal STAR architecture is thus in the $100\text{--}1000$ range. Techniques such as hamming weight phasing for parallel rotation layer synthesis could likely bring down the space-time overhead over transversal STAR closer to $100\times$ limit, at the cost of additional algorithmic ancillas. The cultivation schemes and Cliffords are less efficient for fixed-connectivity architectures~\cite{Gidney2018}, so the space-time advantage for transversal STAR could be on the order of $1000\times$ over the fixed-connectivity architectures. However, further advancements in cultivation are possible, in particular the design of new protocols that interpolate between two known cultivation protocols in the $10^{-7}$--$10^{-8}$ error rate regime. On the algorithmic side, the Trotter error bounds are often empirically found to be somewhat loose~\cite{Childs2018, Martinez-Martinez2024}, which would potentially benefit $T$-synthesis-based approaches through the possibility of utilizing larger Trotter steps; such empirical results are beyond the scope of analysis in this work and left for future studies.

To provide a concrete space-time resource estimate for a characteristic megaquop quantum simulation problem with transversal STAR and alternative architectures, we consider the physical qubit and clock cycle cost to perform one first-order Trotter step of square lattice nearest-neighbor transverse field Ising model dynamics. The details of the estimate, described in Appendix~\ref{sec:space-time_comparison}, follow general considerations described above, with final results reported in Tab.~\ref{tab:resource_summary}. The general estimates provided are directly validated on this example. Surface-code-based transversal STAR architectures achieve $\sim 250 \times$ time savings and $2\times$ space savings over a fixed-connectivity, fully fault-tolerant scheme. We note that this is the case even though a fully fault-tolerant architecture was based on comparatively low-cost magic state cultivation~\cite{Gidney2024}. Compared to the original STAR architecture, the transversal savings amount to $\sim 10\times$ time and $2\times$ space. While the fixed-connectivity STAR can reach a similar megaquop simulation regime, the transversal STAR can perform equivalent simulation with drastically smaller physical resources. This can be improved further, with $\sim 5 \times$ additional space compression, by considering task-codesigned, high-rate-code-based transversal STAR architecture, as described in the next section.

\begin{table*}[ht]
\centering
\renewcommand{\arraystretch}{1.25}
\begin{tabular}{c|c|c|c|c}
\hline \hline
&\makecell{fixed connectivity \\ fully fault-tolerant}& \makecell{fixed connectivity  \\ STAR from Ref.~\cite{Toshio2024}} &   transversal STAR & high-rate transversal STAR \\
\hline
clock cycles per Trotter step & 14460  & 630 & 64 & $\approx$ 60--130$^*$ \\
\hline
physical qubits & 30976 & 30976  & 15488 &  3328 (1664)$^{**}$ \\
\hline \hline
\end{tabular}
\caption{\textbf{Comparison of Ising model dynamics space-time resource requirements across architectures.} The total clock cycle times and physical qubit requirements for one (first-order) Trotter step of the $8\times 8$ square lattice nearest neighbor transverse field Ising model aiming at megaquop scale are shown across four QEC architectures. The architectures include a) fully fault-tolerant fixed connectivity surface-code-based architecture based on magic state cultivation~\cite{Gidney2024}, b) fixed connectivity STAR architecture as described in Ref.~\cite{Toshio2024}, c) surface-code-based transversal STAR architecture described through this paper, and d) high-rate-code based transversal STAR architecture described in Sec.~\ref{sec:disc_qLDPC}. The estimates are based on general considerations presented in Sec.~\ref{sec:disc_utility} and the detailed cost model from Appendix~\ref{sec:space-time_comparison}. The number of logical factories assumed for each of the architecture is equal to the number of physical data qubits (64). *Clock cycle rates of high-rate transversal STAR architecture are given as an approximate range due to uncertainties in its performance that require detailed circuit-level evaluation. **Physical qubit requirements for high-rate transversal STAR are given both as the minimal number of qubits to perform two computations in parallel,  necessary due to the disjoint nature of homological and cohomological qubit sets in the self-dual construction, and as the number of physical qubits per single computation (parentheses).}
\label{tab:resource_summary}
\end{table*}

 \subsection{Near-term performance}
\label{sec:near_term_performance}
 
Finally, in the near term, limited number of qubits naturally pose a question of what is possible with smaller code distances with transversal STAR, including $d=3\text{--}7$ cases presented in Sec.~\ref{sec:cliffordnoise} and~\ref{sec:inj_fid}. The estimate is changed fundamentally as the limiting factor now becomes the Clifford-gate fidelity. We thus introduce the parameter $n_C$ describing the number of Clifford gates per Trotter step and data qubit. Even for simple spin Hamiltonians, $n_C \sim O(10)$ is typical, which includes, on average, two CNOTs involved in the small-angle teleportation with the RUS protocol. The exact $n_C$ number is highly Hamiltonian dependent, correlating with the $L_0$ Hamiltonian norm instead of $L_1$ norm that is relevant for magic cost. While the two norms can be similar in many practical simulation regimes, the number of Cliffords can further proliferate based on the Pauli weight of each term --- this is particularly resource demanding in fermionic quantum simulation, where $n_C$ will grow with the system size even for a local Hamiltonian. Matching the circuit error rate dominated by Clifford error ($p_C$) and the Trotter error rates as described previously, one obtains the following estimates of the simulation volume ($N_sT$) and circuit volume:
\begin{equation}\label{eq:STAR_Clifford limit}
    N_sT\sim \frac{1}{\sqrt{w n_Cp_C}}, \, \quad N_sN_{\mathrm{tr}}\sim\frac{1}{n_Cp_C}.
\end{equation}
The simulation volume is now determined by the inverse root of the Clifford error rate, in contrast to the inverse of the physical error in the magic-limited case, with the estimates being valid for $p_C > p_{\mathrm{ph}}^2$. Note that the exact power again depends on the choice of Trotter formula order, with the first-order Trotter chosen here. The Trotter order choice, however, significantly affects $w$ and in particular $n_C$, but still provides an additional possibility for Trotter-STAR co-design. Comparing the total circuit volume to NISQ approaches, regardless of the Trotter order, one can still obtain the standard scaling advantage with quantum error correction --- circuit volume scales as $ p_{\mathrm{ph}}^{-(d+1)/2}$ versus $p_{\mathrm{ph}}^{-1}$.

\section{Transversal STAR architecture with high-rate QEC codes}
\label{sec:disc_qLDPC}

\begin{figure*}[t]
\centering
\includegraphics[width=1.0\textwidth]{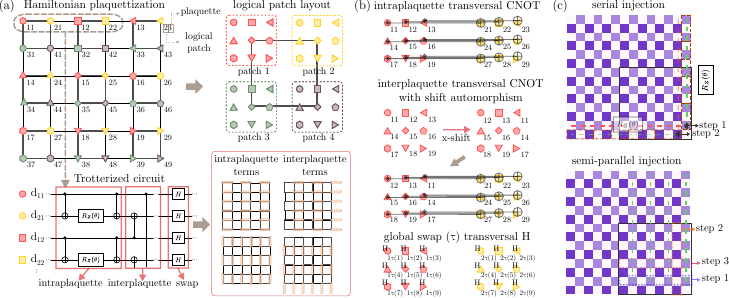}
\caption{\textbf{Transversal qLDPC-STAR for Trotterized quantum simulation.} Transversal STAR realized on high-rate quantum codes can be codesigned to the quantum simulation problem at hand. (a) As an example, we present the nearest-neighbor transverse field Ising model on a square lattice simulated by Trotterization. The locality of the Hamiltonian allows us to plaquettize the lattice into size-four sublattices (top left). The data qubits are labeled by $d_{ij}$ with $j$ denoting the plaquette, and $i$ site order within the plaquette (indicating logical code patch). The Trotter simulation then consists of parallel intraplaquette CNOTs, interplaquette CNOTs, and intraplaquette single-qubit Cliffords and rotations (bottom). The simulation protocol can be realized by matching the qLDPC-STAR code to the plaquettization of the lattices, with each plaquette site corresponding to a distinct code patch, $i$. The logical qubit layout is derived from the plaquette shape (top right). (b) Intraplaquette CNOTs can straightforwardly be realized by interpatch parallel transversal CNOTs. Interplaquette CNOTs require the utilization of codes with shift automorphisms (``x-shift'' label in the example shown) adapted to the periodicity of the lattice. Parallel single-qubit Cliffords can be performed by swap-transversal gates ($H$ shown) that equally permute (permutation $\tau$) the qubits in all code patches. (c) Finally, small-angle rotations can be injected and subsequently teleported by the repeated applications of serial or semi-parallel injection protocols within the qLDPC code patch. Here, an example of injection on [[98, 32, 3]] square hypergraph product code generated as a product of two cyclic classical [7, 4, 3]~\cite{Macwilliams1977} codes is shown --- data qubits are dark purple (two shades, with pivot qubits~\cite{Quintavalle2023} contained in the black box), ancillas light, while logical $X$/$Z$ operators have supports on red/green lines. Semi-parallel $R_Z({\theta})$ injection can be performed by parallel injections on all the qubits with non-overlapping logical $Z$ operator supports.
}
\label{fig:Fig9_qLDPC_STAR}
\end{figure*}

The surface-code transversal STAR architecture described so far realizes sizable space-time volume savings by employing transversal logical gates in the correlated decoding setting. The critical small-angle injection protocol, however, is readily generalizable to generic stabilizer codes, motivating us to further explore the possibility of space savings for STAR operations outside of the surface code paradigm. Quantum low-density parity-check (qLDPC) codes~\cite{Kovalev2013, Breuckmann2021, Lawrence2022, Quintavalle2023, Bravyi2024, Xu2024} can attain high logical encoding rates (e.g., $k\sim n$) while preserving code distance scaling (e.g., $d\sim \sqrt{k}$), making them an excellent candidate for reducing logical operation overhead. Notably, a recent proposal~\cite{Xu2024} has designed a neutral atom hardware-efficient scheme for fault-tolerant computation with qLDPCs. The structure of major qLDPC code families, in particular, allows for efficient syndrome extraction circuits on dynamically reconfigurable neutral atom hardware --- an advantage for qLDPC realization over quantum hardware with fixed connectivity. This flexibility further enables us to consider a wide variety of qLDPC codes without the need for hardware architecture changes. 

Despite the promises of qLDPC codes, a significant open question remains --- how to efficiently operate with such codes. Efficient transversal operations with qLDPC codes form a limited logical gateset. For example, readily accessible transversal entangling operations entangle all the qubits within the two logical patches at once~\cite{Breuckmann2024, eberhardt2024}, introducing significant overhead in algorithmic compilation. The efficient qLDPC operation thus requires careful design for the algorithm at hand, as recently shown for the case of resource state preparation and quantum adder circuits~\cite{Xu2025}. Here, we present a construction requiring only fold (swap)-transversal qLDPC operation and STAR injection primitives that enable effective quantum simulation for carefully co-designed code and simulation problem instances.

\subsection{Parallel transversal gateset for Trotterized quantum simulation}
\label{sec:qldpc_gateset}

The generic class of quantum simulation problems that our construction aims to tackle is the Trotterized simulation of geometrically local lattice spin Hamiltonians. A paradigmatic example of the problem class is the two-dimensional (2D) transverse field Ising model, shown in Fig.~\ref{fig:Fig9_qLDPC_STAR}(a). Common to all such Hamiltonian instances is the ability to cover the lattice with a set of plaquettes, with each plaquette performing an equivalent set of entangling operations to all other plaquettes, and the sum over all plaquette operations covering all the terms in the Hamiltonian. For the 2D transverse Ising case, the minimal plaquette choice corresponds to a periodic tiling with site plaquettes. All the terms in the Hamiltonian can be implemented on the node, as entangling operations within the plaquette (intraplaquette terms), or as entangling operations between neighboring plaquettes (interplaquette terms), with Fig.~\ref{fig:Fig9_qLDPC_STAR}(a) showing an explicit 2D Ising example.

The usefulness of Hamiltonian plaquettization lies in the fact that the Trotter simulation of the Hamiltonian naturally decomposes into gate layers that are ``plaquette-parallel''. In the 2D Ising case, Trotter decomposition contains four layers covering each of the bond directions on the lattice. Two such two-qubit entangling layers can be performed by fully parallel intraplaquette entangling operations, while the remaining two require fully parallel nearest-neighbor interplaquette operations. Single-qubit gate layers also come in two intraplaquette-parallel flavors --- near fully parallel layers of Hadamard gates (depending on the boundary conditions), and layers of arbitrary angle single-qubit rotations on every other qubit in the plaquette. We note that this construction is generic, with exact specifics such as plaquette size, shape, and term decomposition depending on the Hamiltonian locality, but with a typical geometrically local Hamiltonian decomposing into a significant number of relatively small plaquettes. In addition, starting from the minimally-sized plaquettes, plaquettization with bigger plaquettes that each contain several minimal plaquettes is possible, providing significant plaquette-size flexibility. The decomposition also does not require translational invariance in the Hamiltonian, just the notion of structured (geometric) locality, as the single-qubit gate angles can be arbitrary (including zero).

We are now ready to consider the requirements for compiling the plaquettized Hamiltonian simulation with fold(swap)-transversal qLDPC-STAR gatesets. To facilitate this, we assign the logical qubits to individual qLDPC code patches (logical code patches labeled by $i$) such that each logical data qubit ($d_{ij}$) within a given plaquette (plaquettes labeled by $j$) corresponds to a distinct code patch, with patches assigned between plaquettes in a set order. An example of such an assignment for the 2D transverse Ising model with four-site plaquettes is shown in Fig.~\ref{fig:Fig9_qLDPC_STAR}(a)-(b). The plaquette size choice thus dictates the minimal number of logical patches necessary for simulation (four in this case), with the number of encoded qubits, $k$, necessarily commensurate with the number of plaquettes in the lattice. The flexibility to choose plaquettes larger than minimal leads to the ability to adjust the number of logical patches or the number of encoded qubits $k$, facilitating code design. The plaquette simulation construction can then be implemented with the following reduced transversal logical gateset, as depicted in Fig.~\ref{fig:Fig9_qLDPC_STAR}(b)-(c):
\begin{itemize}
    \item Intraplaquette parallel transversal CNOTs (CNOT$_{p1}$, with subscript $p$ labeling the patch-transversal parallel gate) that, by construction, correspond to interpatch parallel transversal CNOTs with qubits in the same order for both patches.
    \item Interplaquette parallel transversal CNOTs (CNOT$_{p2}$) that correspond to interpatch parallel transversal CNOTs with qubit order in one patch shifted with respect to another according to a lattice translation of the underlying plaquette lattice.
    \item Parallel intraplaquette and intrapatch transversal Hadamard gates ($H_p$) and more generally parallel transversal single-qubit Cliffords ($H_p/S_p$).
    \item Arbitrary angle single qubit rotations ($R_{Z_p}(\{\theta_i\})$). These rotations can be injected as equiangular or distinct on the ancilla factory patches. The rotations are teleported to the logical data qubits with parallel transversal logical CNOTs between the data and ancilla patches,  and transversal ancilla patch measurements.
\end{itemize}

\subsection{Parallel transversal CNOT realization with shift automorphisms of the code}
\label{sec:qldpc_cnot}

The physical implementation of the gateset above requires careful qLDPC code design for the plaquette lattice structure at hand. Intraplaquette CNOT$_{p1}$ gates are readily available since all CSS codes support a parallel transversal CNOT between patches when qubit orders in the patches are equivalent~\cite{Calderbank1997}, as presented in Fig.~\ref{fig:Fig9_qLDPC_STAR}(b). The remaining Clifford gates have to rely on a set of code symmetries ensured by construction. For the case of shifted CNOT$_{p2}$ gates, the necessary qubit shifts that would make qubit order in the two patches equivalent correspond to simple primitive shifts of the plaquette lattice. This leads to the requirement that the code itself should have the cyclic property: all codewords and $X$/$Z$ parity checks should map to themselves under the cyclic permutations of the code corresponding to the primitive plaquette lattice translations. In other words, the quantum code ($\mathcal{C}$) is required to have a cyclic automorphism~\cite{eberhardt2024, Breuckmann2024, Xu2025} congruent with the plaquette lattice. If such an automorphism exists, the physical implementation of the necessary logical shift operation between the qubits is straightforward for neutral atom hardware, discussed amongst other automorphisms in Ref.~\cite{Xu2025}. In particular, if the code automorphisms that result in the primitive cyclic shift $\vec{T}$ on the logical codewords $q_L$ results in the bijective mapping of physical qubit $q$ to $f_{\vec{T}}(q)$, the logical shift gate is simply performed by the corresponding permutation of the physical qubits $\textrm{SWAP}_{q_L \rightarrow \vec{T} [q_L]}=\prod_{q \in \mathcal{C}} \textrm{SWAP}_{q \rightarrow f_{\vec{T}}(q) }$~\cite{eberhardt2024}. The choice of the code with the right cyclic automorphism thus allows to realize CNOT$_{p2}$ gates as swap-transversal gates through physical qubit permutations within one of the two patches, followed by parallel transversal CNOT.

It is now natural to consider which classes of quantum codes possess cyclic automorphisms compatible with the expected translation symmetries of the plaquette lattice. The typical simulation problem we design for would result in the 2D plaquette lattice that can be generated with two cyclic shifts. We note that while this construction is the most efficient for 2D local Hamiltonians, it does not preclude higher-dimensional Hamiltonian simulation. For example, a 3D spin Hamiltonian can be embedded in a 2D plaquette lattice, as long as the transverse dimension is kept within the plaquette itself. The transverse (embedded) lattice dimension can be encoded in multiple code patches. With this in mind, hypergraph product codes (HGPs)~\cite{Tillich2014} are a natural choice, as the two-code product structure could be adapted to the plaquette lattice at hand. The HGP construction takes two classical codes with parity check matrices $H_1$ and $H_2$, and makes a quantum CSS code with $X$ and $Z$ parity checks given by~\cite{Tillich2014, Kovalev2013}:
\begin{align}\label{eq:HGP}
    H_X&=(H_1\otimes I_2 \hspace{8pt} I_1\otimes H_2), \\ H_Z&=(\tilde{I}_1\otimes H_2^T \hspace{8pt} H_1^T\otimes \tilde{I}_2).\notag
\end{align}

The cyclic shift automorphism we require can now be directly inherited from the underlying classical codes~\cite{Kovalev2013}. Indeed, if $H_1$ and $H_2$ are both cyclic codes~\cite{Macwilliams1977, berthusen2025}, the resulting quantum HGP code will inherit the cyclic shift automorphism of the two classical codes~\cite{Xu2025}, thus directly realizing the cyclic code shifts along primitive plaquette lattice directions we require. The classical linear binary cyclic LDPC codes are ubiquitous and could be systematically generated through generator polynomials $g(x)$, such that $g(x)h(x)=x^n-1$ over the binary field, where $h(x)$ defines the check polynomial of the code~\cite{Macwilliams1977, Kovalev2013}. An example of a particularly useful classical cyclic class are the Bose–Chaudhuri–Hocquenghem (BCH) codes~\cite{bose1960class, hocquenghem1959codes}. These codes will be generally of the form $[2^m-1, k, d]$, with a constant encoding rate due to the lower bound for $k$ given by $k\geq n-m (d-1)$~\cite{Macwilliams1977}. One potentially useful HGP class construction based on the classical BCH codes was presented in Ref.~\cite{Kovalev2012, Kovalev2013} and is achieved by taking cyclic $H_1=H_2$, resulting in parameters $[[2n^2, 2k^2, d]]$. In order to construct a self-dual CSS code, anticipating requirements for fold-transversal single-qubit Cliffords in Sec.~\ref{sec:qldpc_h}, one would have to require self-orthogonality of the classical code, $H_1H_1^T=0$, a property that is realized in a subclass of BCH codes~\cite{Macwilliams1977}. An example of the code constructions with the BCH representative, $[7,4,3]$ Hamming code, is presented in Fig.~\ref{fig:Fig9_qLDPC_STAR}(c). 

The cyclic shift automorphisms in the self-dual HGP construction generally act within the subsets of either homological or cohomological~\cite{Breuckmann2021} logical code qubits, whose logical operator supports are disjoint. These two classes of logical qubits arise naturally from the hypergraph product structure: homological logical qubits correspond to those whose logical $Z$ operators are supported on the left physical qubit block, which is associated with the tensor product of the column (bit) spaces of the classical codes. In contrast, cohomological logical qubits have logical $Z$ operators supported on the right physical qubit block, stemming from the tensor product of the row (check) spaces of the classical parity-check matrix. Generically, the two subsets differ in size due to the asymmetry in the classical parity-check matrix. This is not a limitation; rather, it means that a typical quantum hardware implementation will execute two quantum simulations in parallel, one on each subset of dual qubits, possibly with differing system sizes. In particular, for the HGP construction based on classical BCH codes with parameters $[[2n^2,2k^2,d]]$, as shown in Fig.~\ref{fig:Fig9_qLDPC_STAR}(c), the equality of homological and cohomological logical qubit spaces is ensured. Finally, we note that for cyclic HGP constructions, the derived shift automorphisms do not close periodically, since the last logical representative in a sequence is not mapped to the first one. This is, however, generally not a significant limitation for simulations with open boundary conditions; the superfluous operation resulting from the shift of the last qubit is uncomputed during the inverse shift. 

Beyond cyclic HGP codes based on classical BCH codes, a related construction can be based on duals of self-orthogonal BCH codes. Among these, binary simplex codes with parameters $[2^{m}-1, m, 2^{m-1}]$ are of particular interest due to good code distances and self-orthogonality for $m \geq 3$~\cite{Macwilliams1977}, and solid encoding rates for intermediate $m$. The HGP construction thus leads to self-dual quantum codes with parameters $[[(2^{m}-1)^2, 2m^2, 2^{2m-1}]]$. Recently, a general search for cyclic HGP codes with good parameters and performance in~\cite{Aydin2025}, has reported simplex-based $[[450,32,8]]$ HGP code as having megaquop suitable memory performance under circuit level noise ($4.5 \times 10^{-7}$ logical error rate per qubit at $p_{\mathrm{ph}}=10^{-3}$). Furthermore, other non-simplex self-orthogonal BCH duals have been reported, demonstrating the construction's general promise.

More generally, the cyclic shift automorphism property we require could be inherited from specific classes of quasicyclic classical codes~\cite{Xu2025}, or a lifted product code construction such as bivariate-bicycle (BB) codes~\cite{Bravyi2024, eberhardt2024}, providing a variety of co-design opportunities.  Among these, recent work~\cite{Xu2025batched} has reported a family of self-dual BB codes that satisfy all symmetry requirements necessary for parallel, transversal STAR operation. In contrast to cyclic HGP codes, these codes have shift automorphisms generated from a single direction only, but the automorphisms are fully periodic. With appropriate lattice plaquetization and associated problem-size congruence, the construction is suitable for quantum simulation in higher dimensions. For example, a typical 2D lattice can be plaquetized with plaquettes spanning the full width (or height) of the lattice. The parameters of the reported self-dual BB codes, such as $[[66,6,8]]$ and $[[78,6,10]]$ are generally suitable for simulation of common system sizes with such plaquettization. In addition, the memory performance reported for $[[66,6,8]]$ is near-megaquop ($3 \times 10^{-6}$ logical error rate per qubit at $p_{\mathrm{ph}}=10^{-3}$), with $[[78,6,10]]$ likely being sufficient to reach the error rate limits of transversal STAR architecture.

\subsection{Fold-transversal single-qubit Cliffords from code dualities}
\label{sec:qldpc_h}

Similarily, the parallel single-qubit Clifford gates, $H_p/S_p$ are not generically fold(swap)-transversal, and a careful co-design of the code properties is needed. In particular, the Hadamard gate can be present in the self-dual quantum CSS codes, codes invariant under the exchange of $Z$ and $X$ parity checks~\cite{eberhardt2024, Breuckmann2024}. The construction of the corresponding swap-transversal Hadamard gate proceeds in a similar fashion as with cyclic automorphisms --- for a $ZX$ duality $\tau$, the swap-transversal Hadamard is a composition of the corresponding duality induced physical swaps and the parallel transversal Hadamard gates, $H_p=\prod_{q \in \mathcal{C}} \textrm{SWAP}_{q \rightarrow \tau^{-1}(q) }\prod_{q \in \mathcal{C}}{H_q}$~\cite{eberhardt2024, Breuckmann2024} [see Fig.~\ref{fig:Fig9_qLDPC_STAR}(b)]. For the HGP construction described above, self-duality is ensured for $H_2=H_1$ and self-orthogonal classical codes with $H_1H_1^T=0$. Thus, the self-orthogonal BCH-based HGP class example by construction supports a swap-transversal Hadamard gate, including the $[[98,32,3]]$ code depicted in Fig.~\ref{fig:Fig9_qLDPC_STAR}(c). 

Hadamard gate alone is sufficient to provide a gateset that can realize classically-hard Hamiltonian simulations, with simulation space limited to real Hamiltonians. For complex Hamiltonians, a fault-tolerant swap-transversal $S_p$ gate is desirable. Although such constructions, akin to the surface code $S$-gate, are more involved, it is plausible that they could follow from underlying code symmetries similar to $H_p$, in particular when the $ZX$-duality of the code is self-inverse, $\tau^2=I$~\cite{Breuckmann2024}. While this condition seems broadly applicable for the self-dual HGP codes, we leave such constructions for future work.

We note that swap-transversal single-qubit Cliffords, such as Hadamards, do affect the order of the qubits in the patches they are applied to, and one might worry that this will affect the calculation, as the follow-up parallel transversal CNOTs are not applied between the right qubits. In a general case, this cannot be simply addressed by logical SWAP gates, as $ZX$-duality does not provide a pure logical SWAP gate, in contrast to shift automorphisms that are easily invertible. In a wide class of problem instances of interest, however, Hadamard matrices are used to change the basis globally between the applications of the specific Hamiltonian terms. Critically, in these cases, swap-transversal Hadamards permute qubit order in all logical qubits equally, allowing for quantum simulation to proceed with no additional logical gates. For example, in the paradigmatic transverse field Ising case, all Hadamard layers in the circuit are global, and Hadamard-related SWAPs are not an issue. A similar property is also valid for all real Hamiltonians where the Hamiltonian Pauli terms between any two CNOT layers in the Trotter decomposition can be connected by a global basis transformation. We note that the global Hadamard layer property does not require translational invariance of the Hamiltonian or periodic boundary conditions, as one can choose to perform global Hadamard layers and then make specific entangling gates identity by the choice of injected angles. 

\subsection{Small-angle injection for high rate codes}
\label{sec:qldpc_injection}

The final component necessary to complete the transversal qLDPC-STAR simulation gateset is the STAR-based injection of  $R_{Z_p}(\{\theta_i\})$ to the factory patches. The STAR small-angle injection protocol and its basic error rate scaling are directly applicable to all CSS codes. The protocol proceeds equivalently to the surface code with post-selection-based fault-tolerant $\ket{+}_L$ state preparation. Generically, for a given logical qubit of a high-rate code, the logical $Z_L$ operator can be chosen as a product of physical $Z$s in the logical operator support, $Z_L=\prod_{j\in \textrm{supp}} Z_j$. The logical $Z$ rotation on the $\ket{+}_L$ can then be straightforwardly done by any combination of physical rotations that cover physical support without overlaps and the stabilizer postselection that chooses the right logical rotation:
\begin{align}\label{eq:injection-qldpc}
&R_{Z_p}(\{\theta_i\})_L \ket{+}_L=\cr &P_{\substack{\mathrm{postselect} \\ \mathrm{to \, logical}}} \left[ \prod_{\{Z...Z\}^j \in \mathrm{cover}(\textrm{supp})}R_{\{Z...Z\}^j}{(\theta^*_i)}\ket{+}_L\right].
\end{align}

The decomposition of the rotation product of the right-hand side directly governs the stabilizer pattern postselection. As long as one makes sure all terms arising in the physical rotation product other than our desired logical rotations do trip certain $X$-stabilizer patterns, one recovers the same error rate scaling of the injection protocol as in the surface code case. This is easily ensured by vast majority of the protocols that cover physical support of the $Z_L$ without overlaps. The structure of the polynomial on the right-hand side does not depend on the underlying properties of the code, only on the parameters of the transversal multi-rotation and the number of qubits in the support of $Z_L$. Thus, the analytical scaling of the small-angle logical error rates and postselection rates is as derived in Ref.~\cite{Toshio2024}. Specifically, the effective logical error rate will be $\propto p_{\mathrm{ph}} |\theta_i|$~\cite{Choi2023, Toshio2024}. The postselection rate ($p_s$) in the zeroth order in physical error rate will depend on the number of rotations used to cover the support ($l$) as $p_s\propto 1-|\theta_i| l$~\cite{Toshio2024}.

The small-angle injection protocol described so far allows for the injection of rotation on a single logical qubit of the qLDPC code classes of interest for transversal STAR. Injecting the whole code patch with an arbitrary pattern of small angles is now straightforward. The simplest case is the serial protocol presented in Fig.~\ref{fig:Fig9_qLDPC_STAR}(c), where each logical rotation is done in the factory patch after the previous rotation protocol is completed, including postselection. For a code with $k$ logical qubits per patch, the total postselection rate of such a factory scales as $p_s\propto 1-l k|\theta|$, assuming equiangular rotations for simplicity, while the protocol time is proportional to $k$. 

The decrease in the postselection rates due to high-rate code injection can be mitigated to some extent by the TMR protocol choice with lower $l$. Alternatively, if the code distance is such that the Clifford error rates are subleading (STAR limit), one could choose to inject into factory patches partially, recovering the postselection rate at the cost of extra protocol time or factory space. Many of the typical code classes one would consider for transversal qLDPC-STAR allow for a significant speedup in the protocol time and thus overall reduction in factory space-time by semi-parallel injection protocol, as depicted in  Fig.~\ref{fig:Fig9_qLDPC_STAR}(c) for $[[98,32,3]]$ HGP code example. In fact, for all HGP codes, one could order the qubits in the code patch in a canonical order on two rectangular grids (one for left and right code qubits)~\cite{Quintavalle2023}, such that each logical qubit has logical operator support only on a column or a row of a grid defined by its physical pivot qubit. In practice, this means that in a typical HGP code encoding $k$ logical qubits, $\propto \sqrt{k}$ of these qubits can be chosen at once without overlapping logical operator support and thus injected in parallel. While the overall scaling of postselection and error rates does not change compared to serial injection, the full injection can thus be performed in $\sqrt{k}$ time less, allowing for faster injection factories. The qubit ordering in the example HGP construction from self-orthogonal BCH codes is canonical by construction, and the semi-parallel injection protocol is manifest in the structure of the code patch, as shown in Fig.~\ref{fig:Fig9_qLDPC_STAR}(c).

\subsection{Transversal qLDPC-STAR summary}
\label{sec:qLDPC_summary}

We have thus completed the necessary transversal qLDPC-STAR gateset that allows for effective, highly parallel, and transversal Trotterized quantum simulation of relevant local spin Hamiltonian classes with high-rate quantum codes. The construction relies critically on careful co-design between the code choice and the simulation problem at hand in order to provide all of the necessary gateset components. The whole construction requiring parallel transversal gates, qubit reordering for shift automorphism SWAPs, and duality-based single-qubit Clifford depends heavily on the dynamical reconfigurability capabilities of the neutral atom platform.

While fixed physical architectures that implement high-rate codes can be constructed~\cite{Bravyi2024}, the large amount of task-specific architecture customizations necessitates dynamical reconfigurability. The code-specific error rates in the circuit setting are left to be evaluated in future work according to the specific simulation problem of interest. Still, estimates of space savings and approximate high-rate transversal STAR architecture speed can be made reliable with currently known codes and assumptions of STAR error rates similar to surface code. In particular, in Tab.~\ref{tab:parameters_estimates} we provide such estimates for the square lattice nearest neighbor Ising model with architecture based on recently discovered self-dual bivariate bicycle codes~\cite{Xu2025batched}. This high-rate transversal STAR implementation provides overall $\sim 5\times$ space savings ($\sim 10 \times$ per individual computation), at a speed comparable to only $\sim 2\times$ that of surface code transversal STAR, showcasing the potential advantages.

Another relevant open question we do not address is the potential to adapt transversal qLDPC-STAR protocols for fermionic simulation. The current construction is not applicable to the fermionic simulation of local Hamiltonians through compact Jordan-Wigner mapping~\cite{Jordan1928}, as the fermionic SWAPs cannot be readily built from our gateset. It is plausible that additional code co-design that takes into account fermionic mappings could solve this issue. Alternatively, one could tap into a significant catalog of fermionic mappings, in particular the Verstraete-Cirac~\cite{Verstraete2005, Evered2025} or other locality-preserving mappings~\cite{Bravyi2002, Derby2021}, and thus recover the transversal qLDPC-STAR simulability with some additional space overhead for the mapping itself.

\section{Outlook}
\label{sec:outlook}

In this paper, we propose and evaluate a transversal STAR architecture~\cite{Akahoshi2024, Toshio2024} based on fault-tolerant logical Clifford gates and non-fault-tolerant, high-fidelity small-angle injection~\cite{Choi2023}. We show that this transversal STAR architecture can utilize fast (fold) transversal logical Clifford operations~\cite{Cain2025, Zhou2024} in the correlated decoding setting, achieving sizable space-time volume savings at both the logical and physical levels over the original STAR proposals~\cite{Akahoshi2024, Toshio2024}. We perform a careful evaluation of the logical noise model for all logical STAR gadgets, based on the hardware-realistic physical noise model, in the surface-code framework. We take special care to properly isolate the logical gadgets from boundary effects and devise a method for the reliable extraction of a logical noise model of an isolated gadget in the correlated decoding setting. 

Our approach is validated by showing explicit logical noise model composability for jointly-decoded logical gadgets. The detailed evaluation of injection and teleportation fidelities presents a clear path towards megaquop-scale~\cite{Preskill2025} quantum simulation based on transversal STAR. We explore the limits of the surface-code-based transversal STAR for quantum simulation, showing that it allows unprecedented simulation volumes ($N_ST\gtrsim600$) compared to NISQ, while at the same time presenting overall $100\text{--}1000\times$ space-time savings over the best-known fully fault-tolerant approaches~\cite{Gidney2024}. Finally, we construct an extension of the transversal STAR-based qLDPC codes~\cite{Breuckmann2021}, thus presenting a promising path for additional space savings. Through co-design between quantum codes and underlying simulation tasks, we show how limited, highly-parallel fold- and swap-transversal qLPDC gateset coupled with STAR small-angle factories can be effectively utilized for quantum simulation of local lattice Hamiltonians.

Our work opens up several avenues for future explorations. The existing neutral atom quantum hardware~\cite{Bluvstein2024, Rodriguez2024} could be employed to experimentally validate the main features of the proposal.  Indeed, our results indicate that the critical angle scaling of logical injection fidelity could be probed with state-of-the-art physical neutral atom fidelities. Further hardware advancements in both physical error rate and the number of qubits will allow smooth progression in terms of available transversal STAR demonstrations and quantum simulation volumes. In parallel, the transversal qLDPC-STAR architecture we propose can further be evaluated in terms of detailed circuit-level simulations of STAR-relevant logical gadgets for promising high-rate code classes~\cite{Yoder2025, Ataides2025, Yang2025}. The close connection between the quantum simulation problem at hand and the STAR simulation power motivates more detailed resource estimates and optimal use case choices for both surface code~\cite{Akahoshi2024b} and qLDPC-based STAR constructions. 

Trotterized quantum simulation, the main STAR application, can be utilized as the main subroutine in a wide range of algorithmic frameworks, each of which can benefit from a detailed evaluation of the transversal STAR approach. Beyond directly expanding the limits of non-equilibrium quantum dynamic simulation~\cite{Haghshenas2025}, particularly promising algorithmic uses of transversal STAR could include quantum-classical hybrid ground state methods that rely on relatively short Trotter dynamics~\cite{Ding_2023, Mi_2024}. Principal among those include quantum subspace methods~\cite{Epperly2022, Shen2023, Yoshioka2024} on their own or in conjuction with sampling-based diagonalization~\cite{Klymko2021, Yu2025, Moreno2025}. Finally, novel algorithmic compilation methods could be synergistic with transversal STAR~\cite{Campbell2019, gunther2025phaseestimationpartiallyrandomized} --- in particular, the recent proposal of randomized compilation-based magic state dilution~\cite{Luthra2025} could explicitly benefit from including high fidelity small-angle magic into the gate basis used for compilation. 

\begin{acknowledgments}
We acknowledge stimulating discussions with Casey Duckering, Nishad Maskara, Gefen Baranes, Shayan Majidy, Alexander Schuckert, Takuya Kitagawa, Jonathan Wurtz, Joseph Carlson, Ionel Stetcu, and Ivan Chernyshev. This work was supported by the NNSA ASC Beyond Moore's Law project (I.C.C. and A.T.S.). R.W. acknowledges support by the Edwin Thompson Jaynes Postdoctoral Fellowship of the Washington University Physics Department. The numerical studies were performed on the high-performance computing system Perlmutter, a NERSC resource, using NERSC award DDR-ERCAP0030190. The LANL designation for this manuscript is LA-UR-25-29344.
\end{acknowledgments}

\appendix
\section{Decoder performance benchmarks}
\label{sec:decoders}

Here, we evaluate how well different open-source decoding strategies handle transversal Clifford circuits encoded in the rotated surface code. Our test circuits consist of transversal initialization, seven repeated applications of a single logical gate ($I$, $H$, $S$, or CNOT) with one syndrome extraction after each gate application, and transversal measurement. We test across code distances $d \in \{3, 5, 7\}$ and varying physical error rates.

These transversal circuits present unique decoding challenges due to the presence of higher-order hyperedges in their decoding graphs. As a baseline, we use the standard minimum-weight perfect matching decoder via PyMatching \cite{Higgott2025}, which is optimized for decoding surface code memory circuits involving only edge-like hyperedges (i.e., hyperedges connecting just two nodes). We also evaluate logical matching \cite{Cain2025b, serraperalta2025}, which extends standard matching to support hypergraph structures in transversal Clifford circuits.

To address the broader challenge of arbitrary hypergraph structures, we test the Minimum-Weight Parity Factor (MWPF) decoder \cite{wu2025mwpf}, which generalizes minimum-weight perfect matching concepts to arbitrary hypergraphs. Additionally, we consider two belief-propagation-based decoders—BP-OSD and BP-LSD—which adapt classical LDPC decoding techniques to the quantum setting. These decoders incorporate post-processing steps to account for quantum-specific features like degeneracy \cite{Roffe_2020, Roffe_LDPC_tools, delfosse2021}.

All decoders (with exception to MWPF) were evaluated using their default, out-of-the-box configurations, with no decoder-specific parameter optimization or circuit-dependent tuning performed. Our goal is to compare baseline performance and computational cost under typical usage conditions rather than to identify optimal decoder settings. We note that the absolute performance of several heuristic decoders, particularly belief-propagation-based approaches, may improve under further parameter optimization; however, a systematic tuning study lies beyond the scope of this work. All decoder benchmarks were run on NERSC's Perlmutter supercomputer (CPU-only nodes). Each node contains two AMD EPYC 7763 processors (128 cores / 256 threads, 2.45 GHz base) with 512 GB DDR4 memory. The operating system was SUSE Linux Enterprise Server 15 SP5.

\subsection{Decoder performance at fixed low physical error rate}
We first compare decoder performance at a fixed low physical error rate, $p_{\mathrm{ph}} = 7.3 \times 10^{-4}$, focusing on the trade-off between logical error suppression and decoding runtime. The top-left panel of Fig.~\ref{fig:Fig10_decoder_comparison} shows the logical error rate versus mean decoding time per shot for different decoders and circuit types.

This comparison highlights the fundamental limitations of standard minimum-weight perfect matching when applied to transversal Clifford circuits containing $S$ or CNOT gates. Due to the introduction of higher-order hyperedges in the decoding graph, standard matching fails for these circuits, exhibiting logical error rates on the order of $10^{-2}$. In contrast, decoders capable of handling hypergraph structures achieve logical error rates in the $10^{-5}$ to $10^{-6}$ range.

We tested three variants of the MWPF decoder corresponding to different cluster node limit (CNL) hyperparameter settings: 0, 50, and 500. Increasing the CNL parameter improves logical error suppression at the cost of substantially increased runtime. The fastest variant, $\text{MWPF}_0$—which is equivalent to Hypergraph Union Find \cite{wu2025mwpf}—remains approximately two orders of magnitude slower than matching-based approaches. Increasing the CNL from 0 to 50 yields negligible logical error rate improvements for most circuit types, except the identity circuit, while increasing to $\text{CNL}=500$ provides the strongest error suppression at an additional two orders of magnitude runtime cost.

Belief-propagation-based decoders (BP-OSD and BP-LSD) exhibit runtimes roughly two orders of magnitude longer than $\text{MWPF}_0$ and $\text{MWPF}_{50}$, while providing no improvement in logical error rate suppression. In fact, these decoders perform worse than MWPF variants for circuits consisting solely of $I$ or $H$ gates.

Notably, logical matching bridges this gap by extending the microsecond-scale performance of standard matching to circuits containing $S$ and CNOT gates. At this low physical error rate, logical matching achieves execution times five orders of magnitude faster than $\text{MWPF}_{500}$ while incurring less than one order of magnitude degradation in logical error suppression.

The top-right panel of Fig.~\ref{fig:Fig10_decoder_comparison} examines how decoding runtime scales with code distance for CNOT circuits at lower physical error rates ($p_{\mathrm{ph}} = 7 \times 10^{-4}$). As the code distance increases from $d=3$ to $d=7$, the runtime gap between decoder families widens substantially. Matching-based approaches, including logical matching, retain microsecond-scale decoding times even at the largest distance studied, whereas MWPF variants exhibit rapidly increasing runtimes. In particular, $\text{MWPF}_{500}$ becomes five orders of magnitude slower than logical matching at $d=7$, as opposed to only three orders of magnitude at $d=3$.

\subsection{Decoder behavior as a function of physical error rate}
\begin{figure}[!htb]
\centering
\includegraphics[width=0.5\textwidth]{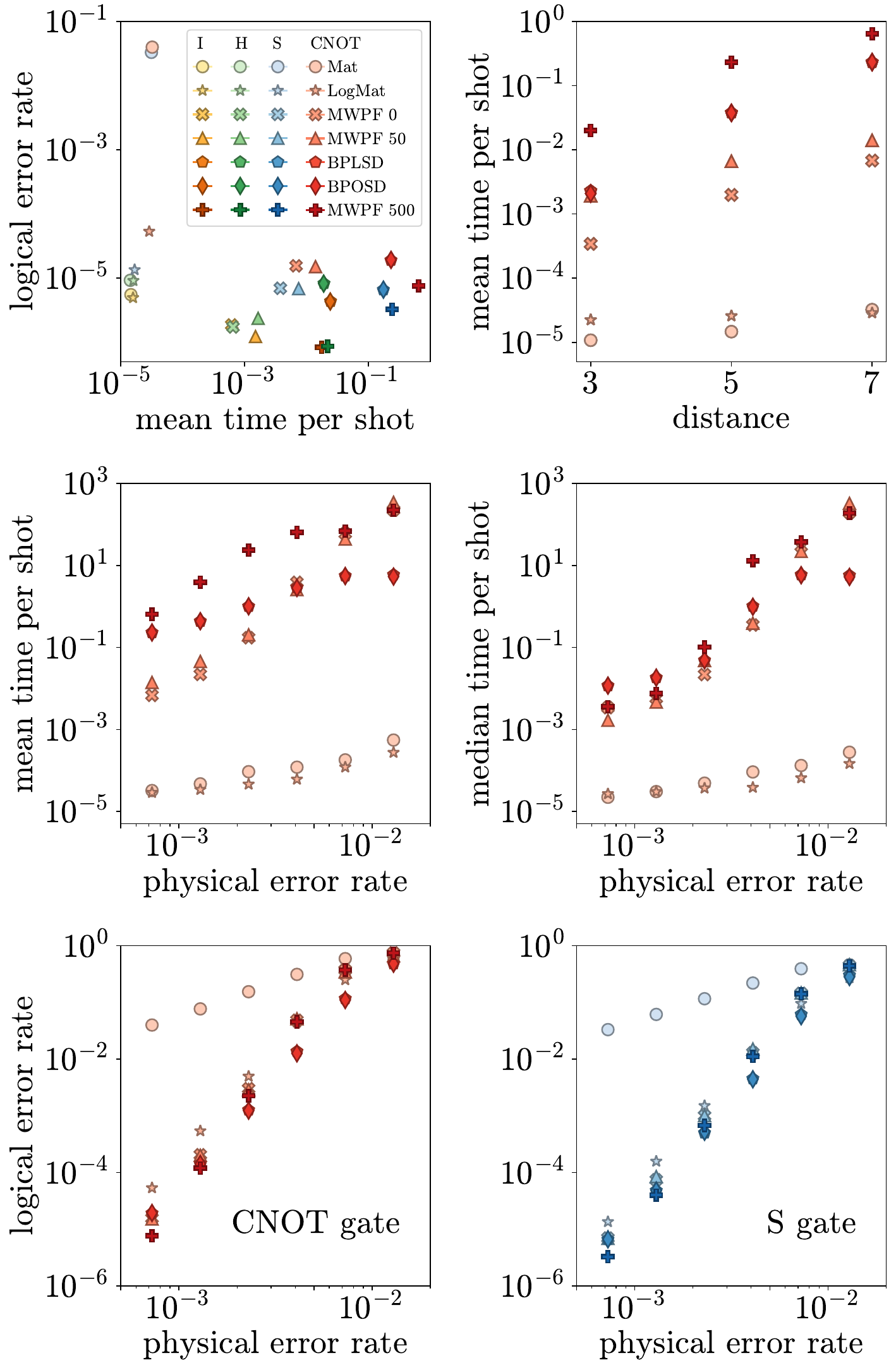}
\caption{ \textbf{Decoder benchmarking for transversal Clifford circuits.} Top left: Logical error rate versus mean decoding time per shot at the lowest tested physical error rate ($p_{\mathrm{ph}} = 7.3 \times 10^{-4}$), revealing a five-order-of-magnitude span in computational cost for approximately one order of magnitude variation in error suppression among hypergraph-capable decoders (excluding standard matching). Each point represents a specific combination of circuit type ($I$, $H$, $S$, CNOT) and decoder. Top right: Mean decoding time per shot as a function of code distance for CNOT circuits ($p_{\mathrm{ph}} = 7.3 \times 10^{-4}$). Logical matching (LogMat) maintains microsecond-scale performance even at distance 7, while MWPF-500 climbs to the second regime. Middle row: Decoder performance versus physical error rate for CNOT circuits at distance 7, showing both mean time per shot (left) and median time per shot (right). All decoders exhibit slower performance near the QEC threshold (around $7.3 \times 10^{-3}$). Bottom row: Logical error rate as a function of physical error rate for all decoders considered, shown for both CNOT (left) and $S$ (right) circuits at code distance $d=7$.}
\label{fig:Fig10_decoder_comparison}
\end{figure}

We next examine how decoder performance evolves as the physical error rate is varied, considering both decoding runtime and logical error suppression. The middle row of Fig.~\ref{fig:Fig10_decoder_comparison} shows the mean and median decoding time per shot as a function of physical error rate for CNOT circuits at code distance $d=7$.

\begin{figure*}
    \centering
    \includegraphics[width=1.0\textwidth]{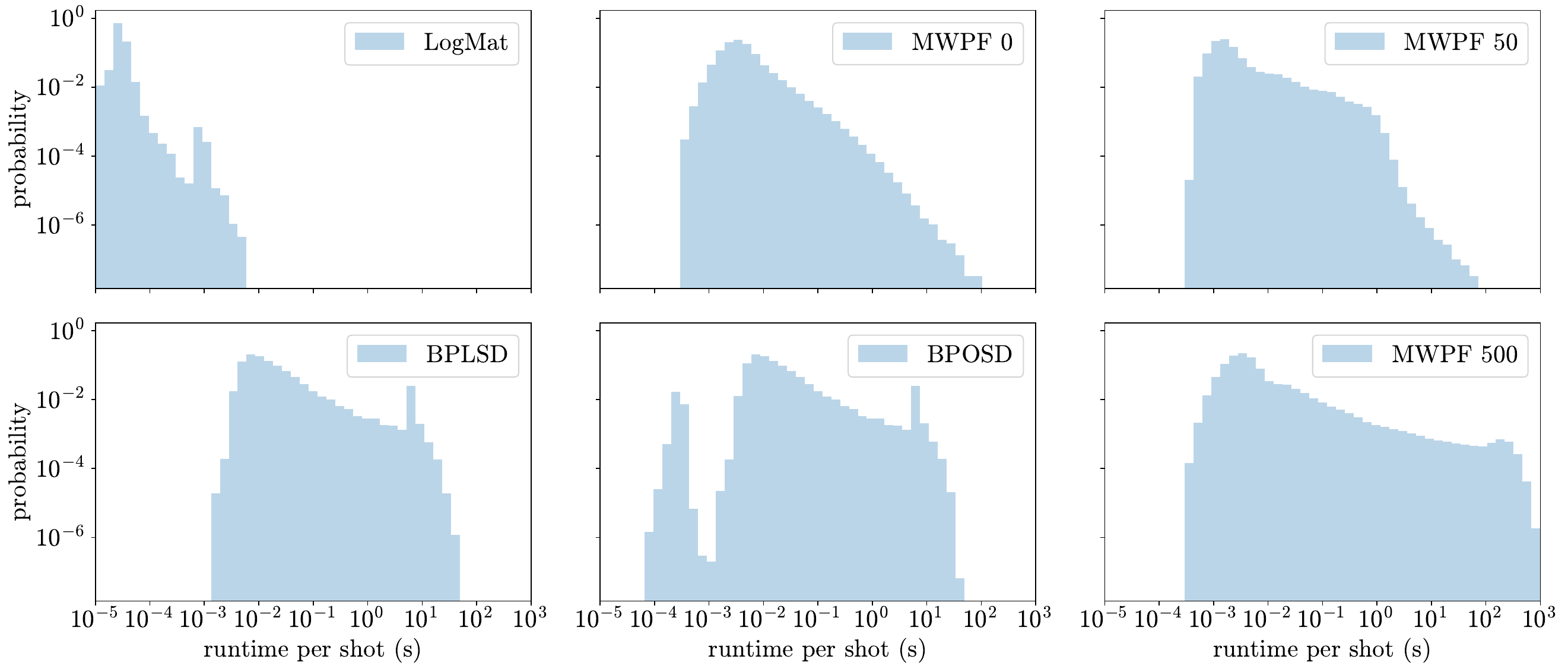}
    \caption{ \textbf{Runtime variability across decoders at low physical error rates.}
 Histogram of per-shot decoding runtimes for all decoders on the CNOT circuit at $d=7$ and physical error rate $p_{\mathrm{ph}} = 7.3 \times 10^{-4}$.  Each histogram was generated using a different total number of shots, chosen so that all decoders achieve comparable relative statistical uncertainty in their estimated logical error rates. In practice, this required approximately $2$–$3 \times 10^{6}$ shots per decoder. The distributions reveal pronounced heavy-tailed behavior for MWPF variants, particularly at high cluster node limits, explaining the observed discrepancy between mean and median decoding times at low error rates.}
    \label{fig:Fig11_decoder_histogram}
\end{figure*}

All decoders exhibit increased decoding time near the QEC threshold (around $7.3 \times 10^{-3}$), with runtime gaps expanding from approximately five orders of magnitude far below threshold to nearly seven orders of magnitude close to threshold. Interestingly, for $\text{MWPF}_{500}$, the median decoding time decreases more rapidly than the mean at low physical error rates. This divergence indicates a heavy-tailed runtime distribution: while most syndrome instances decode quickly when errors are sparse, rare but exceptionally difficult instances disproportionately inflate the mean decoding time. The heavy-tailed nature of these runtime distributions is made explicit in Fig.~\ref{fig:Fig11_decoder_histogram}, which shows the full distribution of per-shot decoding runtimes for all decoders on the same CNOT circuit at $d=7$ and $p_{\mathrm{ph}} = 7.3 \times 10^{-4}$.

The bottom row of Fig.~\ref{fig:Fig10_decoder_comparison} shows the logical error rate as a function of physical error rate for all decoders considered in this study, for both CNOT and $S$ circuits at code distance $d=7$. Across the full range of physical error rates, the qualitative behavior is consistent between the two circuit types. At current physical error rates ($p_{\mathrm{ph}} \approx 7.3 \times 10^{-3}$), all decoders achieve comparable logical error rates. In this regime, belief-propagation-based decoders (BP-OSD and BP-LSD) exhibit marginally lower logical error rates than other approaches. In this regime, logical matching attains logical error rates similar to those of more computationally intensive decoders, including MWPF variants and belief-propagation approaches, while offering decoding times that are orders of magnitude shorter and free of heavy-tailed decoding runtime. Consequently, logical matching emerges as the most suitable choice for real-time decoding under present hardware constraints.

Looking ahead to future operating regimes with lower physical error rates, corresponding to more deeply sub-threshold operation, $\text{MWPF}_{500}$ achieves the lowest logical error rates among all decoders tested. Other MWPF variants and the belief-propagation-based decoders follow closely behind, while logical matching exhibits logical error rates approximately one order of magnitude higher. Despite this gap, logical matching remains substantially faster, preserving its advantage in scenarios where decoding time is a critical constraint. Overall, these trends are consistent across both CNOT and $S$ circuits, indicating that the observed trade-offs are not strongly gate-dependent.

 In such lower-error operating regimes, our results indicate several viable decoding pathways. First, logical matching can be used, incurring only a modest penalty in logical error rate while preserving low and predictable decoding time. Second, MWPF-based decoders become viable as well when coupled with parallelized implementations which may mitigate decoding time concerns \cite{mwpf_parrallel2024}. Third, recently proposed decoding approaches based on machine learning offer an additional alternative, combining competitive logical error suppression with favorable amortized runtime \cite{ML_decoder2025}. Finally, dedicated hardware acceleration, such as GPU- or FPGA-based implementations~\cite{Wu2023, Maurer2025}, provide a complementary route to reducing decoding speed across these approaches. Taken together, these results indicate that decoder speed does not pose a fundamental feasibility barrier under either current or anticipated future operating conditions. 

We emphasize, however, that the absolute runtimes reported here correspond to CPU-based reference implementations. While some decoding implementations can meet the latency requirements for large-scale real-time deployment—typically targeting end-to-end decoding times in the $\sim 100,\mu\mathrm{s}$–$1,\mathrm{ms}$ range—these are not the implementations that achieve the lowest possible logical error rates. The higher-performing decoders considered here attain lower logical error rates at the expense of increased runtime. Bridging this gap—i.e., achieving real-time latency while preserving optimal logical performance—will require further engineering advances, including parallelization and hardware-accelerated implementations, which lie beyond the scope of the present study.

\section{Implementation for transversal Cliffords in surface code}
\label{sec:Cliffords_implementation_details}
Here, we describe our physical implementation of logical gates in the rotated surface code. Our approach implements Clifford operations transversally to harness correlated decoding capabilities, utilizing a complete basis of Clifford gates consisting of $\{H, S, \textrm{CNOT}\}$.

\textbf{Logical $H$} is implemented by applying physical $H$ gates to all code qubits, followed by a 90-degree rotation of the entire code patch. This rotation is necessary to restore the proper relative orientation with respect to other code patches in the circuit, ensuring that subsequent inter-patch operations maintain their intended logical functionality.

\textbf{Logical $S$}: we contrast two different implementations for $S$:
 \paragraph{Intra-SE.}
 This recently proposed method from Ref.~\cite{Chen2024} applies the constituent gates halfway during the syndrome extraction (SE) process. This approach takes advantage of the similarity between the rotated surface code stabilizer pattern during SE and an unrotated surface code one. The implementation consists of two components:
1) Controlled-Z operations: CZ gates are applied between mirror qubits positioned along the diagonal of the code patch. 2) Phase gates: An alternating pattern of $S$ and $S^{\dagger}$ gates is applied to qubits lying directly on the diagonal. This mid-syndrome-extraction approach maintains the distance-preserving properties of the code while implementing the desired logical operation.

\paragraph{Inter-SE.}

\begin{figure}[htb]
\centering
\includegraphics[width=0.5\textwidth]{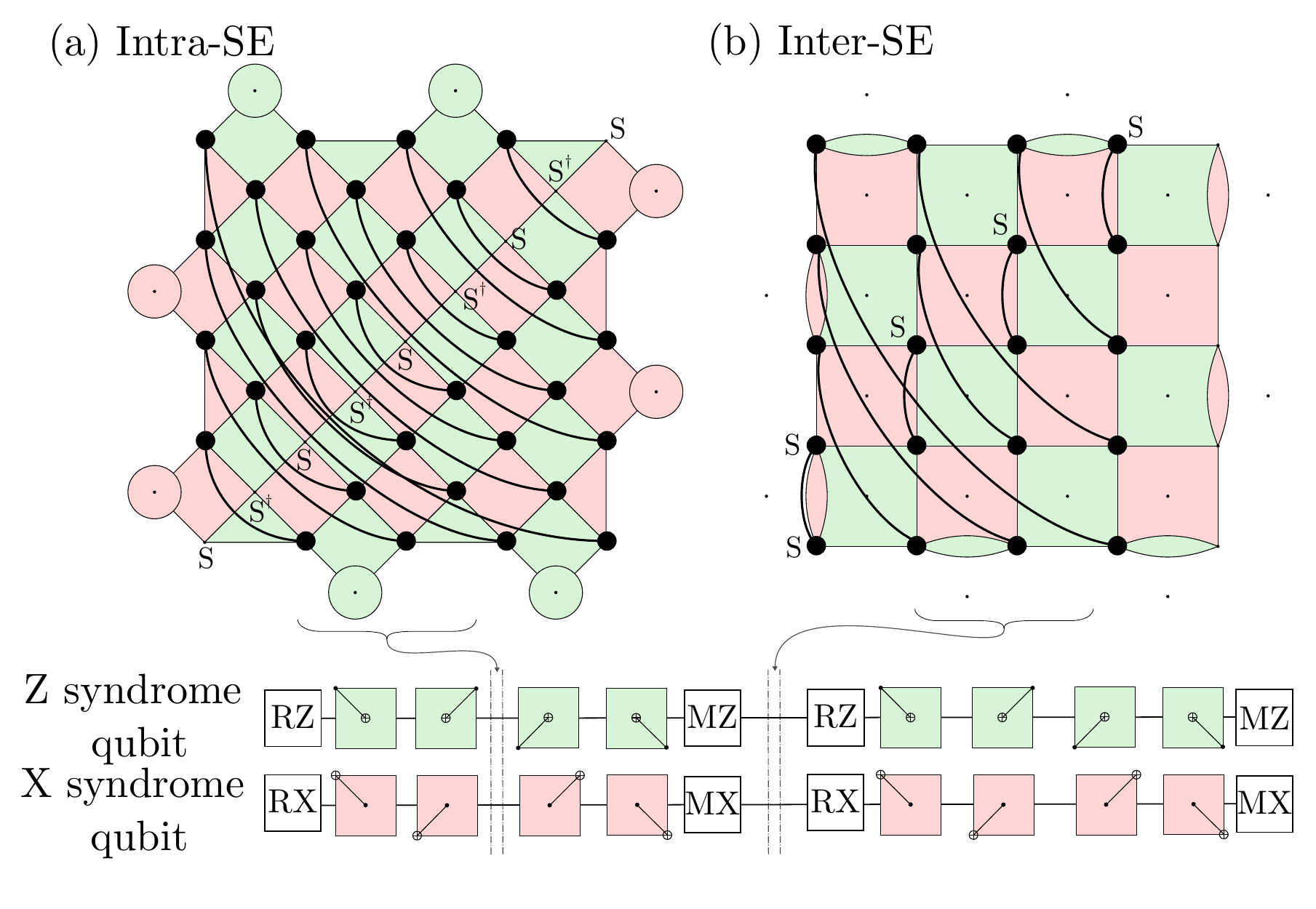}
\caption{ \textbf{Two implementations of logical $S$ gate in rotated surface code patch.} We illustrate detector slice diagrams for implementing the $S$ gate in a distance-5 rotated surface code. Each panel shows the physical gates required for the logical operation—CZ, $S$, and $S^\dagger$ gates—as well as the corresponding detecting regions at the moment the $S$ gate is applied. Light red indicates $X$-type detectors and light green indicates $Z$-type detectors. The detecting regions shown correspond to the baseline scenario without the logical $S$ gate; changes to the detector setup due to the $S$ gate are not depicted. (a) \textit{Intra-SE} CZ gates~\cite{Chen2024} are applied to pairs of qubits mirrored across the diagonal, with alternating $S$, and $S^\dagger$ gates along the diagonal. This implementation occurs mid-cycle during syndrome extraction, after syndrome qubits have coupled to two of their four nearest-neighbor data qubits. The detecting regions include both data and active ancilla syndrome qubits. (b) \textit{Inter-SE} CZ gates are shifted off the main diagonal, accompanied by $S$ gates. This implementation is applied between consecutive syndrome extraction cycles. The bottom panels depict the syndrome extraction cycle for both $X$ and $Z$ syndrome qubits in the bulk: reset, sequential coupling to four nearest-neighbor data qubits (ordered to avoid hook errors), and measurement.}
\label{fig:Fig11_transversal_gates}
\end{figure}

An alternative construction applies CZ gates between the syndrome extraction rounds rather than during them. In this approach, CZ operations are performed across qubits that are mirrored with respect to a shifted diagonal pattern (see Fig.~\ref{fig:Fig11_transversal_gates}).

While this conventional implementation requires fewer CZ gates and operates exclusively on data qubits (avoiding ancilla qubits since it is not performed during syndrome extraction), it reduces the code distance by 1. This distance reduction leads to degraded performance as the code distance scales when compared to the intra-SE implementation. Therefore, we employ the intra-SE implementation method for all logical $S$ gate operations.

\textbf{Logical CNOT} gate between two code patches is implemented transversally by applying physical CNOT gates between each pair of corresponding qubits in the respective patches.

\section{SE rounds optimization in logical gadgets}
\label{sec:SE_benchmark}

We conducted a systematic simulation study to determine the optimal number of syndrome extraction (SE) rounds to include in each logical gadget. In our framework, a logical gadget is defined as the combination of a logical gate operation followed by a specified number of SE rounds. We explored configurations ranging from 1 to $d$ SE rounds per gadget. The analysis was performed across three code distances ($d = 3, 5, 7$) to assess whether the optimal configuration varies with code size.

\begin{figure}[htb]
\centering
\includegraphics[width=0.5\textwidth]{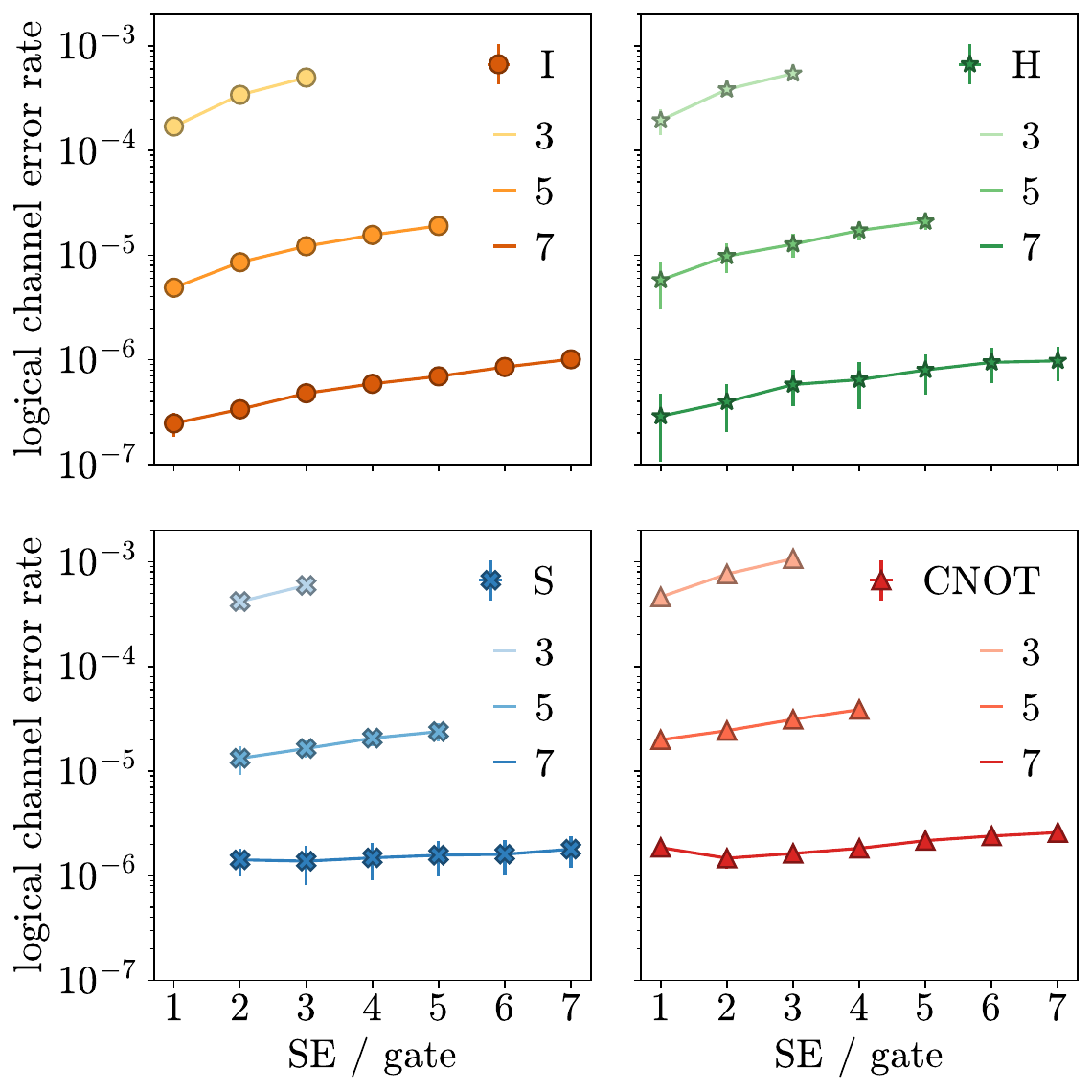}
\caption{ \textbf{Logical error rate versus number of SE rounds per logical gadget}. Results shown for $I$, $H$, $S$, and CNOT gates at code distances $d = 3, 5, 7$ simulated at a projected physical error rate lower limit of $7.3 \times 10^{-4}$. Error rates were estimated using the extraction process described in Section \ref{sec:extraction}. Single SE round per gadget consistently yields the lowest error rates across all configurations, with performance degrading as SE rounds increase. 
}
\label{fig:Fig12_SE_rounds}
\end{figure}

Our simulations reveal that, across all examined logical gate types and code distances, using a single SE round per gadget consistently outperforms configurations with multiple SE rounds, as showcased in Fig.~\ref{fig:Fig12_SE_rounds}. This achieves both lower logical error rates—due to fewer error-prone SE operations—and improved resource efficiency through reduced circuit depth and ancilla measurements. While the single-SE configuration proves optimal across all tested scenarios, the magnitude of improvement varies depending on the logical gate type. For CNOT and $S$ gates, the error suppression gains from using fewer SE rounds are somewhat modest compared to other gate types. We attribute this gate-dependent variation to our suboptimal decoding.

These gadget-level findings reaffirm and extend the circuit-level results established in Ref.~\cite{Cain2025}, demonstrating that minimizing SE rounds per logical gate does not compromise the decoder's ability to correct errors while simultaneously reducing the overhead from syndrome extraction operations.

\section{Physical noise model}
\label{sec:physicalnoise}

This appendix provides detailed specifications for the hardware-derived~\cite{Evered2023, Bluvstein2024, Rodriguez2024} noise model introduced in Section \ref{sec:hardware_noise_model}. We implement the following error channels in Stim~\cite{Gidney2021} to capture the key noise characteristics of neutral atom systems. For the neutral atom platform, there are five fundamental operations required for quantum computation: initialize qubits into stabilizer states, apply single- and two-qubit gates, move qubits spatially, and read out qubits in a specified Pauli basis. At the circuit level, we model the noise of each process using Pauli channels to enable efficient classical simulation. We use Stim for these simulations, modeling each process as follows:

\paragraph{Initialization.} Since Stim cannot introduce errors directly into initialization instructions, we model initialization as a perfect reset gate followed by a depolarizing noise gate with $p=(3/2) \times p_{\text{init}}$. This accounts for the fact that only two of the three Pauli errors affect the initialized state (e.g., for a $\ket{0}$ state, only $X$ and $Y$ errors have an effect).

\paragraph{Single-qubit gate.} Each single-qubit gate is followed by a biased Pauli channel, reflecting the $Z$-bias characteristic of neutral atom hardware errors. We use noise levels corresponding to local single-qubit gates applied to individual qubits, which exhibit higher error rates than global single-qubit operations applied simultaneously across a region, as this better reflects our specific circuit implementation.

\paragraph{Two-qubit gates.} The native two-qubit gate is the controlled-Z (CZ) gate, implemented via Rydberg blockade~\cite{Evered2023}. To execute CZ gates between arbitrary qubit pairs, one qubit from each target pair must first be transported into proximity with its partner. We model each transport operation with a uniform duration of approximately $200 \, \mu$s, regardless of distance—this represents the average transport time during syndrome extraction cycles~\cite{Rodriguez2024}, which account for the majority of CZ operations in typical quantum error correction protocols.

Once target pairs are positioned, they are subjected to a global laser pulse that simultaneously implements CZ gates on all nearby pairs. This parallelization reduces control and execution complexity compared to local addressing of each pair. However, the laser pulse also affects nearby ``spectator" qubits that are not participating in gate operations. Moving these spectators outside the entangling zone (where the laser acts) and returning them afterward could incur extra time and error overhead, and thus our spectator error provides a solid bound to such errors.
Our noise model therefore includes: (i) a two-qubit Pauli channel for each CZ gate pair, (ii) a single-qubit Pauli channel for each spectator qubit affected by the global pulse, and (iii) a $Y$-error channel on all qubits in the entangling zone to model loss errors~\cite{Evered2023, Bluvstein2025b} that occur when atomic traps are temporarily disabled during Rydberg state transitions. 

While all qubits in the entangling zone are affected by laser pulses, we model correlated errors only between pairs of proximate qubits. This is realistic due to good spatial isolation, and thus there is good interaction isolation of distinct pairs of qubits participating in gates. Thus, the architectural choices ensure that correlated errors remain predominantly local, justifying our pairwise correlation model. While some residual correlations may exist beyond nearest neighbors, this pairwise correlation model captures the dominant error mechanisms while remaining computationally tractable for large-scale simulations.

\paragraph{Qubit transport.} As described above, neutral atom qubits must be moved to position target pairs for two-qubit gates. Each transport operation introduces its own error and carries a risk of atom loss from the optical trap. We model transport errors using a biased single-qubit Pauli channel followed by a $Y$-error channel representing the associated loss errors (see main text \ref{sec:hardware_noise_model} for justification of the $Y$-error approximation). 

\paragraph{Readout.} We flip each measurement with a probability $p_{\text{readout}}$.

\begin{table}[ht]
\centering
\caption{physical noise model implemented in Stim for current error rates ($p_{\mathrm{ph}} \approx 7.3 \times 10^{-3}$). This  physical error rate represents the combined error probability associated with implementing a CZ gate, including transport operations and all associated error channels (see Eq.~\eqref{eq:physical_error_rate}).}
\resizebox{\columnwidth}{!}{
\begin{tabular}{lccp{2.8cm}}
\toprule
\textbf{Operation} & \textbf{Noise channel} & \textbf{Strength}  \\
\midrule

initialization & depol. & $p=3\times10^{-3}$  \\

1q gate & depol. & $p=3\times10^{-3}$  \\

atom move & Pauli & $[4\!\times\!10^{-5},\, 4\!\times\!10^{-5},\, 2\!\times\!10^{-4}]^\dag$  \\

atom move loss & $Y$-error  &   $p = 2\!\times\!10^{-5}$    \\

CZ gate & Pauli & $[5\!\times\!10^{-4},\, 6.3\!\times\!10^{-4},\, 1.3\!\times\!10^{-3}]^\ddag$   \\

CZ-spectator & Pauli & $[5\!\times\!10^{-4},\, 5\!\times\!10^{-4},\, 1.3\!\times\!10^{-3}]^\dag$  \\

CZ loss & $Y$-error  &   $p= 1.3\!\times\!10^{-3}$   \\

measurement & bit-flip & $ p = 4\!\times\!10^{-3}$  \\
\bottomrule
\end{tabular}
}

\vspace{10pt}

\dag~$[p_X,p_Y,p_Z]$, \quad
\ddag~the two-qubit Pauli channel has 15 terms. First value is for single $X$ or $Y$ errors on either qubit ($p_{IX}, p_{XI}, p_{IY}, p_{YI}$). Second value is for single $Z$ errors on either qubit $(p_{IZ}, p_{ZI})$. Third value is for the correlated ZZ error $p_{ZZ}$. All other correlated error terms are zero.

\vspace{-4pt}
\label{tab:Tab1_noise_model}
\end{table}

\paragraph{Combined physical error rate.} 
Throughout this work, we parameterize the noise strength by a single combined physical error rate $p_{\mathrm{ph}}$ that captures the total error probability associated with implementing a CZ gate, including the necessary atom transport. As described above, both processes contribute noise and loss channels: transport introduces a single-qubit Pauli channel (total probability $p_{\mathrm{move}}$) and atom loss modeled as a $Y$-error channel ($p_{\mathrm{move,loss}}$), while the CZ gate introduces a two-qubit Pauli channel (total probability $p_{\mathrm{CZ}}$, summing all 15 non-identity terms) and atom loss modeled as a $Y$-error channel ($p_{\mathrm{CZ,loss}}$). We define:
\begin{equation}
    p_{\mathrm{ph}} = p_{\mathrm{CZ}} + 2 \, p_{\mathrm{CZ,loss}} + p_{\mathrm{move}} + p_{\mathrm{move,loss}},
    \label{eq:physical_error_rate}
\end{equation}
where the factor of 2 on the CZ loss term accounts for the fact that loss can occur independently on each of the two qubits during the gate, whereas transport affects only the single qubit being moved. For the parameters in Table~\ref{tab:Tab1_noise_model}, this yields $p_{\mathrm{ph}} \approx 7.3 \times 10^{-3}$. This combined metric is used as the $x$-axis in figures throughout this work (e.g., Fig.~\ref{fig:Fig4_Clifford_composability}).

\subsection{Validation of an effective-\texorpdfstring{$Y$}{Y} loss model}
\label{sec:loss_bound_validation}

In this appendix, we provide the rationale for modeling atom loss using an effective Pauli-$Y$ error channel in the circuit-level noise model described in Sec.~\ref{sec:hardware_noise_model}, and we validate that this approximation yields a tight upper bound on the logical error rate. Our goal is twofold: (i) to establish an implementation-agnostic bound that is independent of the specific loss-detection, qubit-replacement, syndrome-extraction circuit design, or decoding strategy employed in a given neutral-atom architecture, and (ii) to enable fast, scalable numerical simulations of logical performance.

Atom loss in neutral-atom quantum processors occurs predominantly during Rydberg-mediated operations, when optical trapping potentials are temporarily disabled~\cite{Evered2023, Bluvstein2024}. A variety of hardware- and circuit-level strategies have been proposed to detect and mitigate such loss events, differing in (i) how loss is detected, (ii) when detection occurs, (iii) how frequently lost qubits are replaced, (iv) the quantum state in which replacement qubits are reinitialized, and (v) the structure of the syndrome-extraction circuit. Explicitly modeling all such possibilities would require strong assumptions about implementation details that are not essential for assessing achievable logical performance.

To avoid these assumptions while remaining within a Pauli-only noise framework, thereby preserving efficient stabilizer-based simulation, we replace each atom-loss event by an effective Pauli-$Y$ error acting on the affected qubit. This choice is motivated by the principle that errors supplying more syndrome information to the decoder are generally easier to correct, provided the decoder is capable of exploiting this information. Loss events that are detected—either directly or via measurement—typically provide partial or complete location information, constraining the set of possible error locations in space and time~\cite{baranes2025}. Such information reduces decoding ambiguity relative to fully unlocated Pauli errors.

Within the class of Pauli errors, $Y$ errors are maximally informative for CSS codes, as they simultaneously trigger both $X$- and $Z$-type stabilizers. Replacing loss events (which carry partial location information) with $Y$ errors, therefore, yields a pessimistic but controlled upper bound on the logical error rate. This bound is not fully tight in practice, as commonly used approximate decoders do not fully exploit the correlated syndrome information associated with $Y$ errors; nevertheless, it provides a conservative and implementation-independent performance estimate.

\begin{figure}[!htb]
\centering
\includegraphics[width=0.5\textwidth]{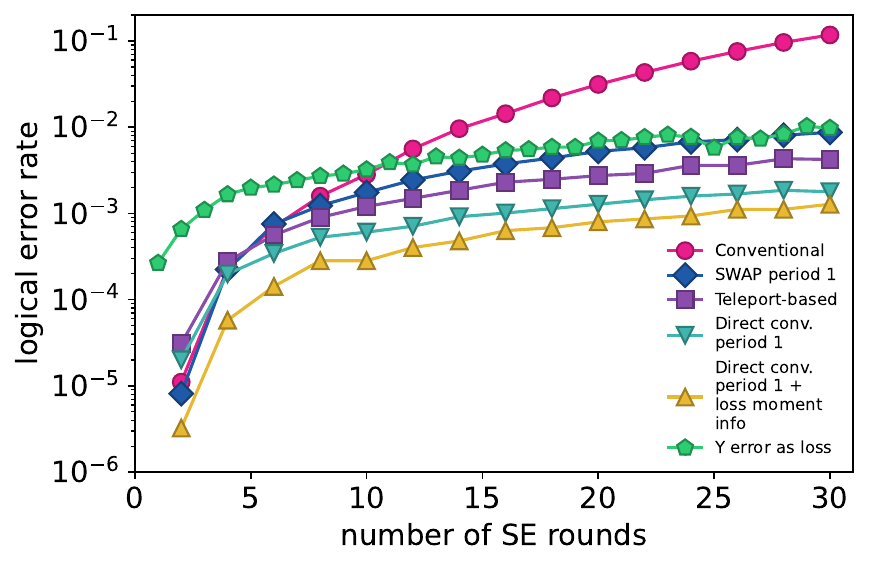}
\caption{\textbf{Validation of the effective-\texorpdfstring{$Y$}{Y} loss model against explicit-loss simulations.}
Logical error rate versus number of syndrome-extraction (SE) rounds for a $d=7$ rotated surface code at physical error rate $p = 1\%$ and loss fraction of $0.5$. Explicit-loss protocols from Ref.~\cite{baranes2025}, decoded using a delayed-erasure decoder, include: SSR-based conventional SE (pink circles), SWAP-based SE with period 1 (blue diamonds), teleportation-based SE (purple squares), direct-conversion SE with period 1 (cyan triangles), and direct-conversion SE with loss-moment information (yellow triangles). Our effective-\(Y\) loss model, implemented within a conventional SE circuit and decoded with an optimal maximum-likelihood decoder (green pentagons), provides an upper bound on the logical error rate achieved by all implementations that actively mitigate loss during the computation, validating its use as a conservative and implementation-agnostic performance estimate. The explicit loss-simulation results are reproduced with permission from Ref.~\cite{baranes2025}.}
\label{fig:loss_comparison}
\end{figure}

To quantitatively validate this bound, we benchmark our effective-$Y$ loss model against a range of explicit-loss simulations reported in Ref.~\cite{baranes2025}. These simulations directly sample atom loss events, loss-detection outcomes, and qubit replacement, and are decoded using erasure-aware delayed-erasure decoders that approximate maximum-likelihood decoding. Ref.~\cite{baranes2025} investigates several loss-aware syndrome-extraction circuits spanning both state-selective readout (SSR)–based and direct-detection paradigms. Figure~\ref{fig:loss_comparison} reproduces the reported comparison of five such protocols, showing the logical error rate as a function of the number of syndrome-extraction rounds.

Among SSR-based approaches, loss detection occurs only during qubit readout, where each measurement outcome is drawn from the set $\{0,1,\text{lost}\}$. Three such circuits are considered:

(i) \emph{Conventional} syndrome extraction, corresponding to the standard rotated surface-code circuit~\cite{Horsman2012}, in which only ancilla qubits are measured each round. In this case, data-qubit loss remains undetected, leading to a rapid increase in the logical error rate as the number of rounds increases.
(ii) \emph{SWAP-based} syndrome extraction, which periodically exchanges data and ancilla qubits, enabling regular measurement of all physical qubits and timely loss detection.
(iii) \emph{Teleportation-based} syndrome extraction, in which the logical state is repeatedly teleported onto fresh code blocks, inherently detecting loss via state-selective readout.

Two additional circuits employ direct erasure conversion, in which atom loss is detected in situ during circuit execution without measuring the affected qubit, and the loss is converted into a flagged erasure error supplied to the decoder. These protocols differ in the temporal resolution of loss detection:
(i) \emph{Direct conversion, period 1}, in which loss detection and qubit replacement are performed at the end of each syndrome-extraction round.
(ii) \emph{Direct conversion with loss-moment information}, in which loss detection is performed after each entangling gate, while qubit replacement occurs only at the end of each round.

For a small number of SE rounds ($\lesssim 10$), all protocols exhibit similar performance, as loss events have limited opportunity to accumulate. As the number of rounds increases, the SSR-based conventional circuit quickly saturates due to undetected data-qubit loss, while SWAP- and teleportation-based circuits maintain controlled logical error growth. The best performance is achieved by direct-conversion circuits, particularly when fine-grained loss-moment information is available.

On top of these results, we overlay simulations using our effective-$Y$ loss model, implemented within a conventional syndrome-extraction circuit and decoded using an optimal maximum-likelihood decoder~\cite{gurobi}. These simulations use the same circuit-level noise model and physical error rates as the explicit-loss simulations of Ref.~\cite{baranes2025}, with atom loss replaced by an effective Pauli-$Y$ error channel. The results in this comparison are for loss fraction of $ p_{\mathrm{loss}}/(p_{\mathrm{loss}} + p_{\mathrm{Pauli}})= 0.5$,
consistent with experimentally observed values in neutral-atom platforms~\cite{Bluvstein2024, Evered2023}. As shown in Fig.~\ref{fig:loss_comparison}, the effective-$Y$ model provides an upper bound on the logical error rates achieved by all loss-aware circuits, with the exception of the SSR-based conventional circuit, which does not effectively mitigate loss. The upper bound on direct loss simulation validates the effective-$Y$ loss model as a computationally efficient, implementation-agnostic tool for large-scale simulations.

In addition to loss during gate operations, atom loss can also occur during qubit readout \cite{Gibbons_readout_loss_2011}. In our simulations, such readout-induced loss is already accounted for within the same Pauli-only framework used throughout the paper, namely via classical bit-flip errors applied to the measurement outcome. A Pauli-$Y$ error occurring immediately prior to measurement flips the outcome of both the $X$- and $Z$-basis measurements employed in our circuits. Equivalently, such a pre-measurement Pauli-$Y$ error is indistinguishable from a classical bit-flip applied to the measurement result. Consequently, modeling readout-related loss as a measurement bit-flip remains fully compatible with the effective-$Y$ loss abstraction, while preserving efficient stabilizer-based simulation.

\subsection{Potential physical error rate improvements}
\label{sec:noise_improvements}

Throughout the work, we assume error rate improvements on the order of $\sim 10 \times$ better than the reported surface code threshold (Fig.~\ref{fig:Fig3_Clifford_noise}) with the hardware-inspired physical noise model of Tab.~\ref{tab:Tab1_noise_model} possible for neutral atom platforms. This assumption is based on the most recent experiments on such hardware, where a path towards similar improvements in error rate is outlined~\cite{Bluvstein2025b}. In fact, in the methods section, the authors of Ref.~\cite{Bluvstein2025b} provide an argument on why operation $\sim 8 \times$ below threshold is possible with known hardware improvements and better calibration. Here, we briefly repeat some of their main arguments for completeness:
\begin{itemize}
    \item Local single-qubit rotation errors in Ref.~\cite{Bluvstein2025b} were already reported at $10^{-3}$. Further increasing the Raman intermediate-state detuning would reduce scattering and miscalibration errors during such rotations, which expect to lead to a local single-qubit rotation gate fidelity of $10^{-4}$.
    \item Idling, rotation gate (especially global), and Pauli errors during atom moves are all tied strongly to the hyperfine coherence times of the atoms. The experiments of Ref.~\cite{Bluvstein2025b} have worked with $\sim 1$s coherence times, while coherence times in excess of $10$s~\cite{Manetsch2025} have already been demonstrated, providing a route for improving all of the coherence-dependent errors. Furthermore, move loss errors can be mitigated by calibration improvements, and AOD power adjustments~\cite{Bluvstein2025b}. The part of the errors coming from background collisions (imperfect vacuum) could be reduced more than $60 \times$ compared to the reported performance of Ref.~\cite{Bluvstein2025b}.
    \item Measurement error with the similar measurement protocols to the ones applied in Ref.~\cite{Bluvstein2025b} have been previously reported at $4 \times 10^{-3}$~\cite{Gross2019}.
    \item Two-qubit gate performance could be improved by several hardware upgrades~\cite{Bluvstein2025b}. In particular, an increase in the laser power and applied magnetic field strength can reduce intermediate state-related undesired couplings and decrease scattering (and ultimately, loss) through faster gate execution. Ref.~\cite{Bluvstein2025b} analysis points out that with $\sim 4 \times$ laser power increase, total two-qubit gate errors (Pauli and loss errors together) can be reduced to $ \sim 1.5\times 10^{-3}$. These improvements also apply to spectator error. The error channels are themselves likely dominated by loss error in this regime.
\end{itemize}

Finally, after these improvements, the majority of the remaining loss channels would correspond to loss during two-qubit gate execution. As discussed in detail in Sec.~\ref{sec:loss_bound_validation}, our estimates are an upper bound on the effect of such a noise channel. Recently proposed~\cite{baranes2025} and implemented~\cite{Bluvstein2022} schemes can utilize the information provided by loss-resolving readout and decoding together with atom-replacement based syndrome extraction protocols to effectively increase the typical surface code threshold by $\sim 2 \times$. Such an increase is equivalent to the decrease in associated physical error rates, providing a direct justification for $\sim 10 \times$ error rate improvements for the largest error channels associated with two-qubit gates.  All such loss-resolving protocols have as an additional benefit the replacement of each qubit with a fresh one after on average $\sim 4$--$8$ clock cycles. This strongly mitigates any performance degradation caused by accumulated atom heating from repeated moves and gates and removes the need for associated recooling steps. The need for constant qubit replacement necessitates continuous reloading of freshly initialized atoms to the array, a technology recently demonstrated at scale with neutral atoms~\cite{Gyger2024, Chiu2025}.

\section{Logical noise model extraction methodology}
\label{app:extraction}

To address the challenges of error modeling in logical circuits using correlated decoding, we develop a systematic approach to extract logical noise parameters for individual gadgets. In our framework, a gadget consists of a logical gate followed by a constant number of syndrome extraction rounds (independent of the code distance)—the fundamental building block of these circuits. While we illustrate this process for CNOT gadgets, the same approach applies to any Clifford logical gate, enabling comprehensive characterization of the complete gate set. Our methodology addresses two key challenges: insufficient syndrome data (gadgets with a constant number of syndrome extraction rounds lack information for reliable independent decoding) and temporal boundary effects (gadgets deep within circuits could behave differently from those at the temporal boundaries).

Our extraction method isolates the noise contribution of each gadget via a three-step process:

\textbf{Step I: Isolate gadget errors via circuit subtraction.} We construct two test circuits [Fig.~\ref{fig:Fig2_noise_circuit_sketch}(b)]: 
\begin{itemize}
    \item \textit{Base circuit}. The base circuit initializes two logical qubits in a specified Pauli basis, performs $d$ rounds of syndrome extraction, applies $d$ noiseless CNOT gates, performs another $d$ rounds of syndrome extraction, and measures in the appropriate basis.
    \item \textit{Full circuit}. The full circuit replaces the noiseless CNOTs with $d$ repetitions of the target gadget (gate plus syndrome extraction).
\end{itemize}
The $d$ SE rounds before and after isolate the gadget block from temporal boundaries, mimicking conditions deep within a circuit. The $d$-fold gadget repetition amplifies the error signal and provides sufficient syndrome data for correlated decoding.

Since both circuits evolve stabilizer states through Clifford operations, the final state is characterized by its stabilizer group. By measuring in the appropriate Pauli basis, we effectively determine which logical stabilizers have been violated. For a two-qubit stabilizer state with two independent stabilizer generators, we represent the logical stabilizer violations as a binary vector $\textbf{s} \in \{0,1\}^2$, where $s_i=1$ indicates that the $i$-th generator was violated. Note that these violations arise from logical errors—the cumulative effect of physical errors throughout the circuit that lead to wrong decoding—rather than individual physical error events.

Through repeated simulations, we estimate the probability of observing $s$ for each circuit $p^{(\text{base})}_{\mathbf{s}}$ and $p^{(\text{full})}_{\mathbf{s}}$. The block error contribution is then isolated via 
\begin{equation}
     p^{(\text{block})}_{\mathbf{s}}  \approx p^{(\text{full})}_{\mathbf{s}} - p^{(\text{base})}_{\mathbf{s}},
\end{equation}
where this approximation holds well for low error rates.

\textbf{Step II: Connect measured probabilities to channel parameters.} We model the measured $d$-gadget block as $d$ ideal gates each followed by a Pauli error channel. For two-qubit gates like CNOT, we use a two-qubit Pauli channel with 15 free parameters; for single-qubit gates, we use a single-qubit Pauli channel with 3 free parameters. Each channel parameter is characterized by:
\begin{itemize}
    \item A Pauli operator $\alpha \in \{I,X,Y,Z\}^{\otimes2 (1)} \backslash {I \otimes I}$.
    \item An index $i \in \{1, \dots, d \}$ indicating it occurs after the $i$-th gate.
\end{itemize}
Let $p_{\alpha,i}$ denote the probability that Pauli error $\alpha$ occurs after gate $i$. Each such error contributes to exactly one pattern of logical stabilizer violations, determined by its commutation relations with the final stabilizer generators. We denote by $\text{syn}(\alpha, i)$ the pattern that results when error $\alpha$ at position $i$ propagates through the remaining gates.

For small error rates, the block error probabilities relate to the channel parameters, $p_{\alpha,i}$, as
\begin{equation}
p^{(\text{block})}_{\mathbf{s}} \approx \sum_{\substack{\alpha,i: \\ \text{syn}(\alpha,i) = \mathbf{s}}} p_{\alpha,i} + O(p^2)
\end{equation}

\textbf{Step III: Extract logical Pauli channel via linear solver.} By repeating the protocol across all stabilizer product states (3 for single-qubit gates, 9 for two-qubit gates), we obtain a sufficient number of linearly independent relations between block error rates, $p^{(\text{block})}_{\mathbf{s}}$, and channel parameters, $p_{\alpha,i}$, to uniquely determine the channel parameters. we employ a Non-Negative Least Squares (NNLS) solver \cite{lawson1974} to robustly extract the Pauli channel parameters while ensuring physical constraints (non-negative probabilities) are satisfied. This approach naturally handles inconsistent systems by finding the least-squares solution that minimizes the residual error. While we assume low error rates to enable linearization, higher-order terms could be included if needed. For our experimental regime, we find empirically that the linear approximation is sufficient.

\section{Logical noise model details}
\label{sec:logical_noise}

This appendix presents the extracted logical noise models for Clifford operations in the surface code. The noise parameters were obtained using the extraction process described in Section \ref{sec:extraction}, applied to logical gadgets consisting of each gate followed by a single syndrome extraction round. The extracted parameters in Table~\ref{tab:Tab2_logical_noise} reveal a notable feature: for single-qubit gates, logical $Y$ errors are suppressed to negligible levels compared to $X$ and $Z$ errors. This behavior is expected from the surface code structure, where the $Y$-distance equals $2d$ (compared to distance $d$ for $X$ and $Z$ errors). Since the smallest weight $Y$-logical operator requires $2d$ independent physical errors while $X$ and $Z$ logical operators require only $d$ errors, $Y$-logical errors are less likely than their $X$ and $Z$ counterparts.

To characterize the scaling behavior of logical error rates with physical error rate and code distance, we fit the CNOT gate data to the standard error suppression model:
\begin{equation}
p_{\text{L}} \approx A \left(\frac{p_{\text{ph}}}{p_{\text{c}}}\right)^{\lceil(d+1)/2 \rceil}
\end{equation}
The fitting procedure yields a threshold error rate of $p_{\text{c}} = 8.5 \times 10^{-3}$ and a prefactor $A = 0.0579$. This threshold value is slightly below the typical surface code thresholds of approximately $1 \times 10^{-2}$ reported in the literature \cite{Stephens2014}. The reduced threshold reflects our use of a realistic, hardware-specific noise model that incorporates the error characteristics of neutral atom hardware, beyond the simplified depolarizing noise models commonly used in previous studies. The fit is used to extrapolate the expected transversal STAR performance in Fig.~\ref{fig:Fig8_STAR_limits}.

\begin{table}[htbp]
\centering
\caption{Extracted logical noise model for Clifford operations in distance-7 surface code. Simulations performed at physical error rate $p_{\text{ph}} \approx 7.3 \times 10^{-4}$ (10$\times$ improvement over current experimental noise rates).}
\vspace{2pt}
\resizebox{\columnwidth}{!}{
\begin{tabular}{lccp{2.8cm}}
\toprule
\textbf{Operation} & \textbf{Noise channel} & \textbf{Strength}  \\
\midrule

$I$ & Pauli & $[1\!\times\!10^{-7}, 0,\, 1\!\times\!10^{-7}]^\dag$   \\

$H$ & Pauli & $[1\!\times\!10^{-7},0,\, 3\!\times\!10^{-7}]^\dag$  \\

$S$ & Pauli & $[4.6\!\times\!10^{-7},\,0,\, 1\!\times\!10^{-6}]^\dag$  \\

$CZ$  & Pauli & $[2\!\times\!10^{-7},\, 7\!\times\!10^{-7},\, 1\!\times\!10^{-7},\, 6\!\times\!10^{-7},\, 2\!\times\!10^{-7}]^\ddag$   \\
\bottomrule
\label{tab:Tab2_logical_noise}
\end{tabular}
}
\dag~$[p_X,p_Y,p_Z]$, \quad
\ddag~Five-element vector representing non-zero error probabilities in a two-qubit Pauli channel. Elements correspond to: [IX, XI, IZ, ZI, ZZ]. All other Pauli error terms are zero.

\vspace{-4pt}

\end{table}

\section{Transversal multi-rotation injection protocol simulation and optimization}

Here, we present the details of the ensemble Clifford simulation method used for transversal multi-rotation injection protocol error rate estimation, as well as supplementary injection error rate data.

\subsection{Ensemble Clifford simulation}
\label{sec:fid_est_TMR}
Simulating the transversal multi-rotation (TMR) protocol requires simulation beyond the Clifford simulation, as each shot of simulation is time-consuming. Instead, based on the method from Ref.~\cite{Toshio2024}, we can estimate the logical error rate using the Clifford simulation. First, after the $q$ rounds of initialization shown in Fig.~\ref{fig:Fig5_injection_circuit_sketch}, the logical code could be left with $E_i$, the physical error occurring during the $q$ rounds of initialization. This process can be sampled using Clifford simulation. For the multi-rotation part, the logical state can be separated into two parts. The first part represents the target state with some error propagated from $E_i$ or an additional $E_r$ from transversal rotations. The other part contains all undesired logical angle states, again with corresponding sampled noise. Each of the wrong-angle components contributes to the coherent noise channel of the TMR injection. Fortunately, since $E_r Z_L E_i$ and $E_r Z_s E_i$ with $Z_s$ denoting the $Z$-error terms in Eq.~\eqref{eq:R_z} have the equivalent syndrome outcomes as the target state and wrong-angle non-Clifford states respectively, one can simulate an ensemble of Clifford states instead by sampling the target and wrong-angle states according to their corresponding probabilities $p^{(t)}_{\theta^*}\equiv \cos^{2k}(\frac{\theta^*}{2})+\sin^{2k}(\frac{\theta^*}{2})$ and $p^{(n)}_{\theta^*,j}\equiv\cos^{2(k-j)}(\frac{\theta^*}{2})\sin^{2j}(\frac{\theta^*}{2})+\sin^{2(k-j)}(\frac{\theta^*}{2})\cos^{2j}(\frac{\theta^*}{2})$. 
After the final $2$ rounds of postselection (PS) or partial error correction (EC) syndrome measurements, the target state and each of the wrong-angle states sample $E_m E_r Z_L E_i$ and $E_mE_r Z_s E_i$ chains, respectively. Note that the number of $Z_s$ strings in the ensemble that one needs to simulate in general grows exponentially with $k$. One can roughly estimate the total success rate based on the numerical experiment and evaluate the corresponding infidelity as
\begin{align}
\label{eq: inf_o}
     1-\mathcal{F}(\theta) \approx \frac{\sum_j^{\left \lfloor\frac{k}{2}\right \rfloor}N_{j}\sin^2\left ( \Delta_j \right )}{N_\text{tot}}
\end{align}
where $N_{\text{tot}}$ ($N_j$) is the number of shots for all states (wrong-angle rotating states) that pass through $2$ rounds of PS or EC syndrome measurements, $\Delta_j=\theta-\theta_j$ is the difference between target angle $\theta$ and wrong-angles $\theta_j$
\begin{align}
      \frac{\theta_j}{2}=
      \resizebox{.85\hsize}{!}{$\sin^{-1}\left ( \frac{(-1)^j\sin^{(k-j)}(\frac{\theta^*}{2})\cos^{j}(\frac{\theta^*}{2})}{\sqrt{\cos^{2(k-j)}(\frac{\theta^*}{2})\sin^{2j}(\frac{\theta^*}{2})+\sin^{2(k-j)}(\frac{\theta^*}{2})\cos^{2j}(\frac{\theta^*}{2})}} \right )$}.
\end{align}
In our protocols, PS denotes postselection on all logical stabilizers in the respective syndrome rounds, while EC (during final $p$ syndrome extraction rounds only) denotes postselection only on stabilizers connected to the physical qubits where TMR is applied, and error correction on the rest.

We directly simulate each state with $N_s$ shots and get their success rates $P_{t}$ and $P_{n,j}$ respectively. Multiplying the success rates by their respective state component probabilities, we can evaluate the total success rate:
\begin{align}
\label{eq:Suc_r}
     P_{\text{tot},\theta^*}=P^{(t)}p^{(t)}_{\theta^*}+  \sum_j^{\left \lfloor\frac{k}{2}\right \rfloor} P^{(n)}_{j}p^{(n)}_{\theta^*,j}.
\end{align}

\begin{figure}[t]
\centering
\includegraphics[width=1\linewidth]{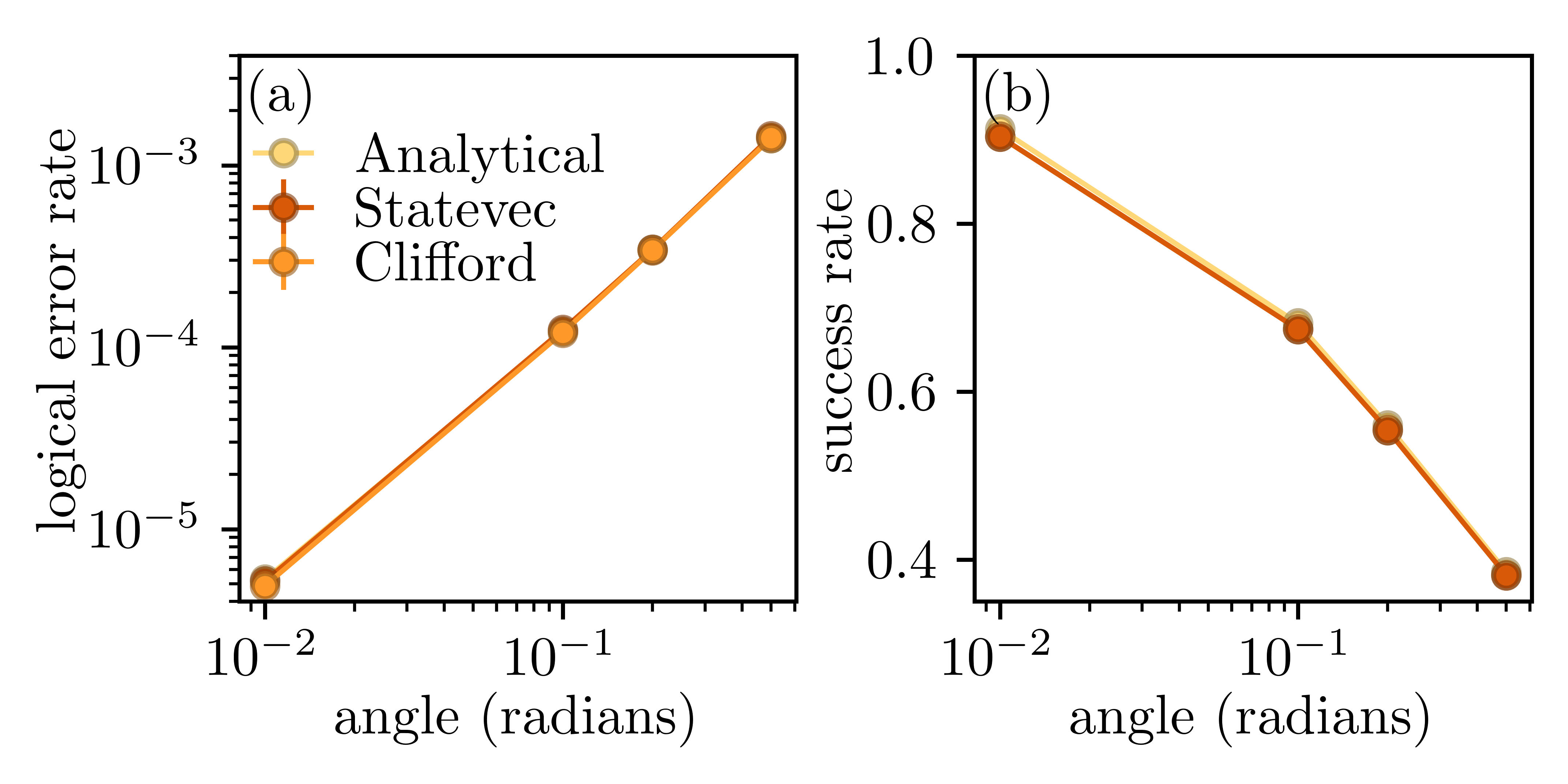}
\caption{ \textbf{Validity of using Clifford circuit simulation to estimate logical error.} The plots manifest the comparison among the Clifford circuit simulation, the statevector, and analytical calculation for $d=3$ and $k=3$ multi-rotation with perfect postselection. (a) The logical error and (b) the success rate versus the rotating angles.
}
\label{fig:Fig13_validating_injection}
\end{figure}

However, the corresponding infidelity in Eq.~\eqref{eq: inf_o} ignores the combination of the residual errors and TMR errors. Thus, we construct a more comprehensive noise model to describe TMR noise. The corresponding noise model is given by
\begin{align}
\begin{split}
\label{eq:rn_model}
      \mathcal{N}_{\theta,\text{tot}}(\rho)=&P^{(t)}_{\theta^*} \mathcal{N}_{t}\left [ \mathcal{R}_{Z,L,\theta}(\rho) \right ]\\&+\sum_j^{\left \lfloor\frac{k}{2}\right \rfloor} P^{(n)}_{\theta^*,j} \mathcal{N}_{j}\left [ \mathcal{R}_{Z,L,\theta_j}(\rho) \right ],
\end{split}
\end{align}
where
\begin{align}
      P^{(t)}_{\theta^*}=\frac{P^{(t)}p^{(t)}_{\theta^*}}{P_{\text{tot}}}, \qquad      P^{(n)}_{\theta^*,j}&=\frac{P^{(n)}_{j}p^{(n)}_{\theta^*,j}}{P_{\text{tot}}}
\end{align}
 are the probabilities of a noisy target state, and a wrong-angle state passing to the end of the protocol, respectively. $\mathcal{N}_{t}$ and $\mathcal{N}_{j}$ are the target angle state and wrong-angle state's noise channels
\begin{align}
\label{eq:t_model}
      \mathcal{N}_{l}(\rho)=\left ( 1-\epsilon_{\text{tot,l}} \right )\rho+\epsilon_{X,l} X\rho X+\epsilon_{Y,l} Y\rho Y+\epsilon_{Z,l} Z\rho Z  
\end{align}
where $l\in \{t,j\}$, and $\epsilon_{\text{tot},l}=\epsilon_{X,l}+\epsilon_{Y,l}+\epsilon_{Z,l}$, while $\mathcal{R}$ is the rotation channel:
\begin{align}
\label{eq:rot_channel}
      \mathcal{R}_{Z,L,\theta}(\rho) = R_{Z,L}(-\theta)\rho R_{Z,L}(\theta).
\end{align}

To estimate the fidelity for preparing the small-angle magic state $\ket{m_\theta}_L$, we plug $\rho=\ket{+}_L\bra{+}_L$ into Eq.~\eqref{eq:rn_model} and compare the corresponding noisy state with $\ket{m_\theta}_L$. Since logical $Y$ errors require the occurrence of both logical $X$ and $Z$ errors, $\epsilon_{Y,l}$ is relatively small compared to $\epsilon_{X,l}$ and $\epsilon_{Z,l}$. Moreover, the logical $X$ error in the state $\ket{m_\theta}_l$ only induces $\epsilon_{X,t}\sin^2(\theta)+\sum_j^{\left \lfloor\frac{k}{2}\right \rfloor}\epsilon_{X,j}\sin^2(\frac{\theta+\theta_j}{2})$ error, which is subleading compared to other contributions to infidelity. Finally, the corresponding infidelity can be approximately given by 
\begin{align}
\begin{split}
\label{eq:inf_3}
      &1-\mathcal{F}\approx P^{(t)}_{\theta^*} \epsilon_{Z,t}\\&+\sum_j^{\left \lfloor\frac{k}{2}\right \rfloor} P^{(n)}_{j,\theta^*}\left [ (1-\epsilon_{\text{tot},j})\sin^2(\Delta_j)+\epsilon_{Z,j}\cos^2(\Delta_j)\right ],
\end{split}
\end{align}
as quoted in Sec.~\ref{sec:fid_Rz}. With the generalized noise model for TMR injection, we are able to further explore the optimal way for TMR injection.

To examine the validity of the Clifford circuit simulation for TMR, we execute the state-vector simulation for the $k=3$ noisy multi-rotation on $d=3$ surface code with the following perfectly full postselection and inverse perfect rotation to calculate the corresponding logical error and the success rate. In addition, in this case, one can analytically estimate the probability of the wrong-angle rotation passing through PS or EC as $P^{(n)}_{j,\theta^*}\approx p_{\mathrm{ph},Z}p^{(n)}_{j,\theta^*}$, with $p_{\mathrm{ph},Z}$, the Pauli $Z$ error rate of the physical rotation gate. Thus, the corresponding infidelity can be estimated as $\approx p_{\mathrm{ph},Z}p^{(n)}_{j,\theta^*}\sin^2(\Delta_j)$.

Fig.~\ref{fig:Fig13_validating_injection} shows the corresponding logical error rate and the success rate that vary with the rotation angle using different methods. The Clifford circuit simulation result is consistent with the statevector simulation and the analytical estimate. The results validate TMR logical error rate estimation through ensemble Clifford circuit simulation.

\begin{figure}[!htb]
\centering
\includegraphics[width=1\linewidth]{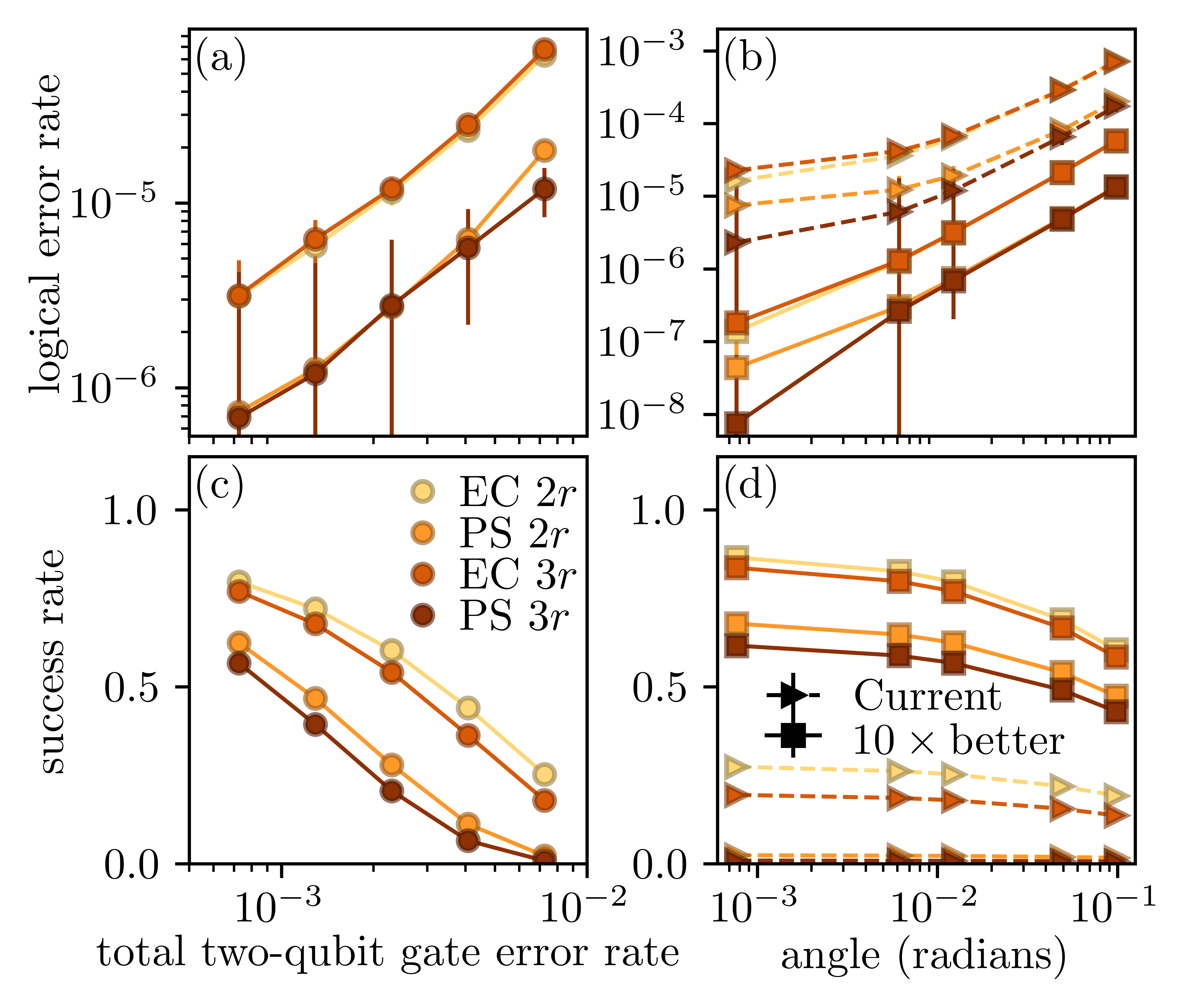}
\caption{\textbf{The transversal multi-rotation injection performance with varying protocol settings.} The upper panel: The logical fidelity of injection for 2 and 3 rounds (represented by 2r and 3r respectively) of initialization with $k=3$ full postselection (PS) or partial error correction (EC). TMR performance is shown as a function of (a)  the noise scale at fixed angle $10^{-2}$ radians (solid lines with circle data points) and (b) the injection angle at a chosen noise scale, with total two-qubit physical error at rate $7.3\times 10^{-3}$ (current noise scale manifested by dashed lines with triangle data points) and $7.3\times 10^{-4}$ (solid lines with square data points). The lower panel: The success rate of injection with $k=3$ varies with (c) the noise scale at fixed rotating angle $10^{-2}$ radians (solid lines) and (d) the rotating angle at fixed noise scale for different rounds of initialization. The error bars for different initialization are generated from the standard error of $40$ Clifford circuit simulations that use $10^{9}$ shots each. The data is generated for the isolated TMR gadget, with the method described in subsection~\ref{sec:iso_TMR}. 
}
\label{fig:Fig14_injection_fidelity_init}
\end{figure}

Here, we note the significance of considering the application-relevant worst-case error metrics for coherent error evaluation during injection, such as trace-distance or full logical channel characterization as noted in Ref.~\cite{Toshio2024}. For example, if only total injection fidelity is used, one can imagine small-angle rotation approximated by $\ket{+}_L$ state, with very low infidelity of $\theta^2/4$. In comparison, the coherent part of the TMR injection error improves on this infidelity by an extra factor of $p_{\mathrm{ph}, Z}$, as $1-\mathcal{F}\approx p_{\mathrm{ph},Z}p^{(n)}_{j,\theta^*}\sin^2(\Delta_j)\sim p_{\mathrm{ph}, Z} \theta^2/4$ , $\sum p^{(n)}_{j,\theta^*}<1$, and typically $\Delta_j\sim \theta/2$. The advantage of TMR injection is thus clear, with the $p_{\mathrm{ph},Z}$ suppression steming from actively trying to rotate the logical state and succeeding with $1-p_{\mathrm{ph},Z}$ probability. The advantage is in practice in highly structured circuits with repeated small-angle gates; there, $\ket{+}_L$ coherent errors can dangerously accumulate.  The trace distance provides a more realistic error estimate of the single injection impact in the scenario with error accumulation~\cite{Toshio2024}, realistic in Trotterized quantum simulation circuits. Indeed, the $\ket{+}_L$ approach has trace distance $\theta/2$ from the desired small-angle $\theta$, while the coherent part of the TMR injection achieves a trace distance of $\sim p_{\mathrm{ph}, Z} \theta/2$. We circumvent the problem of fidelity underestimating the impact of coherent error by evaluating the full error channel for TMR injection and using it in teleportation simulations. We also make sure that the reported final teleportation fidelity reported in Sec.~\ref{sec:teleportation_fid} does not substantially depend on the metric (fidelity or trace distance) used, and is thus representative of the transversal STAR performance in the quantum simulation context.

\subsection{Hyperparameters of transversal multi-rotation protocol}
\label{app:hyperparameters_TMR}

In order to ensure the performance of the TMR injection is optimized given the hardware noise model, it is critical to explore different initialization ($q$) and postselection ($p$) syndrome round settings. To showcase this, here we choose to present the injection data for $q=2$ and $q=3$ rounds of initialization for the $d=5$ surface code and $k=3$ TMR injection with the $p=2$ rounds of partial error correction (EC) or full postselection (PS). We have evaluated a substantially larger set of injection hyperparameters in order to obtain hardware-noise-optimized protocols shown in Sec.~\ref{sec:inj_fid}.

The representative performance dependence on protocol hyperparameters is presented in Fig.~\ref{fig:Fig14_injection_fidelity_init}. Manifestly, the postselection outperforms partial error correction by roughly a factor of $4$--$6$ in the different regimes explored. This comes with a modest impact on success rates, especially in the lower error rate regime, and thus, while a trade-off between EC and PS can be considered depending on the task and physical error rates at hand, PS is preferred closer to STAR performance limits. The extra round of initialization between $q=2$ and $q=3$ does not improve the result as much. The saturation showcases that $q=3$ is enough to achieve equilibrium between syndrome extraction and postselection (or EC) for code distances and regimes considered. The success rate cost of an extra syndrome round is modest, another manifestation of protocol saturation. Decreasing to $q=1$, on the other hand, degrades TMR performance significantly.

The yield of arbitrary-angle magic-state injection in the STAR protocol is governed by postselection and therefore depends exponentially on the underlying physical error rates. The postselection probability factorizes into an initialization-dependent contribution and an angle-dependent contribution. The initialization contribution is dominated by faults during syndrome extraction, when two-qubit gate errors dominate, scales as $\sim (1-p)^{ 4 q d^{2}}$, thus being strongly dependent on hyperparamters $q$ and $p$. At physical error rates comparable to current experiments, this scaling leads to postselection probabilities at the percent level for $d=7$, even before accounting for the additional suppression associated with small-angle injection, in line with reported yields in Fig.~\ref{fig:Fig6_injection_fidelity}. The second, TMR injection factor in the yield depends on the angle, and the $k\sim d$ TMR pieces, with $\sim \theta^{k}$ dependence, together with additional $\sim p^{d}$ correction.

Due to this exponential scaling, a reduction of physical error rates by one order of magnitude drastically increases yields, resulting in $>50\%$ magic-state yields at $d=7$ (see Fig.~\ref{fig:Fig6_injection_fidelity}). As discussed in Sec.~\ref{sec:noise_improvements}, such an improvement is consistent with experimentally motivated hardware upgrades in neutral-atom platforms. Furthermore, the yield can be increased by several methods. For surface code, this includes partial error correction, reducing the number of initialization rounds ($q$), or both, with several representative results showcased in Fig.~\ref{fig:Fig14_injection_fidelity_init}. Finally, the effective yield could be significantly increased in the high-rate code STAR architecture. There, a single successful initialization can serve multiple logical injections (up to $k$), thus amortizing the dominant initialization factor.

\subsection{The effect of noise model on analog rotation performance}
\label{sec:noise_model_TMR}

\begin{figure}[t]
\centering
\includegraphics[width=0.6\linewidth]{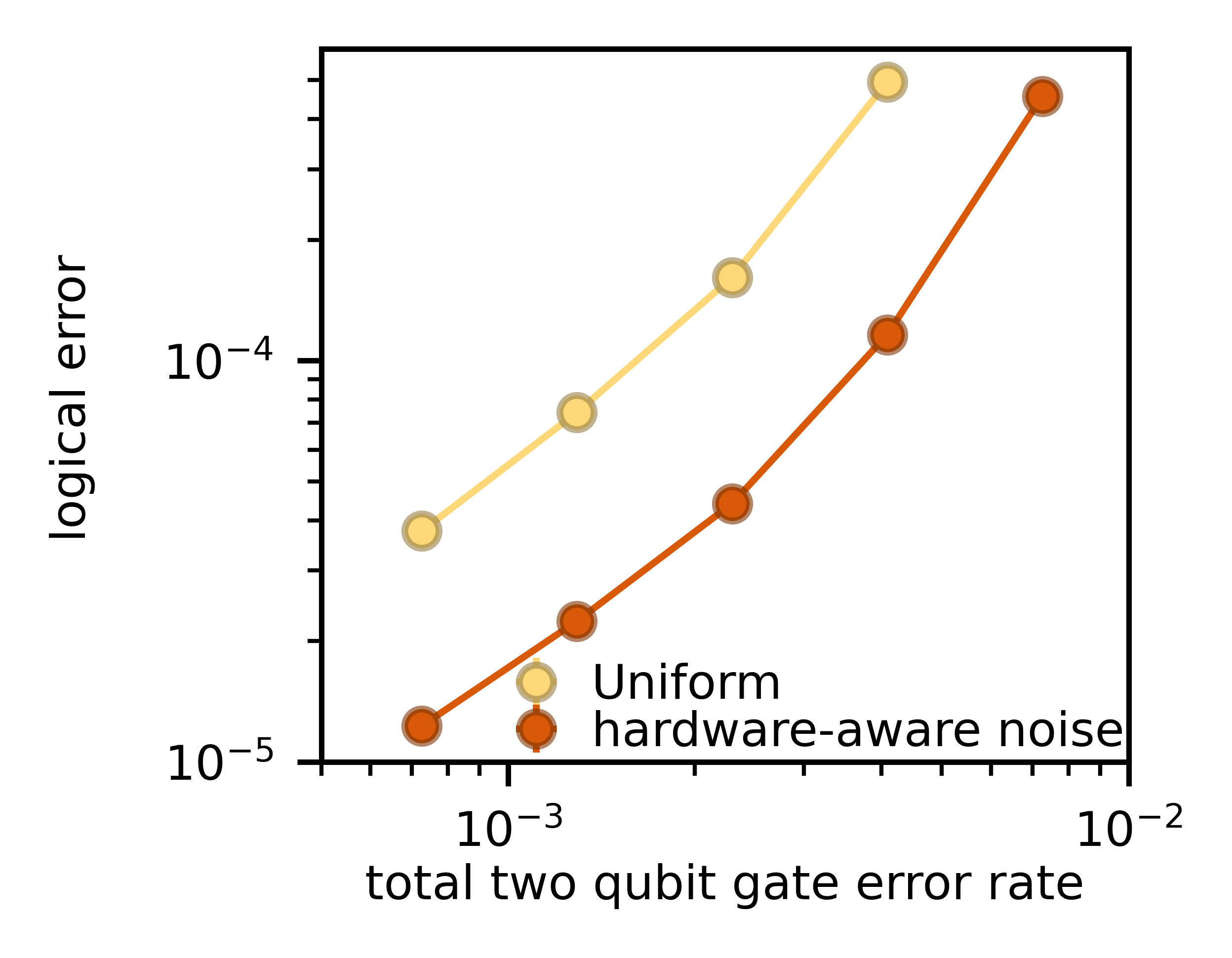}
\caption{ \textbf{The comparison between uniform and hardware-aware noise model.} The logical error rates for end-to-end STAR teleportation are reported as a function of the effective total two-qubit error rate at $\theta=10^{-2}$ radians. The uniform noise model includes uniform depolarizing noise for all gates, measurement error, and initialization error.
}
\label{fig:Fig_uniform}
\end{figure}

A commonly used noise model for evaluating quantum error correction performance, including in the original STAR proposals~\cite{Akahoshi2024, Toshio2024} is the uniform Pauli noise. There, three single-qubit Pauli channels are assigned an error rate of $p/3$ each, all fifteen two-qubit Pauli channels are assigned an error rate of $p/15$, with measurement and initialization also characterized by a $p$-rate flip channel. Here, we compare the performance of the teleportation protocol under this common approximation to the hardware-aware noise model used throughout the paper. The main result is presented in Fig.~\ref{fig:Fig_uniform}. The uniform noise model performs $\sim 3\times$ worse than the hardware-aware noise model. This difference can largely be attributed to the uniform approximation overestimating the size of single-qubit errors. The uniform noise model performs noticeably worse, despite the unfavorable $Z$ bias of the hardware-aware noise and the additional error sources it lacks. $Z$ bias is mitigated by the hyperparameter choices made during TMR injection, in particular, the postselection syndrome round number and distribution, compared to the results of Ref.~\cite{Toshio2024}. Any improvement (or difference) in logical arbitrary-angle injection error rates in the transversal setting over fixed-connectivity STAR is not intrinsic to the transversal architecture itself, but is instead driven by the choice of noise model and the optimization of the injection protocol hyperparameters.            

\subsection{Isolated transversal-multirotation gadget}
\label{sec:iso_TMR}
In order to analyze the limitations imposed by the small-angle injection, we estimate the logical error rate of the isolated transversal-multirotation gadget.  The subtraction-based results could be additionally relevant for early experiments demonstrating TMR protocol error scaling.

\begin{figure}[t]
\centering
\includegraphics[width=1\linewidth]{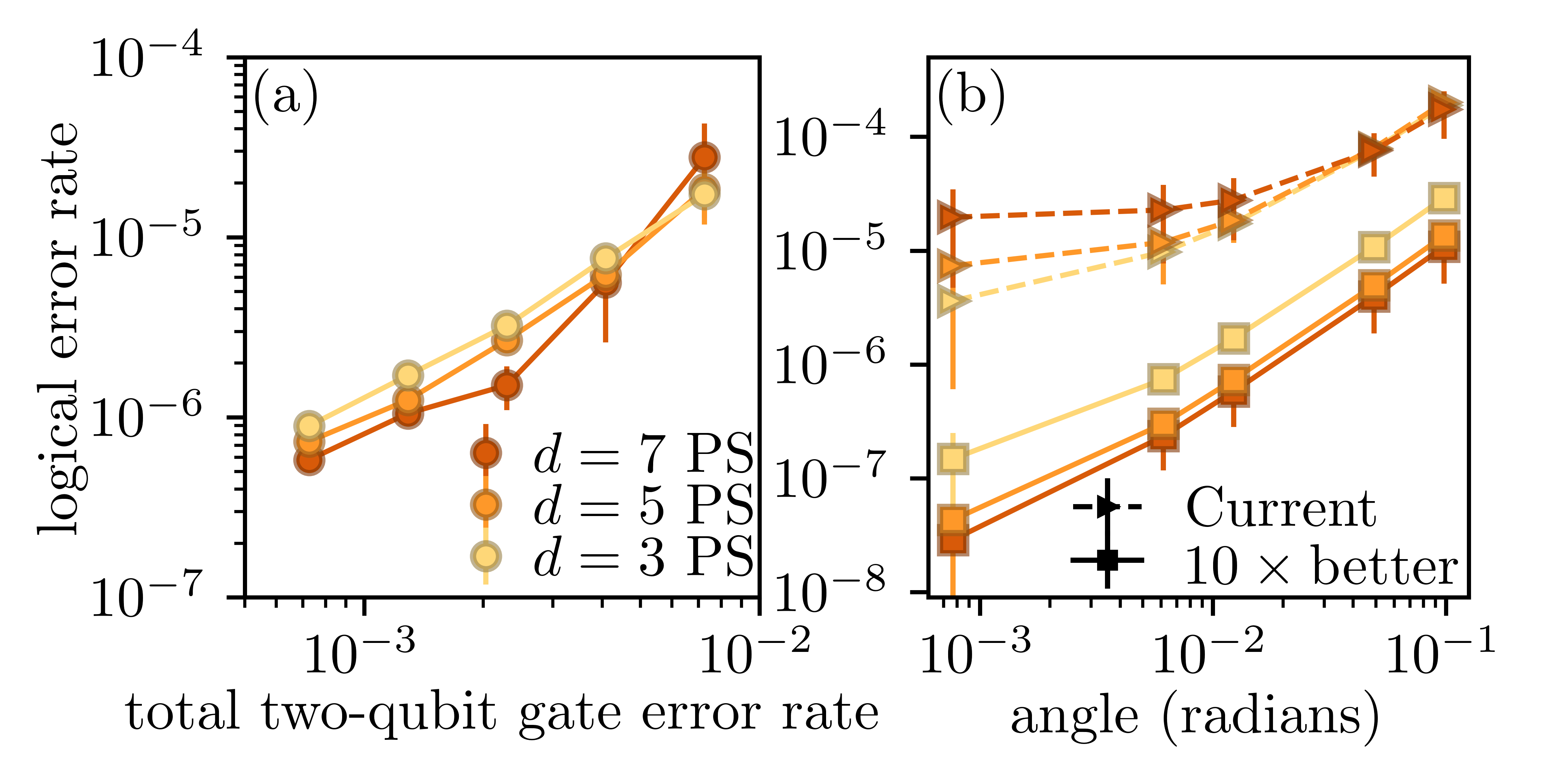}
\caption{ \textbf{The isolated transversal-multirotation gadget logical error rates.} Logical error rate of the isolated TMR gadget is shown for several code distances ($d=3,5,7$) with $k=3$ full postselection (PS) TMR protocol at (a) the fixed angle $10^{-2}$ radians (solid lines with circle data points) as a function of total two-qubit physical error rate and (b) at effective two-qubit physical error rates of $7.3\times 10^{-3}$ (current noise scale manifested by dashed lines with triangle data points) and $7.3\times 10^{-4}$  (full lines with square date points) as a function of angle. The error bars for $d=5,7$ ($d=3$) are generated from the standard error of $40$ ($6$) Clifford circuit simulations that use $10^{9}$ ($10^{10}$) shots each.
}
\label{fig:Fig15_diff_d}
\end{figure}

The postselection-based initialization prevents us from doing a fully isolated subtraction (Sec.~\ref{sec:extraction}, but we add an additional $d-q$ rounds of error correction to the TMR circuit from Fig.~\ref{fig:Fig5_injection_circuit_sketch}, to isolate the gadget from the measurement boundary. In order to estimate the partially-isolated TMR logical error rate, we also use such a padded end-to-end injection circuit and subtract the the circuit with no TMR rotations performed (initialization and final postselection and error correction syndrome rounds only), leading to:
\begin{align}
\label{eq:fid_sub}
     \varepsilon_{r,i,l}=\frac{\varepsilon_{br,i,l}-\varepsilon_{b,i}}{1-2\varepsilon_{b,i}}
\end{align}
where $\varepsilon_{br,i}$ ($\varepsilon_{b,i}$) is the logical error rate of the full (base) circuit for the $i\in(X,Y,Z)$ initial basis, and $\varepsilon_{r,i,l}$ is the logical error rate of TMR for the $i$ basis. The relationships between the Pauli error rates and the logical error rates shown in Eq.~\eqref{eq:fid_sub} are $\varepsilon_{Z,l}=\epsilon_{r,X,l}+\epsilon_{r,Y,l}$, $\varepsilon_{X,l}=\epsilon_{r,Y,l}+\epsilon_{r,Z,l}$, and $\varepsilon_{Y,l}=\epsilon_{r,X,l}+\epsilon_{r,Z,l}$.

Fig.~\ref{fig:Fig15_diff_d} visualizes the success rate of the isolated TMR gadget for different code distances. The logical error rates at $d=7$ have a similar order of magnitude as for the end-to-end gadget, showing that our initialization and postselection settings are well optimized.  The dependence on the code distance, however, is drastically suppressed, showcasing that the code protection is critical in the initialization and postselection syndrome measurement rounds. The TMR protocol is, as expected, the limiting factor in the ultimate performance of arbitrary angle injection. However, at larger physical error rates and smaller code distances, relevant for current hardware, the code protection during syndrome extraction is the limiting factor. Still, the experiments performing the equivalent of the subtraction simulated here show potential to probe and validate basic principles of TMR injection, with the neutral atom hardware available at present~\cite{Rodriguez2024, Bluvstein2025b}.

\section{Comparison between  architectures for 2D transverse field Ising model simulation}
\label{sec:space-time_comparison}

In Table~\ref{tab:resource_summary}, we summarize the physical space-time costs of performing a single Trotter step in a 2D transverse field Ising model simulation with different QEC architectures considered in this work: the surface-code transversal STAR, the fixed connectivity STAR~\cite{Akahoshi2024, Toshio2024}, fixed-connectivity cultivation-based fully fault-tolerant surface code implementation~\cite{Gidney2024}, and high-rate-code transversal STAR. Here, we provide the details and assumptions behind these estimates.

\subsection{Task definition and decomposition}

The task used for architecture comparison is a single first-order Trotter step of nearest-neighbor, square lattice, transverse field Ising model simulation (2D TFIM). The lattice size assumed is $N_q=8\times 8$ spins. The Trotter step is decomposed into two global ($N_q$ gates each) $H$ gate layers, four global ($N_q/2$ gates each) CNOT layers, one global ($N_q$ gates each) rotation layer, and 4 rotation layers acting on half of the qubits ($N_q/2$ gates each). This decomposition can be used directly for estimating STAR architecture resources. For the fully fault-tolerant architecture, each rotation is synthesized. The gate fidelity and synthesis cost requirements are based on the main text estimates relevant to the megaquop simulation of a local Hamiltonian, aiming for $N_{\mathrm{tr}}\approx 10^4$ Trotter steps. We assume direct Ross-Selinger~\cite{Ross2016} synthesis, with the number of $T$ gates required for angle synthesis at STAR-comparable accuracy given by  $N_{\mathrm{syn}} \approx -3\log_2{(\alpha p_{\mathrm{ph}}/\sqrt{l_1})}$. With simulated $\alpha \approx 1.5$ and the Hamiltonian norm per site for 2D TFIM $l_1=2+h/J$, where $h/J$ is the ratio of transverse field to Ising coupling, and $p_{\mathrm{ph}}=10^{-3}$, we get typical $N_{\mathrm{syn}}\approx 30$ $T$ gates per rotation. The gate fidelities required for the task, given the number of Clifford and synthesis layers, are $\sim 10^{-6}$ for analog rotation, and $\sim 10^{-7}$ for Cliffords and $T$ gates.

\subsection{Architecture layout, logical gadgets, and space costs}
\label{sec:estimate_gadgets}

The gate fidelities required can be reached with a surface code of $d=11$ at $p_{\mathrm{ph}}=10^{-3}$ (see Appendix~\ref{sec:logical_noise}), which is the code distance assumed for surface-code transversal STAR, fixed-connectivity STAR, and fully fault-tolerant architectures. All architectures are evaluated for the number of factory qubits equal to the number of data qubits $N_{fq}=N_q$, representing the high-factory parallelism regime. The time and space cost estimates can be appropriately scaled for different factory space allocations for a sizeable range ($\sim 5 \times$) around $N_{fq}=N_q$.

The transversal surface-code STAR architecture assumes a layout of qubits adapted to the 2D nearest-neighbor geometry at hand (see Fig.~\ref{fig:Fig1_STAR}). The shuttling time costs for such a layout are included within the costs of one clock cycle, as the local shuttling required is below the time cost of ancilla atom moves required for syndrome extraction. Thus, each Clifford is assumed to cost 1 clock cycle. Single TMR injection factory attempt has a cost of 4 clock cycles. The factory success rate is assumed to be 50\%, in line with numerical simulations shown in Fig.~\ref{fig:Fig6_injection_fidelity}. Total space cost for STAR, including logical data and factory qubits (ancillas not counted, see Appendix~\ref{sec:noise_improvements}) is thus 15488 physical qubits.

The fixed-connectivity STAR layout requires a significant amount of space overhead to achieve the Hamiltonian connectivity. In the high-factory parallelism regime, two logical routing ancillas are assigned to each data and factory logical qubit, enabling the effective use of factories and providing sufficient space for efficient lattice surgery connectivity, as discussed in detail in Ref.~\cite{Akahoshi2024}. While lower routing overheads are possible (down to $\sim 1.5\times$), they come at the cost of factory and algorithm serialization. With this layout, the $H$ gate can be performed with $3d$ clock cycles, while CNOT with qubits swapped to the right orientation and with clear routing path costs $2d$ cycles~\cite{Akahoshi2024}. To minimize algorithmic routing rearrangements, units of 4 logical qubits, comprising one data qubit, one factory, and one routing qubit, can be placed on a square lattice that matches the Hamiltonian connectivity. This layout enables the necessary routing reordering that happens between each of the entangling layers in the circuit with $\sim 2d$ clock cycles per reordering. Each TMR injection attempt costs 3 clock cycles, which is lower than the transversal STAR due to the possibility of limited reuse of routing ancillas for higher RUS angle injection~\cite{Akahoshi2024}. The factory success rate is again assumed to be 50\%, while each teleportation attempt itself costs an additional $2d$ clock cycles required for the lattice surgery CNOT, and $2d$ cycles for routing lattice surgery.  Total space cost for the fixed-connectivity STAR is thus 30976 physical qubits, due to routing overhead.

The fully fault-tolerant architecture considered shares many similarities with fixed-connectivity STAR. The layout, routing overhead, Clifford operations speed, and total space cost are the same. The difference arises from the $T$ factory speed and synthesis time costs. Cultivation is sufficient at the required $\sim 10^{-7}$ error rate, and we assume fixed-connectivity appropriate cultivation from Ref.~\cite{Gidney2024}. The total space-time volume for reaching this fidelity with $d_1=5$ cultivation protocol at $p_{\mathrm{ph}}=10^{-3}$ is $20000$ qubit rounds~\cite{Gidney2024}. Assuming factory parallelism, including multiple small factories within a single $d=11$ patch during the early stages of cultivation, the average time per produced $T$ state for one factory is estimated at $\sim 80$ clock cycles. Each synthesis requiring $N_{\mathrm{syn}}\approx 30$ $T$ gates is assumed to also cost 30 single-qubit Clifford gates, each equivalent to a Hadamard cost of $3d$ clock cycles. The cost of teleportation for each $T$ gates is $2d$ for CNOT and $2d$ for routing lattice surgery

Finally, we turn to the high-rate transversal STAR architecture specification. The code choice presents multiple options, but in the absence of a detailed performance evaluation for transversal STAR, we perform our estimates with the recently proposed self-dual bivariate-bicycle (BB) codes in mind~\cite{Xu2025batched}. This code family is particularly well suited for the task, given the full set of parallel, transversal Cliffords and the direct cyclic automorphisms: all these operations take $\sim 1$ cycle. The number of cyclic shifts assumed for the algorithm is 2 shifts (forwards and return) for each entangling layer, and thus 8 shift layers per Trotter step. We based our estimates specifically on the code with $[[78,6,10]]$ parameters, from the known codes in this family. Based on extrapolation of memory performance estimates from Ref.~\cite{Xu2025batched}, $d=10$ could be sufficient in terms of reaching $\sim 10^{-7}$ logical error rate per qubit encoded. The number of encoded qubits is not congruent with the lattice size in the problem definition, as the codes with $k$ that is a multiple of 4 would be preferred, but no self-dual BB codes of the required distance have been reported so far. Still, our space estimates, based on $[[78,6,10]]$ parameters, will reach the correct order of magnitude, and the space compression ratio will generally apply to congruent system sizes. In this case, the problem size dictates 1664 physical qubits for the required 128 logical qubits (factories included). However, the parallel and transversal high-rate STAR operations require running two separate but parallel computations in each of the homological and cohomological logical qubit subsets within the code block. With the self-dual BB code, the subset sizes are equal, and the minimal computation requires a total of 3072 physical qubits. In Table~\ref{tab:resource_summary}, we refer to both physical qubit numbers.

\begin{table*}[ht]
\centering
\renewcommand{\arraystretch}{1.25}
\begin{tabular}{c|c|c|c|c}
\hline \hline
&\makecell{fixed connectivity \\ fully fault-tolerant}& \makecell{fixed connectivity  \\ STAR from Ref.~\cite{Toshio2024}} &   transversal STAR & high-rate transversal STAR \\
\hline
$c_1$ & 30$d$ & 30$d$ & 10 & 18  \\
\hline
$s$ & 30 & 2 & 2 & $\approx \lceil \log_2(k)\rceil+ 1$ \\
\hline
$f$ & 1 & 1 & 1 & 1 \\
\hline
$i$ & 80 & 6 & 8 & $\approx [16/k,8]$ \\
\hline
$c_2$ & 7$d$ & 4$d$ & 1 & 1 \\
\hline\hline
\end{tabular}
\caption{\textbf{Values of the time cost model parameters across architectures.} The parameters necessary to evaluate the time cost model of Eq.~\ref{eq:time_cost_model} for one Trotter step of 2D TFIM simulation across different architectures. The values are based on the discussion in Appendix~\ref{sec:estimate_gadgets}, and are used to produce the architecture comparison reported in Tab.~\ref{tab:resource_summary}.}
\label{tab:parameters_estimates}
\end{table*}

The largest uncertainty in terms of the high-rate transversal STAR architecture estimates are the logical error rates and postselection rates of injection factories. Self-dual BB codes enable parallel injection on $k$ qubits due to their non-overlapping logical representatives~\cite{Xu2025batched}. We assume that the logical error rate per produced magic angle in this parallel injection is on par with that of surface code transversal STAR, which, although left for future work, is plausible due to similar code thresholds and $k$-parallelism on the circuit of a similar physical scale. The average factory speed depends critically on the postselection rate. The joint code initialization for $k$ qubits at the time implies that the major part of the postselection rate does not suffer decay due to multiple injections running simultaneously. The part of the postselection that does decay with $k$ is largely governed by the injected angle size and can be mitigated by choosing less disjoint coverings (lower $l$) to some extent, as discussed in Sec.~\ref{sec:qldpc_injection}. Still, given the sensitivity of postselection rates, we take a conservative range $[2,k]$ for the speed of the parallel injection compared to the surface code case. Per angle produced, the number of clock cycles is thus in the range $[8/k, 4]$. Finally, with parallel injection and joint teleportation, additional consideration to be made is the expected number of RUS tries ($t_{\mathrm{RUS}}$) until full-patch teleportation success is achieved. Since each teleportation succeeds with a rate of 50\%,  each qubit has a chance of RUS failure after $t$ cycles of $1/2^t$, and the whole patch succeeds in $t$ cycles with probability $p_s(t)=(1-1/2^t)^k$. The expected cycle length scales asymptotically as $t_{\mathrm{RUS}} =\sum_tp_s(t) \approx \log_2k+ 0.83$, while the expected result for $k=6$ is  $t_{\mathrm{RUS}}\approx 4$, compared to $k=1$ case where $t_{\mathrm{RUS}}\approx 2$. In the more detailed estimate, the expected RUS time and injection postselection rates require joint consideration. However, the overall modest logarithmic overhead in terms of RUS cycle length in the high-rate transversal STAR case will hold.

\subsection{Time cost model}
\label{sec:time_cost_model}

With the time cost of each logical gadget defined, we combine them to estimate the total expected clock cycle time for one Trotter step of 2D TFIM simulation. Our time cost model is as follows:
\begin{equation}\label{eq:time_cost_model}
    T_t=c_1+\frac{3s}{f}(i+c_2),
\end{equation}
where:
\begin{itemize}
    \item $c_1$ (algorithmic Clifford cost) represents the cost of the Clifford gate layers in clock cycles stemming from the simulation circuits, including the routing/shifting costs for each architecture.
    \item $s$ (synthesis cost) counts the expected length of RUS teleportation or $T$ synthesis before the desired rotation on a qubit.
    \item $f$ (factory ratio) denotes the ratio between the number of factory qubits and logical data qubits allocated. All architectures have $f=1$ for comparison, but the cost model can account for different values of $f$.
    \item $i$ (injection cost) is the number of clock cycles it takes on average for a single factory to produce one magic state.
    \item $c_2$ (auxiliary Clifford cost) counts the additional Clifford-related clock cycles during teleportation (CNOT gates and routing), including synthesis-related Cliffords for the fully fault-tolerant case.
\end{itemize}
The values of each relevant parameter across architectures based on the discussion in Sec.~\ref{sec:estimate_gadgets} are presented in Tab.~\ref{tab:parameters_estimates}.


\input{references.bbl}

\end{document}

%% file: references.bbl
%